\documentclass[american,5p]{elsarticle}
\usepackage[T1]{fontenc}
\usepackage[latin9]{inputenc}
\usepackage{geometry}
\usepackage{xcolor}
\usepackage{pdfcolmk}
\usepackage{babel}
\usepackage{mathrsfs}
\usepackage{amsmath}
\usepackage{amssymb}
\usepackage{graphicx}
\PassOptionsToPackage{normalem}{ulem}
\usepackage{ulem}
\usepackage[unicode=true,
 bookmarks=true,bookmarksnumbered=false,bookmarksopen=false,
 breaklinks=false,pdfborder={0 0 0},pdfborderstyle={},backref=false,colorlinks=true]
 {hyperref}

\makeatletter

\newcommand{\noun}[1]{\textsc{#1}}
\providecommand{\tabularnewline}{\\}
\providecolor{lyxadded}{rgb}{0,0,1}
\providecolor{lyxdeleted}{rgb}{1,0,0}
\DeclareRobustCommand{\lyxsout}[1]{\ifx\\#1\else\sout{#1}\fi}

\usepackage{pslatex}
\renewcommand{\boldsymbol}[1]{\pmb{#1}} 
\usepackage{mathrsfs}
\usepackage{mathtools, cuted}
\usepackage{fancyhdr}
\usepackage{babel}

\@ifundefined{showcaptionsetup}{}{%
 \PassOptionsToPackage{caption=false}{subfig}}
\usepackage{subfig}
\makeatother

\begin{document}
\begin{frontmatter}

\title{Grand-potential-based phase-field model of dissolution/precipitation: lattice Boltzmann simulations of counter term effect on porous medium}

\author[CEA]{\noun{Téo Boutin}}

\ead{teo.boutin@cea.fr}

\address[CEA]{Université Paris-Saclay, CEA, Service de Thermo-hydraulique et de Mécanique des Fluides, 91191, Gif-sur-Yvette, France.}

\author[CEA]{\noun{Werner Verdier}}

\ead{werner.verdier@cea.fr}

\author[CEA]{\noun{Alain Cartalade}\corref{cor1}}

\ead{alain.cartalade@cea.fr}

\cortext[cor1]{Corresponding author. Tel.:+33 (0)1 69 08 40 67}

\begin{abstract}

Most of the lattice Boltzmann methods simulate an approximation of
the sharp interface problem of dissolution and precipitation. In such
studies the curvature-driven motion of interface is neglected in the
Gibbs-Thomson condition. In order to simulate those phenomena with
or without curvature-driven motion, we propose a phase-field model
which is derived from a thermodynamic functional of grand-potential.
Compared to the free energy, the main advantage of the grand-potential
is to provide a theoretical framework which is consistent with the
equilibrium properties such as the equality of chemical potentials.
The model is composed of one equation for the phase-field $\phi$
coupled with one equation for the chemical potential $\mu$. In the
phase-field method, the curvature-driven motion is always contained
in the phase-field equation. For canceling it, a counter term must
be added in the $\phi$-equation. For reason of mass conservation,
the $\mu$-equation is written with a mixed formulation which involves
the composition $c$ and the chemical potential. The closure relationship
between $c$ and $\mu$ is derived by assuming quadratic free energies
for the bulk phases. The anti-trapping current is also considered
in the composition equation for simulations with null solid diffusion.
The lattice Boltzmann schemes are implemented in \texttt{LBM\_saclay},
a numerical code running on various High Performance Computing architectures.
Validations are carried out with analytical solutions representative
of dissolution and precipitation. Simulations with or without counter
term are compared on the shape of porous medium characterized by microtomography.
The computations have run on a single GPU-V100.

\end{abstract}

\begin{keyword}

Phase-field model, Grand-potential, Lattice Boltzmann method, Dissolution/Precipitation,
porous media, \texttt{LBM\_saclay} code.

\end{keyword}

\end{frontmatter}

\section{\label{sec:Introduction}Introduction}

The Lattice Boltzmann Equation (LBE) \citep{Book_LBM2017} is an attractive
method to simulate flow and transport phenomena in several areas of
science and engineering. Because of its local collision term and its
ease of implementation of the bounce-back method, the LBE has been
extensively applied in porous media literature for simulating two-phase
flows and transport at pore scale \citep{Pan-Luo-Miller_CF2006,Genty-Pot_TRT_TiPM2013,Pot_etal_AdWR2015}
(see \citep{Review_LBM_PorousMed_IJHMT2019} for a recent review).
When the surface of separation $\Gamma_{sl}$ between solid ($s$)
and liquid ($l$) does not depend on time, it is sufficient to identify
the nodes located at the interface and to apply the bounce-back method.
However, when physico-chemical processes occur on the surface of solid,
such as those involved in matrix dissolution or pore clogging, it
is necessary to consider the free-boundary problem because the interface
position $\Gamma_{sl}(t)$ is now a function of time. The general
sharp interface model of dissolution and precipitation without fluid
flows writes:

\begin{subequations}

\begin{align}
\frac{\partial c}{\partial t} & =D_{\Phi}\boldsymbol{\nabla}^{2}c & \text{in }\Gamma_{\Phi}(t)\label{eq:ADE}\\
(c-c_{s})v_{n} & =-D_{l}\boldsymbol{\nabla}c\cdot\boldsymbol{n}\bigr|_{l}+D_{s}\boldsymbol{\nabla}c\cdot\boldsymbol{n}\bigr|_{s} & \text{on }\Gamma_{sl}(t)\label{eq:CL1_ConservMass}\\
G(c) & =-d_{0}\kappa-\beta v_{n} & \text{on }\Gamma_{sl}(t)\label{eq:CL2_EquivGibbsThomson}
\end{align}
Eq. (\ref{eq:ADE}) is the mass conservation of solute in bulk domains
$\Gamma_{\Phi}(t)$ (where $\Phi=s,\,l$), $c$ is the composition,
$D_{\Phi}$ is the diffusion coefficient of liquid ($\Phi=l$) and
solid ($\Phi=s$). Although $D_{s}$ is supposed to be zero in most
of the dissolution studies, two diffusion coefficients $D_{l}$ and
$D_{s}$ are considered for mathematical reasons. In Section \ref{subsec:Discussion_Matched-Asymptotics},
we will see the necessity of using an anti-trapping current in the
phase-field model when $D_{s}=0$. Two conditions hold at the interface
$\Gamma_{sl}(t)$. The first one (Eq. (\ref{eq:CL1_ConservMass}))
is the balance of advective and diffusive fluxes where $v_{n}$ is
the normal velocity of interface. In that equation the right-hand
side is the difference of diffusive fluxes between liquid and solid,
$\boldsymbol{n}$ is the unit normal vector of interface pointing
into the liquid, and $c_{s}$ is the composition of the solid phase.
The second interface equation (Eq. (\ref{eq:CL2_EquivGibbsThomson}))
is the Gibbs-Thomson condition that relates the driving force $G(c)$
(left-hand side) to the interface motion (right-hand side). In literature,
the most common form of $G(c)$ is proportional to the difference
between the interface composition $c_{i}$ and the solid composition
$c_{s}$: $G(c)\propto(c_{i}-c_{s})$. Two terms contributes to the
interface motion: the first one is the curvature-driven motion $-d_{0}\kappa$
where $\kappa$ is the curvature and $d_{0}$ is a capillary length
coefficient. The second term is the normal velocity $-\beta v_{n}$
where $\beta$ is a kinetic coefficient representing the dissipation
of energy.

\end{subequations}

In literature, the lattice Boltzmann methods often simulate an approximation
of that sharp interface problem. In \citep{Kang_etal_WRR2007}, the
equilibrium distribution functions are designed to fulfill the mass
conservation at the interface (Eq. (\ref{eq:CL1_ConservMass})). However,
the method only simulates an approximation of the Gibbs-Thomson condition
because the curvature term is neglected ($-d_{0}\kappa=0$). For instance
in \citep{KANG20141049}, Eq. (\ref{eq:CL2_EquivGibbsThomson}) is
replaced by an evolution equation of the volume fraction of solid:
the time variation of the mineral volume $V$ is related to the reaction
flux by $\partial_{t}V=-V_{m}A(c-c_{s})$ \citep[Eq. (5)]{KANG20141049}
where $V_{m}$ is the molar volume of mineral and $A$ is the product
of solid area times a kinetic coefficient. The model has been applied
recently in \citep{Zhang_etal_ChemEngSci2022} for studying the influence
of pore space heterogeneity on mineral dissolution. When the surface
tension of the material can be neglected, then the assumption $-d_{0}\kappa=0$
hold. But in most cases $-d_{0}\kappa\neq0$ and the accurate position
of interface $\Gamma_{sl}(t)$ must be computed while maintaining
the two conditions Eqs (\ref{eq:CL1_ConservMass})-(\ref{eq:CL2_EquivGibbsThomson})
at each time-step.

Alternative methods exist for simulating the interface tracking problem.
In the ``phase-field method'', a phase index $\phi\equiv\phi(\boldsymbol{x},\,t)$
is introduced to describe the solid matrix if $\phi=0$ (solid) and
the pore volume if $\phi=1$ (liquid). The phase index varies continuously
between those two extreme values ($0\leq\phi\leq1$) i.e. the method
considers the interface as a diffuse zone. That diffuse interface
is characterized by a diffusivity coefficient $M_{\phi}$ and an interface
width $W$. The interface, initially a surface, becomes a volumic
region of transition between liquid and solid. The model is composed
of two coupled Partial Derivative Equations (PDEs) defined on the
whole computational domain. The first equation describes the dynamics
of the phase-field $\phi$ and the second one describes the dynamics
of composition $c\equiv c(\boldsymbol{x},\,t)$. Those two PDEs recover
the sharp interface problem Eqs. (\ref{eq:ADE})--(\ref{eq:CL2_EquivGibbsThomson})
when $W\rightarrow0$. The Gibbs-Thomson condition Eq. (\ref{eq:CL2_EquivGibbsThomson})
is replaced by the phase-field equation which contains implicitly
the curvature term $-d_{0}\kappa$. By ``implicitly'' we mean that
the phase-field models always include the curvature-driven motion
when they derive from a double-well potential.

Various phase-field models have already been proposed for simulating
the processes of precipitation and dissolution \citep{Xu-Meakin_PhaseField-Preci_JChemPhys2008,Mai_etal_CorroSci2016,Bringedal_etal_MMS2020,Gao_etal_JCAM2020}.
The main feature of those works is the model derivation from a free
energy functional $\mathscr{F}[\phi,\,c]$. The phase-field models
that derive from such a functional have been successfully applied
for solid/liquid phase change such as those encountered in crystal
growth (e.g. \citep{Karma-Rappel_PRE1998} for pure substance and
\citep{Echebarria_etal_PhysRevE.70.061604,Ramirez_etal_BinaryAlloy_PRE2004}
for dilute binary mixture). For those applications, the functional
$\mathscr{F}[\phi,\,T]$ depends on the phase-field $\phi$ and the
temperature $T$, which is an intensive thermodynamic variable. In
spite of those successes for solid/liquid phase change, an issue occurs
for models involving composition. The composition is an extensive
thermodynamic quantity and the models do not necessarily insure the
equality of chemical potentials at equilibrium. In order to fulfill
that condition, the Kim-Kim-Suzuki (KKS) model \citep{KKS_model_PhysRevE.60.7186}
introduces two fictitious compositions $c_{s}(\boldsymbol{x},\,t)$
and $c_{l}(\boldsymbol{x},\,t)$ in addition to the global composition
$c(\boldsymbol{x},\,t)$. The two PDEs are formulated in $\phi$ and
$c$ and the source term of $\phi$ depends on $c_{s}$ and $c_{l}$.
With a Newton method, those two compositions are explicitly computed
inside the interface by imposing the equality of chemical potential
\citep[p. 126]{Provatas-Elder_Book_2010}. That model has been applied
for dissolution in \citep{Mai_etal_CorroSci2016,Gao_etal_JCAM2020}.

A formulation based on the grand-potential thermodynamic functional
$\Omega[\phi,\,\mu]$ avoids that supplementary numerical stage. That
approach, proposed in \citep{Plapp_PhysRevE.84.031601}, yields a
phase-field model that is totally equivalent to the KKS model. That
theoretical framework contains the construction of common tangent
and insures the equality of chemical potential at equilibrium. In
the same way as they are derived from $\mathscr{F}[\phi,\,c]$, the
PDEs are established by minimizing $\Omega[\phi,\,\mu]$. Hence, we
retrieve the same features in the definition of $\Omega$. The density
of grand-potential is composed of two terms. The first one, noted
$\omega_{int}(\phi,\,\boldsymbol{\nabla}\phi)$, contains the standard
double-well potential and the gradient energy term of the interface.
The second one, noted $\omega_{bulk}(\phi,\,\mu)$ is an interpolation
of bulk grand-potentials $\omega_{\Phi}(\mu)$. Those latter come
from the Legendre transform of free energy densities $f_{\Phi}(c)$.
The main dynamical variables of $\Omega$ are the phase-field $\phi$
and the chemical potential $\mu\equiv\mu(\boldsymbol{x},\,t)$. The
chemical potential is the conjugate variable to $c$, and like temperature,
it is an intensive thermodynamic quantity. Whereas it is inappropriate
to make an analogy between $T$ and $c$ when deriving models, an
analogy can be done between $T$ and $\mu$. Thus, the asymptotics
are quite similar for establishing the equivalence between the sharp
interface models and the phase-field ones. That theoretical framework
is already extended to study multi-component phase transformation
\citep{Choudhury-Nestler_PRE2012}. It has been applied for dendritic
electro-deposition in \citep{Cogswell_GrdPot_PRE2015}. The capability
of grand-potential phase-field models to simulate spinodal decomposition
is presented in \citep{Aagesen_etal_GrdPot_PRE2018}. In reference
\citep{Simon_etal_GrdPot_CompMatSci2020} effects are presented of
introducing elasticity with different interpolation schemes in the
grand-potential framework.

Contrary to solidification, the curvature term $-d_{0}\kappa$ is
often neglected in models of dissolution and precipitation. For instance
in \citep{Xu-Meakin_PhaseField-Preci_JChemPhys2008}, the Gibbs-Thomson
condition simply relates the normal velocity $v_{n}$ proportionally
to $(c_{i}-c_{s})$. In \citep{Mai_etal_CorroSci2016} the normal
velocity is only equal to the Tafel's equation \citep[Eqs. (2)-(3)]{Mai_etal_CorroSci2016}.
In \citep{Bringedal_etal_MMS2020}, the curvature term appears in
the sharp interface model but the coefficient in front of the curvature
is considered very small. However, the curvature-driven motion plays
a fundamental role in the Ostwald ripening \citep{Book_Ratke-Voorhees_2002}.
The Ostwald ripening is the dissolution of matter that occurs at regions
with small radius of curvature. After diffusion of solute through
the liquid, a re-precipitation occurs at regions with larger radius
of curvature. The phenomenon originates from the difference of chemical
potentials between solid grains of different sizes which is proportional
to the surface tension and inversely proportional to the radius (i.e.
the curvature $\kappa$). The larger grains are energetically more
favorable than smaller ones which disappear in favor of bigger ones.
The same process occurs for two-phase systems composed of two immiscible
liquids. The drop of pressure is proportional to the ratio of the
surface tension over the radius (Laplace's law). The smallest droplets
disappear whereas the larger ones growth.

As already mentioned, that motion is always contained in the phase-field
model. If it is undesired in the simulations, it is necessary to add
a counter term $-M_{\phi}\kappa\bigl|\boldsymbol{\nabla}\phi\bigr|$
in the phase-field equation as proposed in the pioneer work \citep{Folch_etal_PRE1999}.
The counter term has been included for interface tracking in the Allen-Cahn
equation in reference \citep{Sun-Beckermann_JCP2007}. For two-phase
flows a ``conservative Allen-Cahn'' equation has been formulated
in \citep{Chiu-Lin_JCP2011} and coupled with the incompressible Navier-Stokes
equations. For dissolution, the same term has been considered in the
phase-field equation of \citep{Xu-Meakin_PhaseField-Preci_JChemPhys2008}.
Here, the effect of the counter term is presented on the dissolution
of a 2D porous medium. That term has an impact on the shape of the
porous medium and the heterogeneity of composition inside the solid
phase.

\begin{table*}
\begin{centering}
\textbf{Nomenclature of physical modeling}\\
\textbf{}\\
\par\end{centering}
\begin{centering}
\begin{tabular}{llll}
\hline 
\textbf{Symbol} & \textbf{Definition} & \textbf{Dimension} & \textbf{Description}\tabularnewline
\hline 
\textbf{Thermodynamics} &  &  & \tabularnewline
$\Omega[\phi,\,\mu]$ &  & {[}E{]} & Grand-potential functional\tabularnewline
$\mu(\boldsymbol{x},\,t)$ &  & {[}E{]}.{[}mol{]}$^{-1}$ & Chemical potential\tabularnewline
$C(\phi,\,\mu)$ &  & {[}mol{]}.{[}L{]}$^{-3}$ & Global concentration depending on $\phi$ and $\mu$\tabularnewline
$\omega_{int}(\phi,\,\boldsymbol{\nabla}\phi)$ & Eq. (\ref{eq:Interface_Grand-potential}) & {[}E{]}.{[}L{]}$^{-3}$ & Grand-potential density of interface\tabularnewline
$\omega_{dw}(\phi)$ & see Tab. \ref{tab:Interpolation-functions-used} & {[}--{]} & Double-well potential of minima $\phi_{s}$ and $\phi_{l}$\tabularnewline
$\omega_{bulk}(\phi,\,\mu)$ & Eq. (\ref{eq:Grand-potential_density}) & {[}E{]}.{[}L{]}$^{-3}$ & Interpolation of bulk grand-potential density\tabularnewline
$V_{m}$ &  & {[}L{]}$^{3}$.{[}mol{]}$^{-1}$ & Molar volume\tabularnewline
$\chi$ & $=\partial C(\phi,\,\mu)/\partial\mu$ & {[}mol{]}$^{2}$.{[}L{]}$^{-3}$.{[}E{]}$^{-1}$ & Generalized susceptibility\tabularnewline
$\Phi$ &  & {[}--{]} & Index for bulk phases: solid $\Phi=s$ and liquid $\Phi=l$\tabularnewline
$\mathcal{M}_{\phi}$ &  & {[}L{]}$^{3}$.{[}E{]}$^{-1}$.{[}T{]}$^{-1}$ & Mobility coefficient of the interface\tabularnewline
$\phi_{0}(\boldsymbol{x})$ &  & {[}--{]} & Hyperbolic tangent solution\tabularnewline
$\zeta$ &  & {[}E{]}.{[}L{]}$^{-1}$ & Coefficient of gradient energy term\tabularnewline
$H$ &  & {[}E{]}.{[}L{]}$^{-3}$ & Height of double-well function\tabularnewline
$\sigma$ & $=(1/6)\sqrt{2\zeta H}$ & {[}E{]}.{[}L{]}$^{-2}$ & Surface tension\tabularnewline
$f_{\Phi}(c)$ &  & {[}E{]}.{[}L{]}$^{-3}$ & Free energy density of bulk phases\tabularnewline
$m_{\Phi}$ &  & {[}--{]} & Compositions for which $f_{\Phi}$ is minimum\tabularnewline
$\omega_{\Phi}(\mu)$ & $=f_{\Phi}-\mu C$ & {[}E{]}.{[}L{]}$^{-3}$ & Grand-potential density of each bulk phase\tabularnewline
$\epsilon_{\Phi}$ &  & {[}E{]}.{[}L{]}$^{-3}$ & Curvature of quadratic free energies\tabularnewline
$\mathscr{E}$ & $=\sqrt{\epsilon_{l}\epsilon_{s}}$ & {[}E{]}.{[}L{]}$^{-3}$ & Reference volumic energy for dimensionless quantities\tabularnewline
$\Delta f^{min}$ & $=f_{s}^{min}-f_{l}^{min}$ & {[}E{]}.{[}L{]}$^{-3}$ & Difference of minimum values of free energy densities\tabularnewline
$\overline{\omega}_{\Phi}$ & $=\omega_{\Phi}/\mathscr{E}$ & {[}--{]} & Dimensionless grand-potential of bulk phases\tabularnewline
$\overline{f}_{\Phi}$ & $=f_{\Phi}/\mathscr{E}$ & {[}--{]} & Dimensionless free energy densities\tabularnewline
\hline 
\textbf{Phase-field model} &  &  & \tabularnewline
$\phi(\boldsymbol{x},\,t)$ &  & {[}--{]} & Phase-field $\phi_{s}\leq\phi\leq\phi_{l}$\tabularnewline
$\phi_{s},\,\phi_{l}$ &  & {[}--{]} & values of $\phi(\boldsymbol{x},\,t)$ in bulk phases: $\phi_{s}=0$
and $\phi_{l}=1$\tabularnewline
$W$ & $=\sqrt{8\zeta/H}$ & {[}L{]} & Interface width of $\phi$-equation\tabularnewline
$M_{\phi}$ & $=\mathcal{M}_{\phi}\zeta$ & {[}L{]}$^{2}$.{[}T{]}$^{-1}$ & Diffusivity of $\phi$-equation\tabularnewline
$\lambda$ & $=8\mathscr{E}/H$ & {[}--{]} & Coupling coefficient of $\phi$-equation\tabularnewline
$\boldsymbol{n}(\boldsymbol{x},\,t)$ & $=\boldsymbol{\nabla}\phi/\bigl|\boldsymbol{\nabla}\phi\bigr|$ & {[}--{]} & Unit normal vector of interface\tabularnewline
$c(\phi,\,\mu)$ & $=V_{m}C(\phi,\,\mu)$ & {[}--{]} & Global composition depending on $\phi$ and $\mu$\tabularnewline
$\overline{\mu}(\boldsymbol{x},\,t)$ & $=\mu/(\mathscr{E}V_{m})$ & {[}--{]} & Dimensionless chemical potential\tabularnewline
$\overline{\mu}^{eq}$ & $=\Delta\overline{f}^{min}/\Delta m$ & {[}--{]} & Equilibrium chemical potential of interface Eq. (\ref{eq:PotChim_Equilibirum})\tabularnewline
$c_{\Phi}^{co}$ &  & {[}--{]} & Coexistence (or equilibrium) compositions of each phase\tabularnewline
$c^{co}(\phi)$ & Eq. (\ref{eq:Coexistence_Interpol}) & {[}--{]} & Interpolation of coexistence compositions $c_{s}^{co}$ and $c_{l}^{co}$\tabularnewline
$D_{\Phi}$ &  & {[}L{]}$^{2}$.{[}T{]}$^{-1}$ & Diffusion coefficient of solid ($\Phi=s$) and liquid ($\Phi=l$)\tabularnewline
$p(\phi)$ & see Tab. \ref{tab:Interpolation-functions-used} & {[}--{]} & Interpolation function of derivative zero for $\phi=0$ and $\phi=1$\tabularnewline
$h(\phi)$ & see Tab. \ref{tab:Interpolation-functions-used} & {[}--{]} & Interpolation function for $c(\phi,\,\mu)$\tabularnewline
$q(\phi)$ & see Tab. \ref{tab:Interpolation-functions-used} & {[}--{]} & Interpolation function for diffusion coefficients\tabularnewline
$\mathscr{S}_{\phi}(\phi,\,\overline{\mu})$ & Eq. (\ref{eq:SourceTerm_Equilibrium}) & {[}--{]} & Source term of phase-field equation\tabularnewline
$\kappa(\boldsymbol{x},\,t)$ & $=\boldsymbol{\nabla}\cdot\boldsymbol{n}$ & {[}L{]}$^{-1}$ & Curvature\tabularnewline
$-M_{\phi}\kappa\bigl|\boldsymbol{\nabla}\phi\bigr|$ &  & {[}T{]}$^{-1}$ & Phenomenological counter term\tabularnewline
$\boldsymbol{j}_{at}(\boldsymbol{x},\,t)$ & Eq. (\ref{eq:Anti-trapping_Equilibrium}) & {[}L{]}.{[}T{]}$^{-1}$ & Phenomenological anti-trapping current\tabularnewline
$a$ & $1/4$ & {[}--{]} & Coefficient of anti-trapping current\tabularnewline
\hline 
\textbf{Sharp interface} &  &  & \tabularnewline
$v_{n}$ & $=\boldsymbol{v}\cdot\boldsymbol{n}$ & {[}L{]}.{[}T{]}$^{-1}$ & Normal velocity of interface\tabularnewline
$d_{0}$ & Eq. (\ref{eq:CapillaryLength}) & {[}L{]} & Capillary length in Gibbs-Thomson condition Eq. (\ref{eq:Gibbs-Thomson_PF})\tabularnewline
$\beta_{\Phi}$ & Eq. (\ref{eq:KineticCoeff}) & {[}L{]}$^{-1}$.{[}T{]} & Kinetic coefficients in Gibbs-Thomson condition for $\Phi=s,\,l$\tabularnewline
$q_{s}$ & $=D_{s}/D_{l}$ & {[}--{]} & Ratio of diffusion\tabularnewline
$\varepsilon$ & $=W/d_{0}$ & {[}--{]} & Small parameter of asymptotic expansions\tabularnewline
$\mathscr{F}_{\Phi}$, $\tilde{\mathscr{F}}_{\Phi}$ $\mathscr{G}_{\Phi}$,
$\mathscr{H}_{\Phi}$ & See Tab. \ref{tab:integrals} & {[}--{]} & Integrals (part 1) of interpolation functions (for $\Phi=s,\,l$)\tabularnewline
$\mathscr{I}$, $\mathscr{K}$, $\mathscr{J}_{\Phi}$ & See Tab. \ref{tab:integrals} & {[}--{]} & Integrals (part 2)\tabularnewline
$\mathbb{E}_{1}$, $\mathbb{E}_{2}$, $\mathbb{E}_{3}$ &  &  & Error terms derived from the asymptotic analysis\tabularnewline
\hline 
\end{tabular}
\par\end{centering}
\caption{\label{tab:List-of-main}Main mathematical symbols with their physical
dimensions. Unit convention: energy {[}E{]}, length {[}L{]}, time
{[}T{]} and mole {[}mol{]}.}
\end{table*}

In this paper, we derive in Section \ref{sec:Grand-potential-phase-field-mode}
a phase-field model based on the grand-potential functional for simulating
the processes of dissolution and precipitation. In Section \ref{subsec:General-equations-on},
the phase-field equation is presented without counter term $-M_{\phi}\kappa\bigl|\boldsymbol{\nabla}\phi\bigr|$
for keeping the curvature-driven motion. Next, in Section \ref{subsec:Equilibrium_Properties},
the counter term is included in the phase-field equation which is
reformulated in conservative form. Although the second main dynamical
variable is the chemical potential $\mu$, we use in this work a mixed
formulation between the composition $c$ and the chemical potential
$\mu$ (Section \ref{subsec:Quadratic_Free_Energies}). The reason
of this choice is explained by a better mass conservation when simulating
the model. For the sake of simplicity, the link to a thermodynamic
database is not considered in this work. The grand-potential densities
of each bulk phase derive from two analytical forms of free energy
densities. We assume they are quadratic with different curvatures
$\epsilon_{l}$ and $\epsilon_{s}$ for each parabola (Section \ref{subsec:Quadratic_Free_Energies}).
Next, Section \ref{subsec:Discussion_Matched-Asymptotics} is dedicated
to a discussion about the relationships of phase-field parameters
$W$, $\lambda$ and $M_{\phi}$ with the sharp interface parameters,
the capillary length $d_{0}$ and the kinetic coefficient $\beta$.
Those relationships will give indications to set the coupling parameter
$\lambda$ in the simulations.

The model is implemented in \texttt{LBM\_saclay}, a numerical code
running on various High Performance Computing architectures. With
simple modifications of compilation flags, the code can run on CPUs
(Central Process Units) or GPUs (Graphics Process Units) \citep{Verdier_etal_CMAME2020}.
The LBM schemes of phase-field model are presented in Section \ref{sec:Lattice-Boltzmann-methods}.
A special care is taken for canceling diffusion in solid phase and
accounting for the anti-trapping current. Validations are carried
out in Section \ref{sec:Validations}. LBM results are compared with
analytical solutions for precipitation and next for dissolution. The
first case is performed for $D_{s}\simeq D_{l}$ (Section \ref{subsec:Validation-Precipitation})
to show the discontinuity of composition on each side of interface.
The second one presents for $D_{s}=0$ (Section \ref{subsec:Validation-Dissolution})
the impact of anti-trapping current on the profiles of composition.
Finally, in Section \ref{sec:Simulations}, we present the dissolution
of a porous medium characterized by microtomography. Two simulations
compare the effect of the counter term on the composition and the
shape of porous medium.

\section{\label{sec:Grand-potential-phase-field-mode}Phase-field model of
dissolution/precipitation}

The purpose of this Section is to present the phase-field model of
dissolution and precipitation. Its derivation introduces a great quantity
of mathematical notations. The reason is inherent to the whole methodology:
the \emph{diffuse interface method}, which originates from \emph{out-of-equilibrium
thermodynamics}, recovers the \emph{sharp-interface model} through
the \emph{matched asymptotic expansions}. Each keyword introduces
its own mathematical notations. All those relative to physical modeling
are summarized in Tab. \ref{tab:List-of-main}.

In Section \ref{subsec:General-equations-on}, we remind the theoretical
framework of grand-potential $\Omega$, and we present the general
evolution equations on $\phi$ and $\mu$. Section \ref{subsec:Equilibrium_Properties}
reminds the equilibrium properties of the phase-field equation and
introduces the counter term for canceling the curvature-driven motion.
Equations on $\phi$ and $\mu$ require the densities of grand-potential
for each phase $\omega_{s}(\mu)$ and $\omega_{l}(\mu)$. In Section
\ref{subsec:Quadratic_Free_Energies} their expressions are derived
from analytical forms of free energies $f_{s}(c)$ and $f_{l}(c)$.
The phase-field model will be re-written with a mixed formulation
between $\mu$ and $c$ with the compositions of coexistence and the
equilibrium chemical potential. Finally, in Section \ref{subsec:Discussion_Matched-Asymptotics}
a discussion will be done regarding the links between phase-field
model and free-boundary problem.

\subsection{\label{subsec:General-equations-on}General equations on $\phi$
and $\mu$ in the grand-potential theoretical framework}

The grand-potential $\Omega[\phi,\,\mu]$ is a thermodynamic functional
which depends on the phase-field $\phi\equiv\phi(\boldsymbol{x},\,t)$
and the chemical potential $\mu\equiv\mu(\boldsymbol{x},\,t)$, two
functions of position $\boldsymbol{x}$ and time $t$. In comparison,
$\phi(\boldsymbol{x},\,t)$ and the composition $c(\boldsymbol{x},\,t)$
are two main dynamical variables of free energy $\mathscr{F}[\phi,\,c]$.
The functional of grand-potential contains the contribution of two
terms:

\begin{equation}
\Omega[\phi,\,\mu]=\int_{V}\left[\omega_{int}(\phi,\,\boldsymbol{\nabla}\phi)+\omega_{bulk}(\phi,\,\mu)\right]dV\label{eq:Def_Grand-Potential}
\end{equation}
The first term inside the brackets is the grand-potential density
of interface $\omega_{int}(\phi,\,\boldsymbol{\nabla}\phi)$ which
is defined by the contribution of two terms depending respectively
on $\phi$ and $\boldsymbol{\nabla}\phi$:

\begin{equation}
\omega_{int}(\phi,\,\boldsymbol{\nabla}\phi)=H\omega_{dw}(\phi)+\frac{\zeta}{2}\bigl|\boldsymbol{\nabla}\phi\bigr|^{2}.\label{eq:Interface_Grand-potential}
\end{equation}
In Eq. (\ref{eq:Interface_Grand-potential}), the first term is the
double-well potential $\omega_{dw}(\phi)$ and $H$ is its height.
The second term is the gradient energy term which is proportional
to the coefficient $\zeta$. A quick dimensional analysis shows that
the physical dimension of $H$ is an energy per volume unit ({[}E{]}.{[}L{]}$^{-3}$)
and $\zeta$ has the dimension of energy per length unit ({[}E{]}.{[}L{]}$^{-1}$).
Those two contributions are identical for models that are formulated
with a free energy functional $\mathscr{F}[\phi,\,c]$. The mathematical
form of the double-well used in this work will be specified in Section
\ref{subsec:Equilibrium_Properties}.

In Eq. (\ref{eq:Def_Grand-Potential}), the second term $\omega_{bulk}(\phi,\,\mu)$
interpolates the grand-potential densities of each bulk phase $\omega_{s}(\mu)$
and $\omega_{l}(\mu)$ by:

\begin{equation}
\omega_{bulk}(\phi,\,\mu)=p(\phi)\omega_{l}(\mu)+\left[1-p(\phi)\right]\omega_{s}(\mu)\label{eq:Grand-potential_density}
\end{equation}
where $p(\phi)$ is an interpolation function. It is sufficient to
define it (see Section \ref{subsec:Discussion_Matched-Asymptotics})
as a monotonous function such as $p(0)=0$ and $p(1)=1$ in the bulk
phases with null derivatives (w.r.t. $\phi$) $p^{\prime}(0)=p^{\prime}(1)=0$.
In this work we choose

\begin{subequations}

\begin{equation}
p(\phi)=\phi^{2}(3-2\phi)\label{eq:Interpolation_Function}
\end{equation}
and its derivative w.r.t. $\phi$ is

\begin{equation}
p^{\prime}(\phi)=6\phi(1-\phi)\label{eq:Pint_prime}
\end{equation}
With that convention, if $\phi=0$ then $\omega_{bulk}(\mu)=\omega_{s}(\mu)$
and if $\phi=1$ then $\omega_{bulk}(\mu)=\omega_{l}(\mu)$.

\end{subequations}

In this paper, we work with the dimensionless composition $c(\phi,\,\mu)$
describing the local fraction of one chemical species and varying
between zero and one. It is related to the concentration $C(\phi,\,\mu)$
(physical dimension {[}mol{]}.{[}L{]}$^{-3}$) by $c(\phi,\,\mu)=V_{m}C(\phi,\,\mu)$
where $V_{m}$ is the molar volume of ({[}L{]}$^{3}$.{[}mol{]}$^{-1}$).
For both chemical species, the molar volume is assumed to be constant
and identical. In the rest of this paper $V_{m}$ will appear in the
equations for reasons of physical dimension, but it will be considered
equal to $V_{m}=1$ for all numerical simulations.

The concentration $C$ is now a function of $\phi$ and $\mu$. It
is related to the grand-potential by \citep{Plapp_PhysRevE.84.031601}
$C(\phi,\,\mu)=-\delta\Omega/\delta\mu=-\partial\omega_{bulk}(\phi,\,\mu)/\partial\mu$.
The application of that relationship with $\omega_{bulk}(\phi,\,\mu)$
defined by Eq. (\ref{eq:Grand-potential_density}) yields:

\begin{equation}
C(\phi,\,\mu)=p(\phi)\left[-\frac{\partial\omega_{l}(\mu)}{\partial\mu}\right]+\left[1-p(\phi)\right]\left[-\frac{\partial\omega_{s}(\mu)}{\partial\mu}\right]\label{eq:Def_Concentration}
\end{equation}
The concentration $C(\phi,\,\mu)$ is defined by an interpolation
of derivatives of $\omega_{s}(\mu)$ and $\omega_{l}(\mu)$ w.r.t.
$\mu$. Each derivative defines the concentration of bulk phase $C_{s}(\mu)=-\partial\omega_{s}(\mu)/\partial\mu$
and $C_{l}(\mu)=-\partial\omega_{l}(\mu)/\partial\mu$.

In Eq. (\ref{eq:Grand-potential_density}), the grand-potential densities
of each bulk phase $\omega_{l}(\mu)$ and $\omega_{s}(\mu)$ are defined
by the Legendre transform of free energy densities $f_{s}(c)$ and
$f_{l}(c)$:

\begin{equation}
\omega_{\Phi}(\mu)=f_{\Phi}(c)-\mu C\qquad\text{for}\qquad\Phi=s,\,l\label{eq:Legendre_Transform}
\end{equation}
where $\mu=\partial f_{\Phi}/\partial C$. Finally, the phase-field
equations are obtained from the minimization of the grand-potential
functional $\Omega[\phi,\,\mu]$. The most general PDEs write (see
\citep[Eq. (43) and Eq. (47)]{Plapp_PhysRevE.84.031601}):

\begin{subequations}

\begin{align}
\frac{\partial\phi}{\partial t} & =\mathcal{M}_{\phi}\Bigl\{\zeta\boldsymbol{\nabla}^{2}\phi-H\omega_{dw}^{\prime}(\phi)\nonumber \\
 & \qquad\qquad\qquad-p^{\prime}(\phi)\left[\omega_{l}(\mu)-\omega_{s}(\mu)\right]\Bigr\}\label{eq:PhaseField_Eq_Omega}\\
\chi(\phi,\,\mu)\frac{\partial\mu}{\partial t} & =\boldsymbol{\nabla}\cdot\left[\mathcal{D}(\phi,\,\mu)\chi(\phi,\,\mu)\boldsymbol{\nabla}\mu\right]\nonumber \\
 & \qquad\quad-p^{\prime}(\phi)\left[\frac{\partial\omega_{s}(\mu)}{\partial\mu}-\frac{\partial\omega_{l}(\mu)}{\partial\mu}\right]\frac{\partial\phi}{\partial t}\label{eq:Potchem_Eq_Omega}
\end{align}

\end{subequations}

Eq. (\ref{eq:PhaseField_Eq_Omega}) is the evolution equation on $\phi(\boldsymbol{x},\,t)$
which tracks the interface between solid and liquid. The phase-field
equation is derived from $\partial_{t}\phi=-\mathcal{M}_{\phi}\delta\Omega/\delta\phi$
where $\mathcal{M}_{\phi}$ is a coefficient of dimension {[}L{]}$^{3}$.{[}E{]}$^{-1}$.{[}T{]}$^{-1}$.
The equilibrium properties of that equation are reminded in Section
\ref{subsec:Equilibrium_Properties}. The derivative of the double-well
function w.r.t. $\phi$ is noted $\omega_{dw}^{\prime}=\partial\omega_{dw}/\partial\phi$.
Compared to the model of reference \citep{Plapp_PhysRevE.84.031601},
we notice the opposite sign of the last term because our convention
is $\phi=0$ for solid and $\phi=1$ for liquid. In the reference,
$\phi=1$ is solid and $\phi=-1$ is liquid and the interpolation
function $p(\phi)$ is opposite. In order to reveal the diffusivity
coefficient $M_{\phi}=\mathcal{M}_{\phi}\zeta$ of dimension {[}L{]}$^{2}$.{[}T{]}$^{-1}$,
the coefficient $\zeta$ can be put in factor of the right-hand side.
In that case, the second term is multiplied by $H/\zeta$ whereas
the last term is divided by $\zeta$. 

Eq. (\ref{eq:Potchem_Eq_Omega}) is the evolution equation on chemical
potential $\mu(\boldsymbol{x},\,t)$. It is obtained from the conservation
equation $\partial_{t}C(\phi,\,\mu)=-\boldsymbol{\nabla}\cdot\boldsymbol{j}_{diff}$
where the diffusive flux is given by $\boldsymbol{j}_{diff}=-\mathcal{D}(\phi,\,\mu)\chi(\phi,\,\mu)\boldsymbol{\nabla}\mu$.
The time derivative term has been expressed by the chain rule $\partial C(\phi,\,\mu)/\partial t=(\partial C/\partial\mu)\partial_{t}\mu+(\partial C/\partial\phi)\partial_{t}\phi$.
The function $\chi(\phi,\,\mu)$, called the generalized susceptibility,
is defined by the partial derivative of $C(\phi,\,\mu)$ with respect
to $\mu$. For most general cases, the coefficient $\mathcal{D}(\phi,\,\mu)$
is the diffusion coefficient which depends on $\phi$ and $\mu$.
Here we assume that the diffusion coefficients $D_{s}$ and $D_{l}$
are only interpolated by $\phi$, i.e. $\mathcal{D}(\phi,\,\mu)\equiv\mathcal{D}(\phi)$.
Actually, in section \ref{subsec:Quadratic_Free_Energies}, that equation
on $\mu$ will be transformed back to an equation on $C$ (or $c$)
for reasons of mass conservation in simulations. Eq. (\ref{eq:Def_Concentration})
will be used to supply a relationship between $\mu$ and $c$.

For simulating Eqs. (\ref{eq:PhaseField_Eq_Omega}) and (\ref{eq:Potchem_Eq_Omega}),
it is necessary to define the grand-potential densities of each bulk
phase $\omega_{s}(\mu)$ and $\omega_{l}(\mu)$. They both derive
from Legendre transforms (Eq. (\ref{eq:Legendre_Transform})) which
require the knowledge of free energy densities $f_{s}(c)$ and $f_{l}(c)$.
The free energy densities $f_{s}(c)$ and $f_{l}(c)$ depend on the
phase diagram of chemical species (or materials), the temperature
and the number of species involved in the process (binary or ternary
mixtures). When the model is implemented in a numerical code coupled
with a thermodynamic database, those values are updated at each time
step of computation. A method for coupling a phase-field model based
on the grand-potential with a thermodynamic database is proposed in
\citep{Choudhury_etal_COSSMS2015}. A coupling of a phase-field model
with the ``thermodynamics advanced fuel international database''
is presented in \citep{Introini_etal_JNM2021} with \texttt{OpenCalphad}
\citep{Sundman_etal_IMMI2015,Sundman_etal_CompMatSci2015}. In this
work, we assume in Section \ref{subsec:Quadratic_Free_Energies} that
the densities of free energies $f_{s}(c)$ and $f_{l}(c)$ are quadratic.

The variational formulation based on the grand-potential yields to
evolution equations on $\phi$ and $\mu$ (Eqs. (\ref{eq:PhaseField_Eq_Omega})-(\ref{eq:Potchem_Eq_Omega})).
Two ingredients are missing in those equations: the first one is the
counter term $-M_{\phi}\kappa\bigl|\boldsymbol{\nabla}\phi\bigr|$
and the second one is the anti-trapping current $\boldsymbol{j}_{at}$.
In our work, both are not contained in the definition of grand-potential
$\Omega[\phi,\,\mu]$ and have no variational origin. The counter
term has been derived in \citep{Folch_etal_PRE1999}. It is used in
the phase-field equation (Section \ref{subsec:CounterTerm}) to make
vanish the curvature-driven motion. The anti-trapping current has
been derived in \citep{Karma_AntiTrapping_PRL2001}. It is used in
the chemical potential equation (Section \ref{subsec:Anti-trapping-current})
to cancel spurious effects at interface when the diffusion is supposed
to be null in the solid. Their use is justified by the matched asymptotic
expansions carried out on the phase-field model. The links between
the phase-field model and the free-boundary problem will be discussed
in Section \ref{subsec:Discussion_Matched-Asymptotics}.

\subsection{\label{subsec:Equilibrium_Properties}Equilibrium properties of phase-field
equation}

The phase-field equation Eq. (\ref{eq:PhaseField_Eq_Omega}) has the
same structure as those derived from functionals of free energy. Hence,
the equilibrium properties such as the hyperbolic tangent solution
$\phi_{0}$, the interface width $W$ and the surface tension $\sigma$
remain the same. Those equilibrium properties are reminded in Section
\ref{subsec:Hyperbolic-tangent-solution} with one particular choice
of double-well potential $\omega_{dw}(\phi)$. This is done for two
reasons. The phase-field equation is written with ``thermodynamic''
parameters $\zeta$, $H$ and $\mathcal{M}_{\phi}$. The phase-field
equation is re-written with ``macroscopic'' parameters $M_{\phi}$,
$W$ and the dimensionless coupling coefficient $\lambda$ because
they are directly related to the capillary length $d_{0}$ and kinetic
coefficient $\beta$ of sharp interface model. The equilibrium properties
are also necessary for introducing in Section \ref{subsec:CounterTerm}
the kernel function $\bigl|\boldsymbol{\nabla}\phi\bigr|=4\phi(1-\phi)/W$
and the counter term $-M_{\phi}\kappa\bigl|\boldsymbol{\nabla}\phi\bigr|$.

\subsubsection{\label{subsec:Hyperbolic-tangent-solution}Hyperbolic tangent solution
$\phi_{0}$, width $W$ and surface tension $\sigma$}

When the system is at equilibrium, the construction of common tangent
hold and the chemical potential is identical in both phases of value
$\mu^{eq}$. The construction of common tangent is mathematically
equivalent to $\omega_{s}(\mu^{eq})=\omega_{l}(\mu^{eq})$. When the
two phases are at equilibrium, we define the corresponding compositions
of coexistence (or equilibrium) by $c_{s}^{co}=c_{l}(\mu^{eq})$ and
$c_{l}^{co}=c_{l}(\mu^{eq})$ for solid and liquid respectively. Hence,
the last term proportional to $p^{\prime}(\phi)$ in Eq. (\ref{eq:PhaseField_Eq_Omega})
vanishes at equilibrium and the time derivative is zero ($\partial\phi/\partial t=0$).
We recognize the standard equilibrium equation for the interface $\zeta\boldsymbol{\nabla}^{2}\phi-H\omega_{dw}^{\prime}=0$
i.e. in one dimension $\zeta d^{2}\phi/dx^{2}-Hd\omega/d\phi=0$.
After multiplying by $d\phi/dx$, the first term is the derivative
$d/dx$ of $(d\phi/dx)^{2}$ and the second term becomes a derivative
of the double-well w.r.t. $x$. After gathering those two terms inside
the same brackets, it yields:

\begin{equation}
\frac{d}{dx}\left[\left(\frac{d\phi}{dx}\right)^{2}-\frac{2H}{\zeta}\omega_{dw}\right]=0\label{eq:Equil_Eq1D}
\end{equation}
In this work we define the double-well by

\begin{subequations}

\begin{equation}
\omega_{dw}(\phi)=\phi^{2}(1-\phi)^{2}\label{eq:Double-well}
\end{equation}
for which the two minima are $\phi_{s}=0$ and $\phi_{l}=+1$ and
its derivative w.r.t. $\phi$ is:

\begin{equation}
\omega_{dw}^{\prime}(\phi)=2\phi(1-\phi)(1-2\phi)\label{eq:Omega_prime}
\end{equation}
For that form of double-well, the solution of Eq. (\ref{eq:Equil_Eq1D})
is the usual hyperbolic tangent function

\end{subequations}

\begin{equation}
\phi_{0}(x)=\frac{1}{2}\left[1+\tanh\left(\frac{2x}{W}\right)\right]\label{eq:Sol_TanH}
\end{equation}
where the interface width $W$ and the surface tension $\sigma$ are
defined by

\begin{subequations}

\begin{equation}
W=\sqrt{\frac{8\zeta}{H}}\qquad\text{and}\qquad\sigma=\frac{1}{6}\sqrt{2\zeta H}\label{eq:Width_Sigma}
\end{equation}
We can check that the square root of the ratio $\zeta/H$ is homogeneous
to a length as expected for the physical dimension of the width $W$.
Moreover, the square root of the product $\zeta H$ is homogeneous
to an energy per surface unit as expected for the surface tension
$\sigma$. The two relationships Eq. (\ref{eq:Width_Sigma}) can be
easily inverted to yield

\begin{equation}
\zeta=\frac{3}{2}W\sigma\qquad\text{and}\qquad H=12\frac{\sigma}{W}\label{eq:Zeta_Height}
\end{equation}
From Eq. (\ref{eq:Zeta_Height}), the ratio $H/\zeta$ is equal to
$8/W^{2}$. Hence, the factor in front of the double-well in Eq. (\ref{eq:PhaseField_Eq_Omega})
can be replaced by $8/W^{2}$ and the factor of the last term is once
again expressed with $W^{2}$ i.e. $1/\zeta=8/(W^{2}H)$.

\end{subequations}

As a matter of fact, the double-well function Eq. (\ref{eq:Double-well})
is a special case of other popular choices of double-well. For example
in two-phase flows of immiscible fluids, the double-well is $\omega_{dw}(\phi)=(\phi_{l}-\phi)^{2}(\phi-\phi_{s})^{2}$
\citep{Lee-Lin_JCP2005} with $\phi_{s}\leq\phi\leq\phi_{l}$ for
which the two minima are $\phi_{l}$ and $\phi_{s}$. For that form
of double-well, the equilibrium solution is $\phi_{0}(x)=0.5\left[\phi_{l}+\phi_{s}+(\phi_{l}-\phi_{s})\tanh\left(2x/W\right)\right]$,
the surface tension is $\sigma=(1/6)(\phi_{l}-\phi_{s})^{3}\sqrt{2\zeta H}$
and the interface width is $W=\left[1/(\phi_{l}-\phi_{s})\right]\sqrt{8\zeta/H}$.
Eqs. (\ref{eq:Sol_TanH}) and (\ref{eq:Width_Sigma}) can be recovered
by setting $\phi_{l}=1$ and $\phi_{s}=0$. Another popular choice
of double-well is $\omega_{dw}(\phi)=(\phi^{\star}-\phi)^{2}(\phi+\phi^{\star})^{2}$
\citep{Zheng_etal_LargeDensityRatio_JCP2006} for which the two minima
are $\pm\phi^{\star}$. Once again, that double-well function is a
particular case of the previous one by setting $\phi_{l}=\phi^{\star}$
and $\phi_{s}=-\phi^{\star}$. The equilibrium solution writes $\phi_{0}(x)=\phi^{\star}\tanh(2x/W)$,
the surface tension is $\sigma=(4\phi^{\star3}/3)\sqrt{2\zeta H}$
and the interface width $W=(1/\phi^{\star})\sqrt{2\zeta/H}$. In this
work the choice of Eq. (\ref{eq:Double-well}) is done by simplicity.

\subsubsection{\label{subsec:CounterTerm}Removing the curvature-driven motion in
Eq. (\ref{eq:PhaseField_Eq_Omega})}

Another useful relationship that derives from Eq. (\ref{eq:Equil_Eq1D})
is the kernel function $\bigl|\boldsymbol{\nabla}\phi\bigr|$. The
square root of the term inside the brackets yields $\bigl|\boldsymbol{\nabla}\phi\bigr|=(4/W)\sqrt{\omega_{dw}}$
where the coefficient $2H/\zeta$ was replaced by the interface width
$W$ with Eq. (\ref{eq:Zeta_Height}) ($2H/\zeta=16/W^{2}$). Thus,
with a double-well function defined by Eq. (\ref{eq:Double-well}),
the kernel function writes:

\begin{equation}
\bigl|\boldsymbol{\nabla}\phi\bigr|=\frac{4}{W}\phi(1-\phi)\label{eq:KernelFunction}
\end{equation}

For canceling the curvature-driven interface motion, a counter term
$-M_{\phi}\kappa\bigl|\boldsymbol{\nabla}\phi\bigr|$ is simply added
in the right-hand side of the phase-field equation. The counter term
is proportional to the interface diffusivity $M_{\phi}$, the curvature
$\kappa$ and the kernel function $\bigl|\boldsymbol{\nabla}\phi\bigr|$.
The curvature is defined by $\kappa=\boldsymbol{\nabla}\cdot\boldsymbol{n}$
where $\boldsymbol{n}$ is the unit normal vector of the interface

\begin{equation}
\boldsymbol{n}=\frac{\boldsymbol{\nabla}\phi}{\bigl|\boldsymbol{\nabla}\phi\bigr|}\label{eq:NormalVector}
\end{equation}

In Section \ref{subsec:Case-of-counter}, we check that adding such
a counter term in the phase-field equation cancels the curvature motion
$-d_{0}\kappa$ in the Gibbs-Thomson equation. In order to write the
phase-field equation in a more compact form, we remark that the second
term involving the derivative of the double-well is equivalent to

\begin{subequations}

\begin{equation}
-\frac{8M_{\phi}}{W^{2}}\omega_{dw}^{\prime}(\phi)=-M_{\phi}\boldsymbol{n}\cdot\boldsymbol{\nabla}\bigl|\boldsymbol{\nabla}\phi\bigr|\label{eq:Equiv_Omegaprime_normal}
\end{equation}
provided that the kernel function Eq. (\ref{eq:KernelFunction}) is
used for $\bigl|\boldsymbol{\nabla}\phi\bigr|$. If the counter term
$-M_{\phi}\kappa\bigl|\boldsymbol{\nabla}\phi\bigr|$ is added in
the right-hand side of Eq. (\ref{eq:PhaseField_Eq_Omega}) then

\begin{equation}
-\frac{8M_{\phi}}{W^{2}}\omega_{dw}^{\prime}(\phi)-M_{\phi}\kappa\bigl|\boldsymbol{\nabla}\phi\bigr|=-M_{\phi}\boldsymbol{n}\cdot\boldsymbol{\nabla}\bigl|\boldsymbol{\nabla}\phi\bigr|-M_{\phi}(\boldsymbol{\nabla}\cdot\boldsymbol{n})\bigl|\boldsymbol{\nabla}\phi\bigr|\label{eq:SimplificationWithCounterTerm}
\end{equation}
where the definition of the curvature $\kappa=\boldsymbol{\nabla}\cdot\boldsymbol{n}$
has been applied for the second term. The right-hand side of Eq. (\ref{eq:SimplificationWithCounterTerm})
is $-M_{\phi}\boldsymbol{\nabla}\cdot\left[\bigl|\boldsymbol{\nabla}\phi\bigr|\boldsymbol{n}\right]$
and by using the kernel function $\bigl|\boldsymbol{\nabla}\phi\bigr|=(4/W)\phi(1-\phi)$
the phase-field equation writes

\end{subequations}

\begin{equation}
\frac{\partial\phi}{\partial t}=M_{\phi}\boldsymbol{\nabla}\cdot\left[\boldsymbol{\nabla}\phi-\frac{4}{W}\phi(1-\phi)\boldsymbol{n}\right]-\frac{8M_{\phi}}{W^{2}H}p^{\prime}(\phi)\Delta\omega\label{eq:CAC_withsource}
\end{equation}
where $\Delta\omega=\omega_{l}(\mu)-\omega_{s}(\mu)$. In simulations
of Sections \ref{sec:Validations} and \ref{sec:Simulations} two
versions of the phase-field equation are used: Eq. (\ref{eq:PhaseField_Eq_Omega})
when the curvature-driven motion is desired and Eq. (\ref{eq:CAC_withsource})
when that motion is undesired. When the source term of that equation
is null, and when an advective term $\boldsymbol{\nabla}\cdot(\boldsymbol{u}\phi)$
is considered, Eq. (\ref{eq:CAC_withsource}) is the conservative
Allen-Cahn equation that is applied for interface tracking of two
immiscible fluids \citep{Chiu-Lin_JCP2011,Fakhari_etal_JCP2017}.

\subsection{\label{subsec:Quadratic_Free_Energies}Phase-field model derived
from quadratic free energies}

The source terms of Eqs. (\ref{eq:PhaseField_Eq_Omega}) and (\ref{eq:Potchem_Eq_Omega})
contain the bulk densities of grand-potential $\omega_{l}(\mu)$ and
$\omega_{s}(\mu)$. They need to be specified. Here, we work with
analytical expressions which define explicitly $\omega_{l}$ and $\omega_{s}$
as functions of $\mu$. The main advantage of that choice is to simplify
their expressions by involving several scalar parameters representative
of the thermodynamics. The densities of grand-potential are defined
by the Legendre transform of free energy densities $f_{s}(c)$ and
$f_{l}(c)$. In \citep{Plapp_PhysRevE.84.031601}, several choices
for $f_{\Phi}(c)$ are proposed in order to relate the grand-potential
framework to the well-known models derived from free energy. The simplest
phenomenological approximation is a quadratic free energy for each
phase $\Phi=s$ and $\Phi=l$:

\begin{align}
f_{\Phi}(c) & =\frac{\epsilon_{\Phi}}{2}(c-m_{\Phi})^{2}+f_{\Phi}^{min}\quad\text{for}\quad\Phi=s,\,l\label{eq:Quadratic_FreeEnergy_Liquid}
\end{align}
where $\epsilon_{\Phi}$, of physical dimension {[}E{]}.{[}L{]}$^{-3}$,
are the curvature of each parabola and $m_{\Phi}$ are two values
of composition for which $f_{\Phi}(c)$ are minimum of values $f_{\Phi}^{min}$.
In other words, when the phase diagram (i.e. the free energy versus
composition) is available, it presents two regions (one for each phase)
of smallest free energy $f_{\Phi}^{min}$ corresponding to the composition
$m_{\Phi}$. Eq. (\ref{eq:Quadratic_FreeEnergy_Liquid}) means that
each region is approximated by one parabola, where $\varepsilon_{\Phi}$
is a parameter for improving the curvature fit around each minimum.
As a comparison, the well-known Cahn-Hilliard equation is derived
from one single double-well potential. The Cahn-Hilliard model is
a fourth-order equation where the variable plays the roles of interface
tracking and composition. Here, the single double-well is approximated
by two separated parabolas. The advantage of that splitting is to
facilitate the thermodynamical fit around each minima by using two
functions with their own parameters. The double-well $\omega_{dw}$,
defined in $\omega_{int}$ (Eq. (\ref{eq:Interface_Grand-potential})),
is used for tracking the interface between the bulk phases. With that
approach, the parameters of $\omega_{int}$ control the interface
properties (width and surface tension) whereas the parameters of $f_{\Phi}$
control the thermodynamics. As a drawback, the compositions of solid
and liquid must not be initialized too far from each composition $m_{\Phi}$.
In particular, the spinodal decomposition cannot be simulated without
modification of the model. Let us emphasize that the compositions
$m_{\Phi}$ do not correspond to the coexistence compositions $c_{\Phi}^{co}$
(also called compositions of equilibrium). When a binary system is
considered with $\varepsilon_{s}=\varepsilon_{l}$, the construction
of common tangent yields a simple relationship between $m_{\Phi}$
and $c_{\Phi}^{co}$ (see Section \ref{subsec:Discussion-regarding-composition}).
But this is not true for more general cases, in particular for a system
with two phases and three components.

In this Section, all terms of Eqs. (\ref{eq:PhaseField_Eq_Omega})
and (\ref{eq:Potchem_Eq_Omega}) involving $\omega_{\Phi}$ are simplified
with the  hypothesis of Eq. (\ref{eq:Quadratic_FreeEnergy_Liquid}).
First, Section \ref{subsec:Difference-of-grand-potential} deals with
the difference of grand-potential densities $\omega_{l}(\mu)-\omega_{s}(\mu)$
which will be written with the dimensionless chemical potential $\overline{\mu}$
and the thermodynamical parameters $\epsilon_{\Phi}$, $m_{\Phi}$
and $\overline{f}_{\Phi}^{min}$ of Eq. (\ref{eq:Quadratic_FreeEnergy_Liquid}).
Section \ref{subsec:Discussion-regarding-composition} introduces
the coexistence compositions $c_{\Phi}^{co}$ of interface and the
equilibrium chemical potential $\overline{\mu}^{eq}$. The difference
$\omega_{l}(\mu)-\omega_{s}(\mu)$ will be re-expressed with $c_{\Phi}^{co}$,
$\overline{\mu}$ and $\overline{\mu}^{eq}$. In Section \ref{subsec:Chemical-potential}
the composition equation is re-written with a mixed formulation between
$c(\phi,\,\mu)$ and $\overline{\mu}$, and in Section \ref{subsec:Anti-trapping-current}
the anti-trapping current $\boldsymbol{j}_{at}$ will be formulated
as a function of $c_{\Phi}^{co}$. Finally, the complete model is
summarized in Section \ref{subsec:Summary-of-phase-field}.

\subsubsection{\label{subsec:Difference-of-grand-potential}Difference of grand-potential
densities in $\phi$-equation}

We start with the difference $\omega_{l}(\mu)-\omega_{s}(\mu)$ where
the chemical potential is defined by $\mu=\partial f_{\Phi}/\partial C=V_{m}\partial f_{\Phi}/\partial c$
(for $\Phi=s,\,l$). By inverting those relationships to obtain $c$
as a function of $\mu$, the Legendre transforms of each bulk phase
yield the grand-potential densities as function of $\mu$ (see intermediate
steps in \citep{Plapp_PhysRevE.84.031601}):

\begin{align}
\omega_{\Phi}(\mu) & =-\frac{\mu^{2}}{2V_{m}^{2}\epsilon_{\Phi}}-\frac{\mu}{V_{m}}m_{\Phi}+f_{\Phi}^{min}\quad\text{for}\quad\Phi=s,\,l\label{eq:Quadratic_grand-potential_Liquid}
\end{align}
Before going further we set $\Delta f^{min}=f_{s}^{min}-f_{l}^{min}$
and we define the quantity $\mathscr{E}=\sqrt{\epsilon_{s}\epsilon_{l}}$
(dimension {[}E{]}.{[}L{]}$^{-3}$) for introducing the dimensionless
quantities $\overline{\omega}_{\Phi}$, $\overline{\mu}$ and $\Delta\overline{f}^{min}$
by $\omega_{\Phi}=\overline{\omega}_{\Phi}\mathscr{E}$ (with $\Phi=s,\,l$),
$\mu=\overline{\mu}V_{m}\mathscr{E}$ and $\Delta f^{min}=\mathscr{E}\Delta\overline{f}^{min}$.
With those reduced variables, the difference $\Delta\overline{\omega}=\overline{\omega}_{l}(\overline{\mu})-\overline{\omega}_{s}(\overline{\mu})$
writes 

\begin{equation}
\mathscr{E}\Delta\overline{\omega}=\mathscr{E}\left[\frac{(\epsilon_{l}-\epsilon_{s})}{\sqrt{\epsilon_{l}\epsilon_{s}}}\frac{\overline{\mu}^{2}}{2}-(m_{l}-m_{s})\overline{\mu}-\Delta\overline{f}^{min}\right]\label{eq:Dimensionless_OmegaL-OmegaS}
\end{equation}
Finally, if we define the dimensionless coefficient of coupling by
$\lambda=8\mathscr{E}/H$, the last term of Eq. (\ref{eq:PhaseField_Eq_Omega})
writes

\begin{align}
-\frac{8M_{\phi}\mathscr{E}}{W^{2}H}p^{\prime}(\phi)\Delta\overline{\omega} & =-\frac{\lambda M_{\phi}}{W^{2}}\mathscr{S}_{\phi}(\phi,\,\overline{\mu})\label{eq:Dimensionless_sourceTerm}
\end{align}
where for future use we have set $\mathscr{S}_{\phi}(\phi,\,\overline{\mu})\equiv\mathscr{S}_{\phi}$
defined by:

\begin{equation}
\mathscr{S}_{\phi}=p^{\prime}(\phi)\left[\frac{(\epsilon_{l}-\epsilon_{s})}{\sqrt{\epsilon_{l}\epsilon_{s}}}\frac{\overline{\mu}^{2}}{2}-(m_{l}-m_{s})\overline{\mu}-\Delta\overline{f}^{min}\right]\label{eq:SourceTerm}
\end{equation}
When the free energies are quadratic, the coupling term of Eq. (\ref{eq:PhaseField_Eq_Omega})
becomes Eq. (\ref{eq:Dimensionless_sourceTerm}) with $\mathscr{S}_{\phi}(\phi,\,\overline{\mu})$
defined by Eq. (\ref{eq:SourceTerm}). The dimensionless chemical
potential $\overline{\mu}$ appears explicitly in that equation.

\subsubsection{\label{subsec:Discussion-regarding-composition}Coexistence compositions
and chemical potential of equilibrium}

In Eq. (\ref{eq:SourceTerm}), $m_{s}$ and $m_{l}$ are two specific
values of $c$ for which the quadratic free energies $f_{s}$ and
$f_{l}$ are minimum. A close link exists between $m_{\Phi}$ and
the coexistence (or equilibrium) compositions $c_{\Phi}^{co}$. Two
relationships allow deriving them: the first one is the equality of
chemical potential $\mu^{eq}$:

\begin{subequations}

\begin{equation}
\mu^{eq}=V_{m}\left.\frac{\partial f_{s}}{\partial c}\right|_{c=c_{s}^{co}}=V_{m}\left.\frac{\partial f_{l}}{\partial c}\right|_{c=c_{l}^{co}}\label{eq:Egalite_PotChim}
\end{equation}
and the second one is the equality of grand-potential densities

\begin{equation}
\omega_{s}(\mu^{eq})=\omega_{l}(\mu^{eq}).\label{eq:Egalite_grand-potentiel}
\end{equation}
The graphical representation of Eq. (\ref{eq:Egalite_grand-potentiel})
is the standard construction of common tangent. When the curvature
of each parabola are identical $\epsilon_{s}=\epsilon_{l}=\epsilon$,
Eq. (\ref{eq:Egalite_PotChim}) yields $\overline{\mu}^{eq}=c_{s}^{co}-m_{s}=c_{l}^{co}-m_{l}$
and Eq. (\ref{eq:Egalite_grand-potentiel}) yields 

\end{subequations}

\begin{equation}
\overline{\mu}^{eq}=\frac{\Delta\overline{f}^{min}}{\Delta m}\label{eq:Equilibrium_Potchem}
\end{equation}
where $\Delta m=m_{s}-m_{l}$ and $\Delta\overline{f}^{min}$ has
been defined in Section \ref{subsec:Difference-of-grand-potential}.
Finally, those two conditions yield two simple relationships between
$c_{\Phi}^{co}$ and the parameters $m_{\Phi}$ and $\overline{f}_{\Phi}^{min}$:

\begin{subequations}

\begin{align}
c_{l}^{co} & =m_{l}+\frac{\Delta\overline{f}^{min}}{\Delta m}\label{eq:Cl_coexistence}\\
c_{s}^{co} & =m_{s}+\frac{\Delta\overline{f}^{min}}{\Delta m}\label{eq:Cs_coexistence}
\end{align}

\end{subequations}

In the binary case this couple of coexistence compositions is unique,
and the mathematical model can be re-defined with $c_{\Phi}^{co}$
and $\overline{\mu}^{eq}$. More precisely in Eq. (\ref{eq:SourceTerm}),
$m_{l}$ and $m_{s}$ are replaced with $c_{l}^{co}$ and $c_{s}^{co}$
by using Eqs. (\ref{eq:Cl_coexistence})-(\ref{eq:Cs_coexistence}).
In addition, the ratio $\Delta\overline{f}^{min}/\Delta m$ is simply
replaced by $\overline{\mu}^{eq}$. The source term simplifies to

\begin{equation}
\mathscr{S}_{\phi}=p^{\prime}(\phi)(c_{s}^{co}-c_{l}^{co})(\overline{\mu}-\overline{\mu}^{eq})\label{eq:SourceTerm_Equilibrium}
\end{equation}

Here the source term has been formulated with $c_{s}^{co}$ and $c_{l}^{co}$
provided that $\epsilon_{s}=\epsilon_{l}=\epsilon$. If $\epsilon_{s}\neq\epsilon_{l}$
the relationships between $m_{\Phi}$ and $c_{\Phi}^{co}$ are more
complicated because they are solutions of second degree equations.
That case will be studied in a future work. Finally, we can relate
the compositions $c(\phi,\,\overline{\mu})$ to the chemical potential
$\overline{\mu}$. By using the definition $C_{\Phi}(\mu)=c_{\Phi}(\mu)/V_{m}$
and the dimensionless notation $\overline{\mu}=\mu/V_{m}\mathscr{E}$,
we obtain $c_{l}(\overline{\mu})=\overline{\mu}(\epsilon_{s}/\epsilon_{l})^{1/2}+m_{l}$
and $c_{s}(\overline{\mu})=\overline{\mu}(\epsilon_{l}/\epsilon_{s})^{1/2}+m_{s}$.
Those relationships will be useful in Section \ref{sec:Validations}.

\subsubsection{\label{subsec:Chemical-potential}Mixed formulation and closure relationship
between $c(\phi,\,\overline{\mu})$ and $\overline{\mu}$ in $c$-equation}

Even though the equation on chemical potential (Eq. (\ref{eq:Potchem_Eq_Omega}))
could be directly simulated, we prefer using a mixed formulation that
involves both variables $c$ and $\overline{\mu}$. The time derivative
is expressed with $c$ and the flux is expressed with $\overline{\mu}$.
The advantage of such a formulation, inspired from \citep[p. 62]{Bayle_PhD2020},
is explained by a better mass conservation. With the chain rule, the
PDE on $\mu$ (Eq. (\ref{eq:Potchem_Eq_Omega})) is transformed back
to the diffusion equation $\partial C/\partial t=\boldsymbol{\nabla}\cdot[\chi(\phi,\,\mu)\mathcal{D}(\phi,\,\mu)\boldsymbol{\nabla}\mu]$.
Although, the diffusion coefficient is a function of $\mu$ in general
cases, here we assume that it is only a function of $\phi$ i.e. $\mathcal{D}(\phi)=D_{l}\phi+(1-\phi)D_{s}$.
It is relevant to define $\mathcal{D}(\phi)=D_{l}q(\phi)$ with $q(\phi)=\phi+(1-\phi)(D_{s}/D_{l}$)
because the interpolation function $q(\phi)$ appears naturally during
the asymptotic analysis of Section \ref{subsec:Discussion_Matched-Asymptotics}
when switching to a dimensionless timescale. In addition, the coefficient
$\chi(\phi,\,\mu)$ is defined by $\chi=\partial C(\phi,\,\mu)/\partial\mu$
where $C$ is defined by Eq. (\ref{eq:Composition_Quadratic}) when
the free energies are quadratic. When $\varepsilon_{s}=\varepsilon_{l}=\mathscr{E}$
that coefficient is simply equal to $\chi=1/V_{m}^{2}\mathscr{E}$
(see Eq. (\ref{eq:Composition_Quadratic})). Finally, with $C(\phi,\,\mu)=c(\phi,\,\mu)/V_{m}$
and $\mu=\overline{\mu}V_{m}\mathscr{E}$, the composition equation
writes:

\begin{equation}
\frac{\partial c}{\partial t}=\boldsymbol{\nabla}\cdot\left[D_{l}q(\phi)\boldsymbol{\nabla}\overline{\mu}\right]\label{eq:Mixed-Formulation}
\end{equation}
The closure equation between $\overline{\mu}$ and $c(\phi,\,\mu)$
is simply obtained with Eqs. (\ref{eq:Def_Concentration}) and (\ref{eq:Quadratic_grand-potential_Liquid})
for expressing the composition $c(\phi,\,\mu)$. In Eq. (\ref{eq:Def_Concentration}),
the interpolation function $p(\phi)$ can be replaced by another one
$h(\phi)$. The form of $h(\phi)$ is discussed below. The closure
equation writes:

\begin{align}
c(\phi,\,\mu) & =h(\phi)m_{l}+\left[1-h(\phi)\right]m_{s}+\nonumber \\
 & \qquad\qquad\left\{ h(\phi)\frac{1}{V_{m}\epsilon_{l}}+\left[1-h(\phi)\right]\frac{1}{V_{m}\epsilon_{s}}\right\} \mu\label{eq:Composition_Quadratic}
\end{align}
Next, by inverting Eq. (\ref{eq:Composition_Quadratic}), we find
a relationship that relates the dimensionless chemical potential $\overline{\mu}=\mu/V_{m}\mathscr{E}$
to compositions $c(\phi,\,\mu)$, $m_{s}$ and $m_{l}$:

\begin{align}
\overline{\mu} & =\frac{\sqrt{\epsilon_{s}\epsilon_{l}}}{\epsilon_{s}h(\phi)+\epsilon_{l}\left[1-h(\phi)\right]}\Bigl\{ c(\phi,\,\overline{\mu})-h(\phi)m_{l}\nonumber \\
 & \qquad\qquad\qquad\qquad\qquad\qquad\quad-\left[1-h(\phi)\right]m_{s}\Bigr\}\label{eq:Closure}
\end{align}
Once again, when $\epsilon_{s}=\epsilon_{l}$ that closure can be
re-expressed with $c_{s}^{co}$, $c_{l}^{co}$ and $\overline{\mu}^{eq}$.
In that case, the factor of Eq. (\ref{eq:Closure}) is equal to one,
and we replace $m_{l}$ and $m_{s}$ by Eqs. (\ref{eq:Cl_coexistence})-(\ref{eq:Cs_coexistence})
to obtain:

\begin{subequations}

\begin{equation}
\overline{\mu}=\overline{\mu}^{eq}+c(\phi,\,\overline{\mu})-c^{co}(\phi)\label{eq:PotChim_Equilibirum}
\end{equation}
where:

\begin{equation}
c^{co}(\phi)=c_{l}^{co}h(\phi)+c_{s}^{co}\left[1-h(\phi)\right]\label{eq:Coexistence_Interpol}
\end{equation}
is the interpolation of coexistence compositions.

\end{subequations}

A special care must be taken for choosing the interpolation functions
$q(\phi)$ in Eq. (\ref{eq:Mixed-Formulation}) and $h(\phi)$ in
Eq. (\ref{eq:Closure}). Indeed, the matched asymptotic expansions
show that $q(\phi)$ and $h(\phi)$ are involved in several pairs
of integrals. Each pair of integrals must have identical values for
canceling the spurious terms arising from expansions. The particular
choices $h(\phi)=\phi$ and $q(\phi)=\phi+(1-\phi)q_{s}$ with $q_{s}=D_{s}/D_{l}$
fulfill those requirements. More details are given in Section \ref{subsec:Discussion_Matched-Asymptotics}.
When $D_{s}=0$, the interpolation function $q(\phi)$ is simply equal
to $\phi$. 

\subsubsection{\label{subsec:Anti-trapping-current}Anti-trapping current $\boldsymbol{j}_{at}$
in Eq. (\ref{eq:Potchem_Eq_Omega})}

The anti-trapping current has been proposed in \citep{Karma_AntiTrapping_PRL2001}
in order to counterbalance spurious solute trapping when $D_{s}=0$
or when the ratio of diffusivities $D_{s}/D_{l}$ is very small. The
anti-trapping current is introduced for phenomenological reasons in
the mass balance equation and justified by carrying out the matched
asymptotic expansions. An alternative justification for this current
has been proposed in \citep{Brenner-Boussinot_PRE2012,Fang-Mi_PRE2013}.
Thus, with anti-trapping current, the model becomes equivalent to
the free-boundary problem without introducing other thin interface
effects \citep{Echebarria_etal_PhysRevE.70.061604}. In the framework
of grand-potential, the anti-trapping current is defined by \citep{Plapp_PhysRevE.84.031601}: 

\begin{equation}
\boldsymbol{j}_{at}=a(\phi)W\left[\frac{\partial\omega_{l}(\mu)}{\partial\mu}-\frac{\partial\omega_{s}(\mu)}{\partial\mu}\right]\frac{\partial\phi}{\partial t}\boldsymbol{n}\label{eq:Anti-trapping_Current}
\end{equation}
This current is proportional to the velocity ($\partial_{t}\phi$)
and the thickness $W$ of the interface. It is normal to the interface
and points from solid to liquid. The coefficient $a$ is used as a
degree of freedom to remove the spurious terms arising from the matched
asymptotic expansions. The coefficient $a(\phi)$ depends on the choice
of interpolation functions in the phase-field model. For our choice
it is sufficient to set $a=1/4$ to fulfill the equality of integrals
(see Section \ref{subsec:Case-of-anti-trapping} for more details).
When the quadratic free energies are used, the term inside the brackets
is simplified by deriving Eq. (\ref{eq:Dimensionless_OmegaL-OmegaS})
w.r.t. $\overline{\mu}$. Using the dimensionless quantities, the
anti-trapping current writes:

\begin{equation}
\boldsymbol{j}_{at}=\frac{1}{4}W\left[-\frac{\epsilon_{s}-\epsilon_{l}}{\sqrt{\epsilon_{s}\epsilon_{l}}}\overline{\mu}+m_{s}-m_{l}\right]\frac{\partial\phi}{\partial t}\boldsymbol{n}\label{eq:Anti-Trapping_Quadratic}
\end{equation}
When $\epsilon_{s}=\epsilon_{l}$, the first term inside the brackets
is zero. The coefficients $m_{s}$ and $m_{l}$ are expressed with
the coexistence compositions (Eqs. (\ref{eq:Cl_coexistence})-(\ref{eq:Cs_coexistence}))
and the anti-trapping writes:

\begin{equation}
\boldsymbol{j}_{at}=\frac{1}{4}W\left(c_{s}^{co}-c_{l}^{co}\right)\frac{\partial\phi}{\partial t}\boldsymbol{n}\label{eq:Anti-trapping_Equilibrium}
\end{equation}

The impact of that anti-trapping current will be emphasized in Section
\ref{subsec:Validation-Dissolution}. The chemical potentials and
compositions will be compared on one case of dissolution with $D_{s}=0$.

\subsubsection{\label{subsec:Summary-of-phase-field}Summary of the phase-field
model}

The complete phase-field model is composed of two coupled PDEs which
write:

\begin{subequations}

\begin{align}
\frac{\partial\phi}{\partial t} & =M_{\phi}\boldsymbol{\nabla}^{2}\phi-\frac{8M_{\phi}}{W^{2}}\omega_{dw}^{\prime}(\phi)-\frac{\lambda M_{\phi}}{W^{2}}\mathscr{S}_{\phi}(\phi,\,\overline{\mu})\label{eq:ModelA_EqPhi}\\
\frac{\partial c}{\partial t} & =\boldsymbol{\nabla}\cdot\Bigl[D_{l}q(\phi)\boldsymbol{\nabla}\overline{\mu}-\boldsymbol{j}_{at}(\phi,\,\overline{\mu})\Bigr]\label{eq:ModelA_EqCompos}
\end{align}
where the source term $\mathscr{S}_{\phi}(\phi,\,\overline{\mu})$
is re-written below for convenience:

\begin{equation}
\mathscr{S}_{\phi}(\phi,\,\overline{\mu})=p^{\prime}(\phi)(c_{s}^{co}-c_{l}^{co})(\overline{\mu}-\overline{\mu}^{eq})\label{eq:SourceTerm_Coexistence}
\end{equation}
In $c$-equation, the anti-trapping current $\boldsymbol{j}_{at}$
is defined by Eq. (\ref{eq:Anti-trapping_Equilibrium}).

\end{subequations}

\begin{table*}[t]
\begin{centering}
\begin{tabular}{llllll}
\hline 
\textbf{\small{}Description} &  & \textbf{\small{}Functions} &  &  & \textbf{\small{}Derivatives}\tabularnewline
\hline 
{\small{}Double-well potential of minima $\phi_{s}=0$ and $\phi_{l}=+1$} & {\small{}$\omega_{dw}(\phi)$} & {\small{}$=\phi^{2}\left(1-\phi\right)^{2}$} &  & {\small{}$\omega_{dw}^{\prime}(\phi)$} & {\small{}$=2\phi\left(1-\phi\right)\left(1-2\phi\right)$}\tabularnewline
{\small{}Interpolation of coupling in Eq. (\ref{eq:ModelA_EqPhi})} & {\small{}$p(\phi)$} & {\small{}$=\phi^{2}\left(3-2\phi\right)$} &  & {\small{}$p^{\prime}(\phi)$} & {\small{}$={\displaystyle 6\phi\left(1-\phi\right)}$}\tabularnewline
{\small{}Interpolation of $c(\phi,\,\overline{\mu})$} & {\small{}$h(\phi)$} & {\small{}$=\phi$} &  & {\small{}$h^{\prime}(\phi)$} & {\small{}$=1$}\tabularnewline
{\small{}Equilibrium solution} & {\small{}$\phi_{0}(x)$} & {\small{}$=\frac{1}{2}\left[1+\tanh\left(\frac{2x}{W}\right)\right]$} &  & {\small{}$\frac{\partial\phi_{0}}{\partial x}$} & {\small{}$=\frac{4}{W}\phi_{0}(1-\phi_{0})$}\tabularnewline
{\small{}Interpolation of bulk diffusivities} & {\small{}$\mathcal{D}(\phi)$} & {\small{}$=D_{l}q(\phi)$} &  &  & \tabularnewline
{\small{}Interpolation of $q(\phi)$} & {\small{}$q(\phi)$} & {\small{}$=\phi+\left(1-\phi\right)\frac{D_{s}}{D_{l}}$} &  &  & \tabularnewline
\hline 
\end{tabular}
\par\end{centering}
\caption{\label{tab:Interpolation-functions-used}All functions depending on
$\phi$ in this work.}
\end{table*}

The chemical potential $\overline{\mu}$ appears inside the $\phi$-equation
through the source term (Eq. (\ref{eq:SourceTerm_Coexistence})).
It also appears in the $c$-equation through the laplacian term and
the anti-trapping current $\boldsymbol{j}_{at}$. The closure equation
between $\overline{\mu}$ and $c$ is given by Eqs. (\ref{eq:PotChim_Equilibirum})-(\ref{eq:Coexistence_Interpol}).
The derivatives $p^{\prime}(\phi)$ and $\omega_{dw}^{\prime}(\phi)$
of interpolation function and double-well have been defined by Eqs.
(\ref{eq:Omega_prime}) and (\ref{eq:Pint_prime}). All functions
depending on $\phi$ are summarized in Tab. \ref{tab:Interpolation-functions-used}.

The phase-field equation Eq. (\ref{eq:ModelA_EqPhi}) includes the
curvature-driven motion (see Section \ref{subsec:Results-of-thin-interface}).
In order to cancel it, the following PDE is solved for simulations:

\begin{equation}
\frac{\partial\phi}{\partial t}=M_{\phi}\boldsymbol{\nabla}\cdot\left[\boldsymbol{\nabla}\phi-\frac{4}{W}\phi(1-\phi)\boldsymbol{n}\right]-\frac{\lambda M_{\phi}}{W^{2}}\mathscr{S}_{\phi}(\phi,\,\overline{\mu})\label{eq:ModelA_EqPhi_CounterTerm}
\end{equation}
In that equation the counter term $-M_{\phi}\kappa\bigl|\boldsymbol{\nabla}\phi\bigr|$
has been included in the first term of the right-hand side.

Several scalar parameters appear in that model. The $\phi$-equation
involves the diffusivity coefficient $M_{\phi}$, the interface width
$W$, and the coupling parameter $\lambda$. Those three parameters
have a close link with the capillary length $d_{0}$ and the kinetic
coefficient $\beta$ of the Gibbs-Thomson condition. Their relationships
will be discussed in Section \ref{subsec:Results-of-thin-interface}.
They will indicate us how to set their values for simulations.

The model also requires providing the triplet of values $(c_{s}^{co},\,c_{l}^{co},\,\overline{\mu}^{eq}$).
The phase-field model can simulate the dissolution processes as well
as the precipitation ones. The difference lies in the sign of $(c_{s}^{co}-c_{l}^{co})(\overline{\mu}-\overline{\mu}^{eq})$
in the source term $\mathscr{S}_{\phi}$. If we suppose that $\overline{\mu}(\boldsymbol{x},\,0)=\overline{\mu}^{eq}$
in the solid with $D_{s}=0$, then the processes of dissolution or
precipitation depend on the choice of the initial condition for the
liquid phase. For instance, in the simulations of Sections \ref{sec:Validations}
and \ref{sec:Simulations}, with the convention $c_{s}^{co}-c_{l}^{co}>0$,
the dissolution process occurs when $\overline{\mu}(\boldsymbol{x},\,0)<\overline{\mu}^{eq}$
in the liquid whereas the precipitation occurs when $\overline{\mu}(\boldsymbol{x},\,0)>\overline{\mu}^{eq}$.
In terms of composition, the dissolution occurs when the composition
of liquid is lower than its coexistence value: $c(\boldsymbol{x},\,0)<c_{l}^{co}$.
The precipitation process occurs if its value is greater: $c_{l}^{co}<c(\boldsymbol{x},\,0)<c_{s}^{co}$.

\subsection{\label{subsec:Discussion_Matched-Asymptotics}Discussion on the matched
asymptotic expansions}

The equivalence between the phase-field model and the free-boundary
problem is classically established with the method of ``matched asymptotic
expansions'' \citep{Fife_1988,Caginalp_PRA1989}. The method has
been presented for solid/liquid phase change for identical conductivity
in the solid and the liquid in \citep{Karma-Rappel_PRE1998}, and
for unequal conductivity in \citep{Almgren_SIAM1999,McFadden-Wheeler-Anderson_PhysD2000}.
That approach considers the ratio $\varepsilon=W/d_{0}$ as small
parameter of expansion where $d_{0}$ is the capillary length. This
choice of $\varepsilon$ yields a correction of second order on the
kinetic coefficient $\beta$. That correction makes possible to cancel
$\beta$, if desired, by choosing appropriately the parameters $\lambda$,
$W$ and $M_{\phi}$ of the phase-field model. Based on that analysis,
the anti-trapping current was derived in \citep{Karma_AntiTrapping_PRL2001}.

The matched asymptotic expansions have been applied in \citep{Echebarria_etal_PhysRevE.70.061604}
for dilute binary mixture with anti-trapping current and $D_{s}=0$.
In \citep{Ramirez_etal_BinaryAlloy_PRE2004} the analysis has been
done for coupling with temperature. The case $D_{s}\neq0$ with anti-trapping
current has been studied in \citep{Ohno-Matsuura_PRE2009}. In reference
\citep{Ohno_etal_PRE2016} the method has been applied to investigate
the impact of one additional term in the phase-field equation which
is derived from a variational formulation. Finally, the method has
been applied recently for coupling with fluid flow in \citep{Hester_etal_ProcRSA2020}.
A pedagogical presentation of that method can be found in the Appendix
of \citep{Provatas-Elder_Book_2010} which takes into account the
anti-trapping current with $D_{s}=0$.

\subsubsection{\label{subsec:Results-of-thin-interface}Results of the asymptotic
analysis}

In this paper, the details of the matched asymptotic expansions are
presented in \ref{sec:Matched-asymptotic-expansions} (equations of
order $\varepsilon^{0}$, $\varepsilon^{1}$ and $\varepsilon^{2}$
and their respective solutions $\phi_{j}$ and $\overline{\mu}_{j}$
for $0\leq j\leq2$). The stages and the results remain essentially
the same as those already published in \citep{Echebarria_etal_PhysRevE.70.061604}
and \citep[Appendix A]{Provatas-Elder_Book_2010}. In those references,
the analyses are carried out in the theoretical framework of free
energy with anti-trapping current and $D_{s}=0$. In this Section,
we focus the discussion on the main assumptions and results (Section
\ref{subsec:Results-of-thin-interface}). In our model, the use of
grand-potential simplifies the source term analysis (see \ref{sec:Matched-asymptotic-expansions}).
Two modifications have also an influence on the relationships relating
the phase-field parameters to the interface conditions. The first
one is our choice of interpolation functions (Tab. \ref{tab:Interpolation-functions-used})
which impact the coefficient $a(\phi)$ of the anti-trapping current
$\boldsymbol{j}_{at}$. The second one concerns the counter term $-M_{\phi}\kappa\bigl|\boldsymbol{\nabla}\phi\bigr|$
for canceling the curvature-driven motion. Those two modifications
are discussed respectively in Sections \ref{subsec:Case-of-anti-trapping}
and \ref{subsec:Case-of-counter}.

In \ref{sec:Matched-asymptotic-expansions}, the phase-field model
is expanded with anti-trapping $\boldsymbol{j}_{at}$ with $\epsilon_{s}=\epsilon_{l}$
and $D_{s}\neq D_{l}$. By setting $\partial_{n}\overline{\mu}|_{l}=\partial_{n}c|_{l}$
and $q_{s}\partial_{n}\overline{\mu}|_{s}=\partial_{n}c|_{s}$, the
equivalent sharp interface model writes for $\Phi=s,\,l$:

\begin{subequations}

\begin{align}
\frac{\partial c}{\partial t} & =D_{\Phi}\boldsymbol{\nabla}^{2}c\label{eq:Mass-Conserv_PF}\\
D_{l}\partial_{n}c|_{l}-D_{s}\partial_{n}c|_{s} & =-v_{n}\Delta c^{co}-\mathbb{E}_{2}\Delta\mathscr{H}-\mathbb{E}_{3}\Delta\mathscr{J}\label{eq:Interface_MC_PF}\\
\left(\overline{\mu}_{\Phi}-\overline{\mu}^{eq}\right)\Delta c^{co} & =-d_{0}\kappa-\beta_{\Phi}v_{n}+\nonumber \\
 & \qquad+\mathbb{E}_{1}\left[\Delta\tilde{\mathscr{F}}-\Delta\mathscr{G}_{\Phi}\right]\Delta c^{co}\label{eq:Gibbs-Thomson_PF}
\end{align}
Eq. (\ref{eq:Mass-Conserv_PF}) is the mass balance for each bulk
phase and Eqs. (\ref{eq:Interface_MC_PF})-(\ref{eq:Gibbs-Thomson_PF})
are the two interface conditions, respectively the mass conservation
(or Stefan condition) and the Gibbs-Thomson condition. 

The right-hand sides of the last two equations contain three error
terms: $\mathbb{E}_{1}$ in Eq. (\ref{eq:Gibbs-Thomson_PF}) and $\mathbb{E}_{2}$,
$\mathbb{E}_{3}$ in Eq. (\ref{eq:Interface_MC_PF}). The accurate
form of those error terms are written in \ref{sec:Matched-asymptotic-expansions}.
The $\mathbb{E}$-terms are multiplied by integrals defined in Tab.
\ref{tab:integrals}: $\Delta\mathscr{H}=\mathscr{H}_{l}-\mathscr{H}_{s}$,
$\Delta\mathscr{J}=\mathscr{J}_{l}-\mathscr{J}_{s}$ in Eq. (\ref{eq:Interface_MC_PF})
and $\Delta\tilde{\mathscr{F}}=\tilde{\mathscr{F}}_{l}-\tilde{\mathscr{F}}_{s}$,
$\Delta\mathscr{G}_{\Phi}=\mathscr{G}_{l}-\mathscr{G}_{\Phi}$ in
Eq. (\ref{eq:Gibbs-Thomson_PF}). The integrals vanish with an appropriate
choice of interpolation functions $p(\phi)$, $h(\phi)$ and $q(\phi)$.
Those summarized in Tab. \ref{tab:Interpolation-functions-used} fulfill
the requirements $\mathscr{H}_{l}=\mathscr{H}_{s}$ and $\mathscr{J}_{l}=\mathscr{J}_{s}$.
For satisfying the conditions $\mathscr{F}_{l}=\mathscr{F}_{s}$ and
$\mathscr{G}_{l}=\mathscr{G}_{s}$, we must also consider identical
diffusivities for each phase (i.e. $q_{s}=D_{s}/D_{l}=1)$. When $D_{s}=0$,
the discussion with anti-trapping current (i.e. $a(\phi_{0})\neq0$
in Tab. \ref{tab:integrals}) is detailed in Section \ref{subsec:Case-of-anti-trapping}.

\end{subequations}

\begin{table*}
\begin{centering}
\begin{tabular}{ll|ll}
\hline 
\textbf{\small{}Integral} & \textbf{\small{}Value} & \textbf{\small{}Integral} & \textbf{\small{}Value}\tabularnewline
\hline 
{\footnotesize{}$\mathscr{I}={\displaystyle \int_{-\infty}^{\infty}(\partial_{\xi}\phi_{0})^{2}d\xi}$} & {\scriptsize{}${\displaystyle \frac{2}{3}}$} & {\footnotesize{}${\displaystyle \mathscr{K}=\int_{-\infty}^{\infty}\left\{ \partial_{\xi}p(\phi_{0})\int_{0}^{\xi}\left[\frac{h(\phi_{0})-a(\phi_{0})\partial_{\xi}\phi_{0}}{q(\phi_{0})}\right]dx\right\} d\xi}$} & {\scriptsize{}${\displaystyle \frac{31-30\ln2}{150}}$}\tabularnewline
{\footnotesize{}$\mathscr{F}_{l}={\displaystyle \int_{0}^{\infty}\left[1-\frac{h(\phi_{0})-a(\phi_{0})\partial_{\xi}\phi_{0}}{q(\phi_{0})}\right]d\xi}$} & {\scriptsize{}${\displaystyle \frac{\ln2}{4}}$} & {\footnotesize{}$\mathscr{F}_{s}={\displaystyle \int_{-\infty}^{0}\left[\frac{h(\phi_{0})-a(\phi_{0})\partial_{\xi}\phi_{0}}{q(\phi_{0})}\right]d\xi}$} & {\scriptsize{}${\displaystyle \frac{\ln2}{4}}$}\tabularnewline
{\footnotesize{}${\displaystyle \tilde{\mathscr{F}}_{l}=\int_{0}^{\infty}\left[1-\frac{p(\phi_{0})}{q(\phi_{0})}\right]d\xi}$} & $\frac{\ln(2)-1}{4}$ & {\footnotesize{}$\tilde{\mathscr{F}}_{s}={\displaystyle \int_{-\infty}^{0}\frac{p(\phi_{0})}{q(\phi_{0})}d\xi}$} & $\frac{\ln(2)+1}{4}$\tabularnewline
{\footnotesize{}$\mathscr{G}_{l}={\displaystyle \int_{0}^{\infty}\left[\frac{1}{q(\phi_{0})}-1\right]d\xi}$} & {\scriptsize{}${\displaystyle \left[\frac{1}{q_{s}}-1\right]\frac{\ln(q_{s}+1)}{4}}$} & {\footnotesize{}$\mathscr{G}_{s}={\displaystyle \int_{0}^{-\infty}\left[\frac{1}{q(\phi_{0})}-\frac{1}{q_{s}}\right]d\xi}$} & {\scriptsize{}${\displaystyle \left[\frac{1}{q_{s}}-1\right]\frac{\ln(q_{s}+1)-\ln(q_{s})}{4}}$}\tabularnewline
{\footnotesize{}$\mathscr{H}_{l}={\displaystyle \int_{0}^{\infty}\left[1-h(\phi_{0})\right]d\xi}$} & {\scriptsize{}${\displaystyle \frac{\ln2}{4}}$} & {\footnotesize{}$\mathscr{H}_{s}={\displaystyle \int_{-\infty}^{0}h(\phi_{0})d\xi}$} & {\scriptsize{}${\displaystyle \frac{\ln2}{4}}$}\tabularnewline
{\footnotesize{}$\mathscr{J}_{l}={\displaystyle \int_{0}^{\infty}\left[q(\phi_{0})-1\right]d\xi}$} & {\scriptsize{}${\displaystyle -\frac{\ln2}{4}}$} & {\footnotesize{}$\mathscr{J}_{s}={\displaystyle \int_{0}^{-\infty}\left[q(\phi_{0})-q_{s}\right]d\xi}$} & {\scriptsize{}${\displaystyle -\frac{\ln2}{4}}$}\tabularnewline
\hline 
\end{tabular}
\par\end{centering}
\caption{\label{tab:integrals}Definition of integrals involved in Eqs. (\ref{eq:Interface_MC_PF}),
(\ref{eq:Gibbs-Thomson_PF}), (\ref{eq:CapillaryLength}) and (\ref{eq:KineticCoeff}).
Their values are computed with $q_{s}=0$ (except $\mathscr{G}_{l}$
and $\mathscr{G}_{s}$ discussed in \ref{subsec:Discussion-on-error})
and the interpolation functions defined in Tab. \ref{tab:Interpolation-functions-used}.
Here $\xi=x/W$ and $\phi_{0}$ is defined by Eq. (\ref{eq:Sol_TanH}).}
\end{table*}

As expected, the term in the left-hand side of the Gibbs-Thomson condition
(Eq. (\ref{eq:Gibbs-Thomson_PF})), appears in the source term $\mathscr{S}_{\phi}$
of $\phi$-equation (Eq. (\ref{eq:SourceTerm_Coexistence})). Let
us emphasize that the index $\Phi$ appears in Eq. (\ref{eq:Gibbs-Thomson_PF})
because the condition is not necessarily the same for each side of
the interface: the kinetic coefficient $\beta_{l}$ can be different
of $\beta_{s}$ (see Eqs. (\ref{eq:Gibbs-Thomson_Liquid}) and (\ref{eq:Gibbs-Thomson_Solid})
in \ref{subsec:phi-eq_Oeps2}). More precisely, the capillary length
$d_{0}$ and the kinetic coefficient $\beta_{\Phi}$ are related to
$W$, $\lambda$ and $M_{\phi}$ of the phase-field equation by: 

\begin{subequations}

\begin{align}
d_{0} & =\mathscr{I}\frac{W}{\lambda}\label{eq:CapillaryLength}\\
\beta_{\Phi} & =\frac{W\mathscr{I}}{M_{\phi}\lambda}\left[1-\lambda\frac{M_{\phi}}{D_{l}}\frac{\mathscr{K}+\mathscr{F}_{\Phi}}{\mathscr{I}}\left(\Delta c^{co}\right)^{2}\right]\label{eq:KineticCoeff}
\end{align}
The two different values of $\beta_{s}$ and $\beta_{l}$ come from
the integral $\mathscr{F}_{\Phi}$ in Eq. (\ref{eq:KineticCoeff}).
A single value $\beta_{l}=\beta_{s}=\beta$ is obtained provided that
$\mathscr{F}_{l}=\mathscr{F}_{s}=\mathscr{F}$. The integrals $\mathscr{F}_{\Phi}$,
$\mathscr{I}$ and $\mathscr{K}$ of Eqs. (\ref{eq:CapillaryLength})-(\ref{eq:KineticCoeff})
are defined in Tab. \ref{tab:integrals}.

\end{subequations}

For validations of Section \ref{sec:Validations}, the comparisons
between the numerical simulations of phase-field model and the analytical
solutions of Stefan's problem are carried out by considering $\beta=0$.
The particular value of $\lambda$ that fulfills that requirement
is noted $\lambda^{\star}$ and writes:

\begin{equation}
\lambda^{\star}=\frac{D_{l}}{M_{\phi}\left(\Delta c^{co}\right)^{2}}\frac{\mathscr{I}}{\mathscr{K}+\mathscr{F}}\label{eq:Lambda-Star}
\end{equation}

Finally, Eq. (\ref{eq:CapillaryLength}) relates the capillary length
$d_{0}$ to the interface width $W$ and the coupling coefficient
$\lambda$. The counter term $-M_{\phi}\kappa\bigl|\boldsymbol{\nabla}\phi\bigr|$
must be considered in the phase-field equation when $d_{0}$ is negligible
in the Gibbs-Thomson Eq. (\ref{eq:Gibbs-Thomson_PF}). The capillary
length $d_{0}$ is directly related to the surface tension $\sigma$.
Indeed, from its definition Eq. (\ref{eq:Width_Sigma}), we have $\sigma=(1/6)\sqrt{2\zeta H}$
and we use the relationships $1/\zeta=8/(W^{2}H)$ and $H=2\mathscr{E}/\lambda$
to find $\sigma=\left[(2/3)W/\lambda\right]\mathscr{E}$. The term
inside the brackets is Eq. (\ref{eq:CapillaryLength}) with $\mathscr{I}=2/3$
i.e. $\sigma=d_{0}\mathscr{E}$. If the surface tension of the system
can be neglected, then the counter term must be considered in $\phi$-equation.
Its impact is illustrated in the simulation of Section \ref{sec:Simulations}.

\subsubsection{\label{subsec:Case-of-anti-trapping}Analysis of anti-trapping current
$\boldsymbol{j}_{at}$ in $c$-equation}

When $q_{s}=D_{s}/D_{l}\neq1$, the model does not satisfy the conditions
$\mathscr{F}_{l}=\mathscr{F}_{s}$ and $\mathscr{G}_{l}=\mathscr{G}_{s}$
(see Tab. \ref{tab:integrals} with $a(\phi_{0})=0$). The reason
is that the diffusive behavior is not symmetric anymore inside the
interface. Adding an anti-trapping current $\boldsymbol{j}_{at}$
(Eq. (\ref{eq:Anti-trapping_Current}) inside the composition equation
becomes necessary to correct this asymmetry. In the most general cases,
the coefficient $a(\phi)$ is a function of $\phi$ which adds a supplementary
freedom degree in the model to cancel $\Delta\mathscr{F}$. In this
work, the asymptotic expansions were performed with $\boldsymbol{j}_{at}$
in order to determine the correct form of $a(\phi)$. From Tab. (\ref{tab:integrals})
we can see that $a(\phi)$ is involved in three integrals $\mathscr{F}_{l}$,
$\mathscr{F}_{s}$ and $\mathscr{K}$. Computing the integrals with
the functions $p(\phi)=\phi^{2}(3-2\phi)$, $h(\phi)=\phi$ and $q(\phi)=\phi+(1-\phi)q_{s}$
yields $a=(1-q_{s})/4$. When $D_{s}=0$, that condition simplifies
to $a=1/4$. Another way to derive that value is to consider that
the integrands of condition $\mathscr{F}_{l}=\mathscr{F}_{s}$ (see
Tab. \ref{tab:integrals}) must be identical to those of condition
$\mathscr{H}_{l}=\mathscr{H}_{s}$ yielding the relationship $\left[h(\phi_{0})-a(\phi_{0})\partial\phi_{0}/\partial\xi\right]/q(\phi_{0})=h(\phi_{0})$.
The value $a=1/4$ directly arises from that equality (see \ref{subsec:Discussion-on-error}).
When $q_{s}\neq1$, the condition $\mathscr{G}_{l}=\mathscr{G}_{s}$
cannot be satisfied with the current model, even with the anti-trapping
current. However, the spurious term $\mathbb{E}_{1}$ in Eq. (\ref{eq:Gibbs-Thomson_PF})
vanishes when $D_{s}=0$. Finally, the sharp interface model is recovered
for $D_{s}=D_{l}$ and when $D_{s}=0$ with anti-trapping. When $D_{s}/D_{l}\sim1$,
the error term $\mathbb{E}_{1}\Delta\mathscr{G}_{\Phi}$ remains very
small as confirmed by Section \ref{subsec:Validation-Precipitation}.

\subsubsection{\label{subsec:Case-of-counter}Analysis of counter term $-M_{\phi}\kappa\bigl|\boldsymbol{\nabla}\phi\bigr|$
in $\phi$-equation}

The references \citep{Folch_etal_PRE1999} and \citep{Jamet-Misbah_PRE2008}
have proved that adding the counter term $-M_{\phi}\kappa\bigl|\boldsymbol{\nabla}\phi\bigr|$
cancels the curvature motion $-d_{0}\kappa$ in the Gibbs-Thomson
condition. As a matter of fact, the curvature-driven term $-d_{0}\kappa$
arises from the asymptotic expansions of standard $\phi$-equation
(Eq. (\ref{eq:ModelA_EqPhi})). More precisely it arises from the
expansion of two terms: the laplacian term and the double-well one.
In references \citep{Folch_etal_PRE1999} and \citep{Jamet-Misbah_PRE2008}
such an analysis has been performed directly by adding $-M_{\phi}\kappa\bigl|\boldsymbol{\nabla}\phi\bigr|$
to those two terms.

Here, the phase-field equation (Eq. (\ref{eq:ModelA_EqPhi_CounterTerm}))
differs slightly of $\phi$-equations of those references. It has
been reformulated in Section \ref{subsec:CounterTerm} by using the
kernel function $\bigl|\boldsymbol{\nabla}\phi\bigr|=(4/W)\sqrt{\omega_{dw}}$
(Eq. (\ref{eq:KernelFunction})) and the chain rule of the divergence
operator. The manipulations made to obtain Eq. (\ref{eq:ModelA_EqPhi_CounterTerm})
conserves the structure of order zero of the phase-field equation
and the Gibbs-Thomson condition. The results of the asymptotic expansions
of Eq. (\ref{eq:ModelA_EqPhi_CounterTerm}) were found to be equivalent
to those of the previous references: the equation guarantees canceling
the curvature motion of the interface.

\section{\label{sec:Lattice-Boltzmann-methods}Lattice Boltzmann methods}

The phase-field model of section \ref{subsec:Summary-of-phase-field}
is implemented in \texttt{LBM\_saclay}, a 3D numerical code written
in C++ language. The main advantage of this code is its portability
on all major HPC architectures (especially GPUs and CPUs). It has
already been used to study two-phase flows with phase change in the
framework of the phase-field method in reference \citep{Verdier_etal_CMAME2020}.
Section \ref{subsec:LBM-notations} introduces the main notations
and the lattice. The LBM schemes for $\phi$-equation are presented
in Section \ref{subsec:LBM_phi} and those for $c$-equation in Section
\ref{subsec:LBM_c}.

\subsection{\label{subsec:LBM-notations}LBM notations}

Several standard lattices have already been implemented in top-level
files of \texttt{LBM\_saclay}. The two-dimensional lattices are D2Q5
and D2Q9 and the three-dimensional ones are D3Q7, D3Q15 and D3Q19.
The lattice speed is $s=\delta x$/$\delta t$ where $\delta x$ and
$\delta t$ are respectively the space- and time-steps. Among all
those lattices, we only use in this work the standard D2Q9 one. It
is defined by nine directions of displacement, each one of them is
indexed by $k=0,\,\ldots,\,N_{pop}$ with $N_{pop}=8$. The nine vectors
are $\boldsymbol{e}_{0}=(0,\,0)^{T}$, $\boldsymbol{e}_{1}=(1,\,0)^{T}$,
$\boldsymbol{e}_{2}=(0,\,1)^{T}$, $\boldsymbol{e}_{3}=(-1,\,0)^{T}$,
$\boldsymbol{e}_{4}=(0,\,-1)^{T}$, $\boldsymbol{e}_{5}=(1,\,1)^{T}$,
$\boldsymbol{e}_{6}=(-1,\,1)^{T}$, $\boldsymbol{e}_{7}=(-1,\,-1)^{T}$
and $\boldsymbol{e}_{8}=(1,\,-1)^{T}$. The lattice velocities are
defined by $\boldsymbol{\xi}_{k}=s\boldsymbol{e}_{k}$ and the lattice
weights are $w_{0}=4/9$, $w_{1,...,4}=1/9$ and $w_{5,...,8}=1/36$.
The lattice coefficient is noted $\xi_{s}^{2}=s^{2}/3$.

\begin{table*}
\begin{centering}
\textbf{Nomenclature for lattice Boltzmann}\\
\textbf{}\\
\par\end{centering}
\begin{centering}
\begin{tabular}{llll}
\hline 
\textbf{Symbol} & \textbf{Definition} & \textbf{Dimension} & \textbf{Description}\tabularnewline
\hline 
$\boldsymbol{e}_{k}$ & Sec. \ref{subsec:LBM-notations} & {[}--{]} & Vectors of displacement on the lattice\tabularnewline
$k$ & $0\leq k\leq N_{pop}$ & {[}--{]} & Index for each direction of propagation\tabularnewline
$N_{pop}$ & $N_{pop}=8$ for D2Q9 & {[}--{]} & Total number directions\tabularnewline
$\delta x$ &  & {[}L{]} & Spatial discretization\tabularnewline
$\delta t$ &  & {[}T{]} & Time discretization\tabularnewline
$s$ & $=\delta x/\delta t$ & {[}L{]}.{[}T{]}$^{-1}$ & Lattice speed\tabularnewline
$\boldsymbol{\xi}_{k}$ & $=s\boldsymbol{e}_{k}$ & {[}L{]}.{[}T{]}$^{-1}$ & Velocities associated to the vectors of displacement\tabularnewline
$\xi_{s}^{2}$ & $=s^{2}/3$ & {[}L{]}$^{2}$.{[}T{]}$^{-2}$ & Lattice coefficient\tabularnewline
$g_{k}(\boldsymbol{x},\,t)$, $h_{k}(\boldsymbol{x},\,t)$ &  & {[}--{]} & Distribution function for $\phi$ and for $c$\tabularnewline
$g_{k}^{eq}$, $h_{k}^{eq}$ & Eq. (\ref{eq:Geq_Allen-Cahn_Eq-1}) and Eq. (\ref{eq:Heq_Composition}) & {[}--{]} & Equilibrium distribution functions in LBE for $g_{k}$ and $h_{k}$\tabularnewline
$\overline{\tau}_{g}$, $\overline{\tau}_{h}$ &  & {[}--{]} & Collision rate in LBE for $g_{k}$ and $h_{k}$\tabularnewline
$\mathcal{G}_{k}$, $\mathcal{H}_{k}$ & Eq. (\ref{eq:TermG}) and Eq. (\ref{eq:SourceTerm_LBE_h}) & {[}T{]}$^{-1}$ & Source terms in LBE for $g_{k}$ and $h_{k}$\tabularnewline
$w_{k}$ &  & {[}--{]} & Weights (constant) for each LBE\tabularnewline
$\gamma$ & $\overline{\tau}_{h}=3\delta t/(\gamma\delta x^{2})$ & {[}L{]}$^{-2}$.{[}T{]} & Parameter related to $\overline{\tau}_{h}$ in LBE for $h_{k}$\tabularnewline
\hline 
\end{tabular}
\par\end{centering}
\caption{\label{tab:Nomenclature_LBM}Main mathematical symbols for the lattice
Boltzmann schemes (LBE: lattice Boltzmann equation).}
\end{table*}

On that lattice, two distribution functions $g_{k}(\boldsymbol{x},\,t)$
and $h_{k}(\boldsymbol{x},\,t)$ are defined, for updating respectively
the phase-field $\phi(\boldsymbol{x},\,t)$ and the composition $c(\boldsymbol{x},\,t)$
at each time-step. No distribution function is introduced for the
chemical potential $\overline{\mu}(\boldsymbol{x},\,t)$. Here $\overline{\mu}$
is simply an additional macroscopic field which is kept in memory
for updating $c(\boldsymbol{x},\,t)$. A LBM using $\overline{\mu}$
as main variable instead of $c$ could have been possible. Indeed,
the mathematical form of Eq. (\ref{eq:Potchem_Eq_Omega}) is similar
to the supersaturation equation of reference \citep{Cartalade_etal_CAMWA2016}.
But here, we use $c$ as main computational variable for reason of
mass conservation.

The evolution of distribution functions $g_{k}$ and $h_{k}$ obeys
the ``discrete velocity lattice Boltzmann equations'' with a collision
approximated by the BGK operator. With that form of collision, each
distribution function relaxes toward an equilibrium $g_{k}^{eq}$
and $h_{k}^{eq}$ proportionally to collision times $\tau_{g}$ and
$\tau_{h}$. For each LBE, the source terms are noted $\mathcal{G}_{k}$
and $\mathcal{H}_{k}$. The space and time discretizations are performed
by method of characteristics. The BGK collision operators and the
source terms are integrated with the trapezoidal rule, a method of
second-order accuracy. In order to keep an explicit algorithm, the
variable changes of $g_{k}$ and $h_{k}$ are defined by $\tilde{g}_{k}=g_{k}+(\delta t/2\tau_{g})(g_{k}-g_{k}^{eq})-\mathcal{G}_{k}(\delta t/2)$
and $\tilde{h}_{k}=h_{k}+(\delta t/2\tau_{h})(h_{k}-h_{k}^{eq})-\mathcal{H}_{k}(\delta t/2)$.
The ratios $\tau_{g}/\delta t$ and $\tau_{h}/\delta t$ are the dimensionless
collision rates respectively noted $\overline{\tau}_{g}$ and $\overline{\tau}_{h}$.
All details of that variable change can be found in \citep[Appendix C]{Verdier_etal_CMAME2020}.

\subsection{\label{subsec:LBM_phi}LBM for $\phi$-equation}

The lattice Boltzmann method for the phase-field equation acts on
the distribution function $\tilde{g}_{k}$. The evolution equation
is:

\begin{align}
\tilde{g}_{k}^{\star} & =\tilde{g}_{k}-\frac{1}{\overline{\tau}_{g}+1/2}\left[\tilde{g}_{k}-\tilde{g}_{k}^{eq}\right]+\mathcal{G}_{k}\delta t\label{eq:LBE_PhaseField_Eq}
\end{align}
where $\tilde{g}{}_{k}^{\star}\equiv\tilde{g}_{k}(\boldsymbol{x}+\boldsymbol{\xi}_{k}\delta t,\,t+\delta t)$
and the variable change $\tilde{g}_{k}^{eq}=g_{k}^{eq}-\delta t\mathcal{G}_{k}/2$
has been used. The equilibrium distribution function $g_{k}^{eq}$
is defined by:

\begin{equation}
g_{k}^{eq}=\phi w_{k}\label{eq:Geq_Allen-Cahn_Eq-1}
\end{equation}
for which its moments are $\phi$ (moment of order zero), $\boldsymbol{0}$
(order one) and $\phi\boldsymbol{I}$ (order two) where $\boldsymbol{I}$
is the identity tensor of second-order. The diffusivity coefficient
is related to the collision rate by $M_{\phi}=\overline{\tau}{}_{g}\xi_{s}^{2}\delta t$.
The source term $\mathcal{G}_{k}$ contains two contributions:

\begin{equation}
\mathcal{G}_{k}=w_{k}\left(\mathcal{G}^{st}+\mathcal{G}^{curv}\right)\label{eq:TermG}
\end{equation}
The first one $\mathcal{G}^{st}$ involves the source term $\mathscr{S}_{\phi}(\phi,\,\overline{\mu})$
defined by Eq. (\ref{eq:SourceG_ST}). The second one $\mathcal{G}^{curv}$
is either equal to the double-well term $\mathcal{G}^{dw}$ or equal
to the counter term $\mathcal{G}_{k}^{ct}$. The three source terms
are defined by:

\begin{subequations}

\begin{align}
\mathcal{G}^{st} & =-\frac{\lambda M_{\phi}}{W^{2}}\mathscr{S}_{\phi}(\phi,\,\overline{\mu})\label{eq:SourceG_ST}\\
\mathcal{G}^{dw} & =-\frac{8M_{\phi}}{W^{2}}\omega_{dw}^{\prime}(\phi)\label{eq:SourceG_dw}\\
\mathcal{G}_{k}^{ct} & =\frac{4}{W}\phi(1-\phi)\boldsymbol{\xi}_{k}\cdot\boldsymbol{n}\label{eq:SourceG_CT}
\end{align}

\end{subequations}

The choice between $\mathcal{G}^{dw}$ or $\mathcal{G}_{k}^{ct}$
depends on the curvature-driven motion term i.e. the version of the
phase-field equation we wish to simulate. For simulating Eq. (\ref{eq:ModelA_EqPhi}),
the curvature term must contain the double-well $\omega_{dw}(\phi)$.
In that case $\mathcal{G}^{curv}$ is equal to Eq. (\ref{eq:SourceG_dw}).
If the curvature-driven motion is undesired, the term must involve
the kernel function $\bigl|\boldsymbol{\nabla}\phi\bigr|=(4/W)\phi(1-\phi)$
with the normal vector $\boldsymbol{n}$. In that case $\mathcal{G}^{curv}$
is equal to Eq. (\ref{eq:SourceG_CT}).

After the stages of collision and streaming, the new phase-field is
obtained by the zeroth-order moment of $\tilde{g}_{k}$ which must
be corrected with the source term $\mathcal{G}_{k}$:

\begin{equation}
\phi=\sum_{k}\tilde{g}_{k}+\frac{\delta t}{2}\sum_{k}\mathcal{G}_{k}.\label{eq:Moment_Phi}
\end{equation}

The unit normal vector $\boldsymbol{n}$ requires the computation
of gradients of $\phi$. The gradients are discretized by using the
method of directional derivatives. The method has already demonstrated
its performance in hydrodynamics in order to reduce parasitic currents
for two-phase flow problems \citep{Lee_Parasitic_CAMWA2009,Lee-Liu_DropImpact_LBM_JCP2010}.
The directional derivative is the derivative along each moving direction
on the lattice. The Taylor expansions at second-order of a differentiable
scalar function $\phi(\boldsymbol{x})$ at $\boldsymbol{x}+\boldsymbol{e}_{k}\delta x$
and $\boldsymbol{x}-\boldsymbol{e}_{k}\delta x$ yields the following
approximation of directional derivatives:

\begin{subequations}

\begin{equation}
\boldsymbol{e}_{k}\cdot\boldsymbol{\nabla}\phi\bigr|_{\boldsymbol{x}}=\frac{1}{2\delta x}\left[\phi(\boldsymbol{x}+\boldsymbol{e}_{k}\delta x)-\phi(\boldsymbol{x}-\boldsymbol{e}_{k}\delta x)\right]\label{eq:DerivDirec_Grad}
\end{equation}

The number of directional derivatives is equal to the number of moving
directions $\boldsymbol{e}_{k}$ on the lattice i.e. $N_{pop}$. The
gradient is obtained by:

\begin{equation}
\boldsymbol{\nabla}\phi\bigr|_{\boldsymbol{x}}=3\sum_{k}w_{k}\boldsymbol{e}_{k}\left(\boldsymbol{e}_{k}\cdot\boldsymbol{\nabla}\phi\bigr|_{\boldsymbol{x}}\right).\label{eq:Grad}
\end{equation}

\end{subequations}

The two components of gradient $\partial_{x}\phi$ and $\partial_{y}\phi$
are computed by the moment of first-order of each directional derivative
$\boldsymbol{e}_{k}\cdot\boldsymbol{\nabla}\phi\bigr|_{\boldsymbol{x}}$.

\subsection{\label{subsec:LBM_c}LBM for $c$-equation}

The basic LB algorithm for composition equation works on a new distribution
function $h_{k}$. The specificity of Eq. (\ref{eq:ModelA_EqCompos})
is the mixed formulation between $c$ and $\overline{\mu}$. The closure
relationship is given by Eq. (\ref{eq:Closure}). The equilibrium
distribution $h_{k}^{eq}$ must be designed such as its moment of
zeroth-order is $c$ and its moment of second-order is $\boldsymbol{I}\overline{\mu}$.
That equation is quite close to the Cahn-Hilliard (CH) equation with
a simpler closure (Eq. (\ref{eq:Closure})) which does not involve
the laplacian of $c$ (case of CH equation). The numerical scheme
can be inspired from what is done for CH equation for two-phase flows
of two immiscible fluids \citep{Zheng_etal_LargeDensityRatio_JCP2006,Fakhari-Rahimian_PRE2010}.
For anti-trapping current $\boldsymbol{j}_{at}$, the methods are
the same as those presented in \citep{Cartalade_etal_CAMWA2016} for
crystal growth applications of binary mixture.

In the usual BGK operator, the diffusion coefficient $\mathcal{D}(\phi)$
is related to the relaxation time $\overline{\tau}_{h}(\phi)$ with
the relationship $\mathcal{D}(\phi)=(1/3)\overline{\tau}_{h}(\phi)\delta x^{2}/\delta t$.
However, the interpolation of diffusion $\mathcal{D}(\phi)=D_{l}\phi$
means that the diffusion is null in the solid phase. In that case,
the relaxation time would be equal to $0$ leading to the occurrence
of instabilities in the algorithm. In order to overcome the instability,
the diffusive term is reformulated with the chain rule by $\boldsymbol{\nabla}\left[\mathcal{D}(\phi)\overline{\mu}\right]=\mathcal{D}(\phi)\boldsymbol{\nabla}\overline{\mu}+\mathcal{D}^{\prime}(\phi)\overline{\mu}\boldsymbol{\nabla}\phi$
with $\mathcal{D}^{\prime}(\phi)=D_{l}$. Eq. (\ref{eq:ModelA_EqCompos})
becomes:

\begin{equation}
\frac{\partial c}{\partial t}+\boldsymbol{\nabla}\cdot\left[\overline{\mu}\mathcal{D}^{\prime}(\phi)\boldsymbol{\nabla}\phi\right]+\boldsymbol{\nabla}\cdot\boldsymbol{j}_{at}=\boldsymbol{\nabla}^{2}\left[\mathcal{D}(\phi)\overline{\mu}\right]\label{eq:Concentration_DissolutionModel_Reformulated}
\end{equation}
Moreover, the laplacian term is reformulated as $\boldsymbol{\nabla}^{2}\left[\mathcal{D}(\phi)\overline{\mu}\right]=(1/\gamma)\boldsymbol{\nabla}^{2}\left[\gamma\mathcal{D}(\phi)\overline{\mu}\right]$
where $\gamma$ is a supplementary parameter allowing a better control
of the relaxation rate. When the parameters $M_{\phi}$ and $\mathcal{D}(\phi)$
presents a ratio of several order of magnitude, it is useful to set
$\gamma=1/M_{\phi}$. The stability condition of the relaxation rates
will be the same for both LBE. The discrete lattice Boltzmann equation
writes

\begin{equation}
\tilde{h}_{k}^{\star}=\tilde{h}_{k}-\frac{1}{\overline{\tau}_{h}+1/2}\left[\tilde{h}_{k}-\tilde{h}_{k}^{eq}\right]+\mathcal{H}_{k}\delta t\label{eq:LBE_Composition}
\end{equation}
where $\tilde{h}_{k}^{\star}\equiv\tilde{h}_{k}(\boldsymbol{x}+\boldsymbol{\xi}_{k}\delta t,\,t+\delta t)$
and $\tilde{h}_{k}^{eq}=h_{k}^{eq}-\mathcal{H}_{k}\delta t/2$. The
equilibrium distribution function writes:

\begin{align}
h_{k}^{eq} & =\begin{cases}
c(\phi,\,\overline{\mu})-(1-w_{0})\gamma\mathcal{D}(\phi)\overline{\mu}(\boldsymbol{x},\,t) & \text{if }k=0\\
w_{k}\gamma\mathcal{D}(\phi)\overline{\mu}(\boldsymbol{x},\,t) & \text{if }k\neq0
\end{cases}\label{eq:Heq_Composition}
\end{align}
The first line of Eq. (\ref{eq:Heq_Composition}) corresponds to a
moment of order zero that is equal to $c$. The second line corresponds
to a second-order moment equal to $\mathcal{D}(\phi)\overline{\mu}\boldsymbol{I}$.
The anti-trapping current $\boldsymbol{j}_{at}$ and the term $\mathcal{D}^{\prime}(\phi)\overline{\mu}\boldsymbol{\nabla}\phi$
appear in the source term $\mathcal{H}_{k}\equiv\mathcal{H}_{k}(\boldsymbol{x},\,t)$
defined by:

\begin{equation}
\mathcal{H}_{k}=\gamma w_{k}\boldsymbol{\xi}_{k}\cdot\left[\overline{\mu}\mathcal{D}^{\prime}(\phi)\boldsymbol{\nabla}\phi+\boldsymbol{j}_{at}(\phi,\,\overline{\mu})\right]\label{eq:SourceTerm_LBE_h}
\end{equation}
The relaxation rate $\overline{\tau}_{h}$ is related to $\gamma$
by $\overline{\tau}_{h}=3\delta t/(\gamma\delta x^{2})$. After the
stages of collision and streaming the composition $c(\phi,\,\overline{\mu})$
is updated by:

\begin{equation}
c=\sum_{k}\tilde{h}_{k}\label{eq:Moment0_LBM_Concent}
\end{equation}

The moment of zeroth-order of Eq. (\ref{eq:SourceTerm_LBE_h}) is
null. For this reason, $\mathcal{H}_{k}$ does not appear in the calculation
of $c$. Once the new composition is known, the chemical potential
$\overline{\mu}$ is computed by Eq. (\ref{eq:PotChim_Equilibirum})
and used in equilibrium function (Eq. (\ref{eq:Heq_Composition})).
The anti-trapping current $\boldsymbol{j}_{at}$ is computed by Eq.
(\ref{eq:Anti-trapping_Equilibrium}) where the normal vector $\boldsymbol{n}$
and the time derivative $\partial\phi/\partial t$ are required. The
normal vector has already been computed in Section \ref{subsec:LBM_phi}.
The time derivative of $\phi$ is computed by an explicit Euler scheme
of first-order. Hence, the LBE on $\tilde{h}_{k}$ must be solved
after the LBE on $\tilde{g}_{k}$. At first time-step the term $\partial\phi/\partial t$
in Eq. (\ref{eq:Anti-Trapping_Quadratic}) is obtained by $\phi(\boldsymbol{x},\,\delta t)-\phi(\boldsymbol{x},\,t=0)/\delta t$
where $\phi(\boldsymbol{x},\,t=0)$ is the initial condition and $\phi(\boldsymbol{x},\,\delta t)$
is the phase-field after the first time-step.

Another formulation is possible for $\boldsymbol{j}_{at}$ and $\overline{\mu}\mathcal{D}^{\prime}(\phi)\boldsymbol{\nabla}\phi$.
They could have been included inside an alternative equilibrium distribution
function $h_{k}^{eq,\,alt}$ with $\mathcal{H}_{k}=0$. In that case,
the scheme writes

\begin{subequations}

\begin{equation}
\tilde{h}_{k}^{\star}=\tilde{h}_{k}-\frac{1}{\overline{\tau}_{h}+1/2}\left[\tilde{h}_{k}-h_{k}^{eq,\,alt}\right]\label{eq:LBE_Concent_Alt}
\end{equation}
with $h_{k}^{eq,\,alt}$ defined by

\begin{equation}
h_{k}^{eq,\,alt}=\begin{cases}
c(\phi,\,\overline{\mu})-(1-w_{0})\gamma\mathcal{D}(\phi)\overline{\mu} & \text{if }k=0\\
w_{k}\left[\gamma\mathcal{D}(\phi)\overline{\mu}+\frac{\boldsymbol{\xi}_{k}\cdot\left(\boldsymbol{j}_{at}+\overline{\mu}\mathcal{D}^{\prime}\boldsymbol{\nabla}\phi\right)}{\xi_{s}^{2}}\right] & \text{if }k\neq0
\end{cases}\label{eq:Seq_Concent_Alt}
\end{equation}
The computational stages for $\overline{\mu}$ and $\boldsymbol{j}_{at}$
remain the same as those presented above.

\end{subequations}

\section{\label{sec:Validations}Validations}

The implementation of lattice Boltzmann schemes is validated with
several analytical solutions. The solutions are obtained from the
classical Stefan's problem. We present one case of precipitation in
Section \ref{subsec:Validation-Precipitation} for $D_{l}\simeq D_{s}$
and one case of dissolution in Section \ref{subsec:Validation-Dissolution}
for $D_{s}=0$. The domain is one-dimensional with $x$ varying between
$[-L_{x},\,L_{x}]$ where $L_{x}=0.25$. The initial configuration
states an interface position located at $x_{i}(0)=0$ with a solid
phase on the left side (interval $[-L_{x},\,0[$) and a liquid phase
on the right side (interval $]0,\,L_{x}]$). For the phase-field model,
the first test is simulated without anti-trapping. Next, the second
one is simulated successively with and without $\boldsymbol{j}_{at}$
to present its impact on the profiles of composition and chemical
potential. For each validation, the relative $L^{2}$-errors, defined
by $\left\Vert \vartheta^{LBM}-\vartheta^{as}\right\Vert _{2}/\left\Vert \vartheta^{as}\right\Vert _{2}$,
are indicated in the caption of figures. The function $\vartheta$
corresponds to $c$ or $\overline{\mu}$. The errors are computed
only over the range shown on each graph and the superscript $as$
means ``analytical solution''.

The LBM simulations are carried out on a 2D computational domain varying
between $[-L_{x},\,L_{x}]\times[\ell_{y},\,L_{y}]$ with $\ell_{y}=0$
and $L_{y}=0.0036$. The D2Q9 lattice is used with $N_{x}\times N_{y}$
nodes with $N_{x}=5000$ and $N_{y}=36$. The space- and times-steps
are $\delta x=10^{-4}$ and $\delta t=5\times10^{-9}$. The initial
conditions for $\phi$-equation and $c$-equation are two hyperbolic
tangent functions: $\phi(x,\,0)$ is initialized by Eq. (\ref{eq:Sol_TanH})
and $c(x,\,0)$ by

\begin{equation}
c(x,\,0)=\frac{1}{2}\left[c_{l}^{\infty}+c_{s}^{\infty}+(c_{l}^{\infty}-c_{s}^{\infty})\tanh\left(\frac{2x}{W}\right)\right]\label{eq:C_init}
\end{equation}
where $c_{s}^{\infty}$ and $c_{l}^{\infty}$ are the compositions
of bulk far from interface. For horizontal walls at $y=0$ and $y=L_{y}$,
the boundary conditions are periodic. For vertical walls at $x=\pm L_{x}$,
the boundary conditions are imposed with the bounce-back method. A
preliminary test was carried out to check that solutions of both $\phi$-equations
(Eqs. (\ref{eq:ModelA_EqPhi}) and (\ref{eq:ModelA_EqPhi_CounterTerm}))
are identical on that one-dimensional case.

\subsection{\label{subsec:Validation-Precipitation}Validation with $D_{l}\simeq D_{s}$}

We first check the LBM implementation with two coefficients of diffusion:
$D_{s}=0.9$ and $D_{l}=1$. The analytical solution of such a problem
can be found in \citep[Chap. 12]{Hahn-Ozisik_2012}. However, in that
reference, the mathematical formulation of this problem is done by
using the temperature as main variable. The equivalent intensive quantity
in our model is the chemical potential. The validations using that
quantity will be presented in next section. Here, we prefer use the
solutions of reference \citep{Maugis_etal_ActaMater1997} which are
written in terms of compositions. The numerical implementation must
reproduce correctly the discontinuity of compositions at interface.

\begin{figure*}[t]
\begin{centering}
\subfloat[\label{fig:Interface-position_Ds-Dl}Interface position $x_{i}(t)$
increasing with time because of precipitation process: the interface
moves towards positive values of $x$.]{\begin{centering}
\includegraphics[angle=-90,scale=0.33]{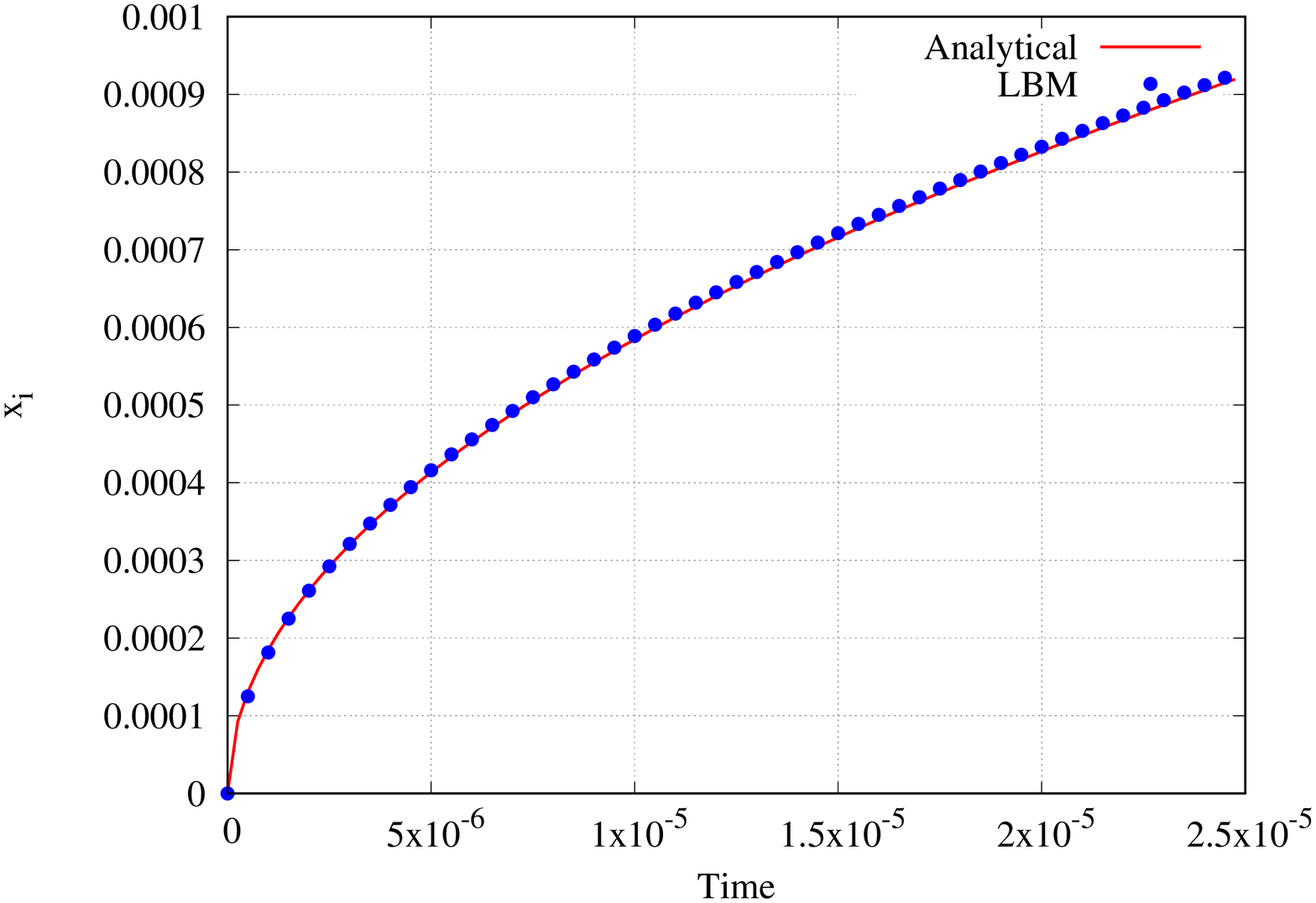}
\par\end{centering}
}~~~~\subfloat[{\label{fig:Composition_Ds-Dl}Compositions of solid and liquid w.r.t
$x$ at $t_{1}=2.5\times10^{-5}$ and $t_{2}=5\times10^{-4}$. The
two vertical lines (black) indicate the interface positions $x_{i}(t_{1})$
(close to 0) and $x_{i}(t_{2})$. Zoom between $[-0.04,\,0.04]$.
The $L^{2}$-errors are $4.8\times10^{-3}$ for $t_{1}$ and $t_{2}$.}]{\begin{centering}
\includegraphics[angle=-90,scale=0.33]{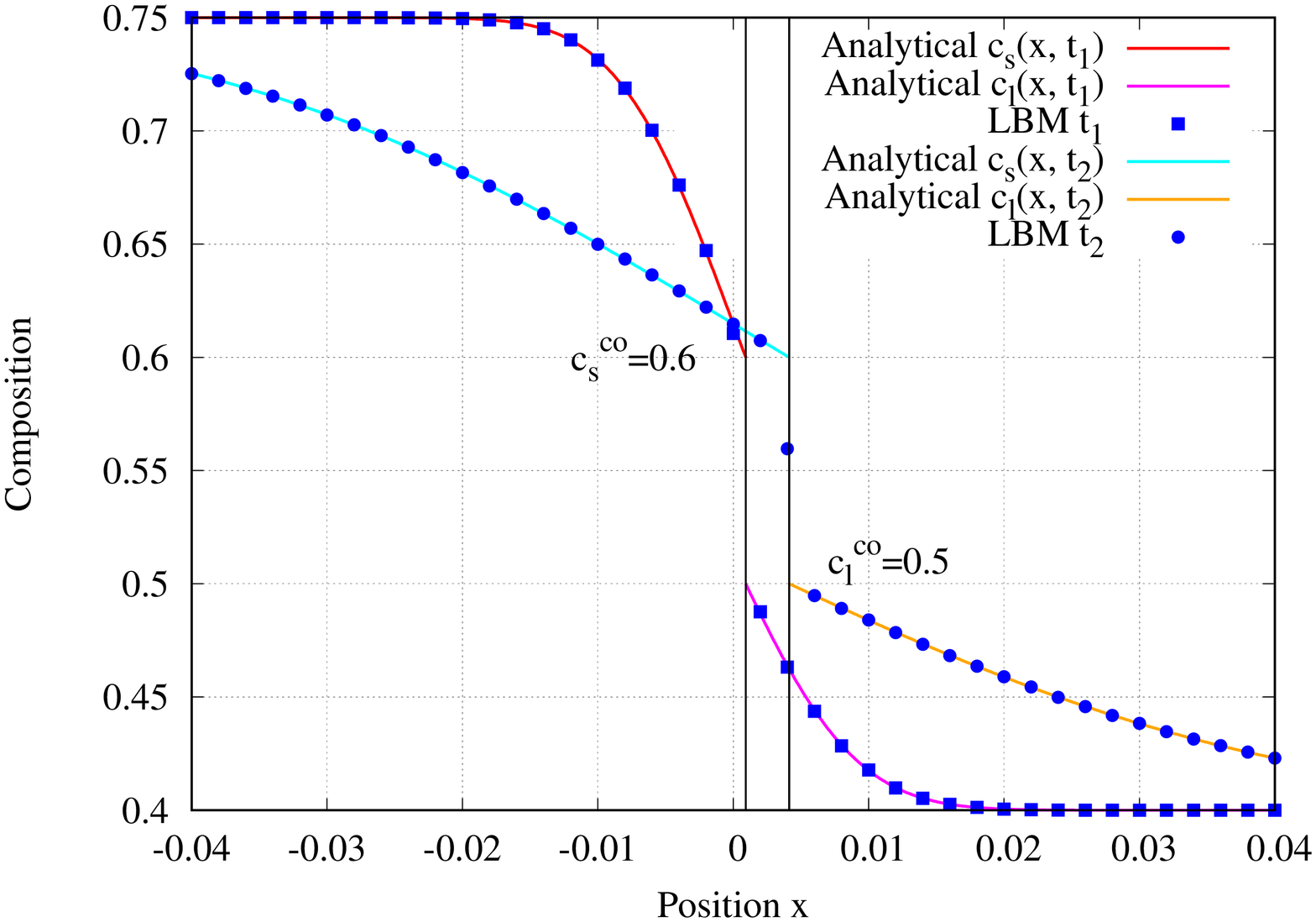}
\par\end{centering}
}
\par\end{centering}
\caption{\label{fig:Validation_Ds-Dl}Analytical solutions Eqs. (\ref{eq:x_i_Ds-Dl})--(\ref{eq:Compos_Liq})
(lines) compared to LBM (symbols) for one case of precipitation with
$D_{s}=0.9$ and $D_{l}=1$. The system is initialized with an interface
located at $x=0$. The solid and liquid are respectively on left-
and right-side.}
\end{figure*}

In \citep{Maugis_etal_ActaMater1997}, the solutions are derived for
a ternary case. For binary case, the transcendental equation reduces
to:

\begin{subequations}

\begin{align}
-\frac{1}{2}\alpha\Delta m{}^{2} & =\Delta\overline{f}^{min}\left[u_{s}(-\alpha)+u_{l}(\alpha)\right]+\nonumber \\
 & \qquad\Delta m\left[\left(m_{s}-c_{s}^{\infty}\right)u_{s}(-\alpha)+\right.\nonumber \\
 & \qquad\qquad\left.\left(m_{l}-c_{l}^{\infty}\right)u_{l}(\alpha)\right]\label{eq:Transcendental_Eq}
\end{align}
where the function $u_{\Phi}(\alpha)$ is defined by

\begin{equation}
u_{\Phi}(\alpha)=\sqrt{\frac{D_{\Phi}}{\pi}}\frac{e^{-\alpha^{2}/4D_{\Phi}}}{\text{erfc}\left(\alpha/2\sqrt{D_{\Phi}}\right)}\quad\text{for }\Phi=s,\,l\label{eq:solution_u}
\end{equation}

\end{subequations}

The compositions far from the interface are $c_{s}^{\infty}=0.75$
and $c_{l}^{\infty}=0.4$. For $m_{s}=0.2$, $m_{l}=0.1$, $\Delta m=0.1$
and $\Delta\overline{f}^{min}=0.04$, the root of the transcendental
equation is $\alpha=0.184841$. The three solutions are the interface
position $x_{i}(t)$, the composition of solid $c_{s}(x,\,t)$ and
the composition of liquid $c_{l}(x,\,t)$. The interface position
writes as a function of $\alpha$ and $t$:

\begin{subequations}

\begin{equation}
x_{i}(t)=\alpha\sqrt{t}\label{eq:x_i_Ds-Dl}
\end{equation}
Since $\alpha>0$, the interface moves from $x_{i}(0)=0$ towards
positive values of $x$, meaning that a precipitation process occurs.
The two analytical solutions in the solid and the liquid write:

\begin{align}
c_{s}^{as}(x,\,t) & =c_{s}^{\infty}+(c_{s}^{co}-c_{s}^{\infty})\frac{\text{erfc}\left[-x/2\sqrt{D_{s}t}\right]}{\text{erfc}\left[-\alpha/2\sqrt{D_{s}}\right]}\label{eq:Compos_Solid}\\
c_{l}^{as}(x,\,t) & =c_{l}^{\infty}+(c_{l}^{co}-c_{l}^{\infty})\frac{\text{erfc}\left[x/2\sqrt{D_{l}t}\right]}{\text{erfc}\left[\alpha/2\sqrt{D_{l}}\right]}\label{eq:Compos_Liq}
\end{align}
where Eq. (\ref{eq:Compos_Solid}) is defined for $x\in[-L_{x},\,x_{i}(t)[$
and Eq. (\ref{eq:Compos_Liq}) for $x\in]x_{i}(t)],\,L_{x}]$. On
the whole domain, the composition $c$ is discontinuous at interface
$x_{i}(t)$, of value $c_{s}^{co}=0.6$ on solid side and $c_{l}^{co}=0.5$
on liquid side. From those values, each profile of composition diffuses
until $c_{s}^{\infty}$ for $x\rightarrow-L_{x}$ and $c_{l}^{\infty}$
for $x\rightarrow L_{x}$.

\end{subequations}

The simulations are performed without anti-trapping current. The interpolation
of diffusion coefficients is simply done by $\mathcal{D}(\phi)=\phi D_{l}+(1-\phi)D_{s}$.
In $\phi$-equation the parameters are $M_{\phi}=1.2$, $W=1.2\times10^{-3}$
and $\lambda^{\star}=277$. The comparisons between the analytical
solutions and the LBM simulation are presented on Fig. \ref{fig:Validation_Ds-Dl}.
As expected from the theory, the interface position $x_{i}(t)$ is
an increasing function of time (Fig. \ref{fig:Interface-position_Ds-Dl}).
On the profiles of composition (Fig. \ref{fig:Composition_Ds-Dl}),
the jump on each side of the interface is also well-reproduced by
the numerical model. The coexistence values $c_{s}^{co}=0.6$ and
$c_{l}^{co}=0.5$ remain the same at two times $t_{1}=2.5\times10^{-5}$
and $t_{2}=5\times10^{-4}$. The LBM simulations fit perfectly with
the analytical solutions.

The two solutions Eqs. (\ref{eq:Compos_Solid})--(\ref{eq:Compos_Liq})
can be easily expressed in terms of chemical potential $\overline{\mu}_{s}^{as}(x,\,t)$
and $\overline{\mu}_{l}^{as}(x,\,t)$. For instance, we add $-m_{s}$
on both sides of Eq. (\ref{eq:Compos_Solid}) and add $m_{s}-m_{s}$
inside the term $(c_{s}^{co}-c_{s}^{\infty})$. Thanks to Eqs. (\ref{eq:Cl_coexistence})--(\ref{eq:Cs_coexistence})
we obtain $(c_{s}^{co}-c_{s}^{\infty})\equiv(\overline{\mu}^{eq}-\overline{\mu}_{s}^{\infty})$
for solid and $(c_{l}^{co}-c_{l}^{\infty})\equiv(\overline{\mu}^{eq}-\overline{\mu}_{l}^{\infty})$
for liquid. When expressed in terms of chemical potential, the solution
does not present a jump at interface $x_{i}(t)$. The single value
is $\overline{\mu}^{eq}$ and each profile diffuse from that value
until $\overline{\mu}_{s}^{\infty}$ when $x\rightarrow-L_{x}$ (solid)
and $\overline{\mu}_{l}^{\infty}$ when $x\rightarrow L_{x}$ (liquid).
The next section presents a validation using $\overline{\mu}$ as
main variable for discussing the analogy with temperature and comparing
with solidification problems.

\subsection{\label{subsec:Validation-Dissolution}Validation with $D_{s}=0$:
effect of anti-trapping current}

The analytical solution of the one-sided diffusion is presented in
\citep[Sec. 12-1]{Hahn-Ozisik_2012}. Now a direct analogy is done
between the temperature of that reference and the chemical potential
of our model. The two solutions are $x_{i}(t)$, the interface position,
and $\overline{\mu}_{l}^{as}(x,\,t)$ the chemical potential of liquid.
The chemical potential of solid is set equal to the equilibrium value
$\overline{\mu}^{eq}=0.4$. Its value remains constant during the
simulation because $D_{s}=0$. The transcendental equation of that
problem writes:

\begin{equation}
\alpha e^{\alpha^{2}}\text{erfc}(\alpha)+\frac{(\overline{\mu}^{eq}-\overline{\mu}_{l}^{\infty})}{(c_{s}^{co}-c_{l}^{co})\sqrt{\pi}}=0\label{eq:Cartesian_Transcendental}
\end{equation}
where $\alpha$ is the root of this equation, and $\overline{\mu}_{l}^{\infty}$
is the value of chemical potential far from the interface. By analogy
with problems of phase change (solidification or melting), the equilibrium
chemical potential $\overline{\mu}^{eq}$ plays the role of melting
temperature. The term $\Delta c^{co}=c_{s}^{co}-c_{l}^{co}$ can be
compared to the latent heat. For phase change problems, that quantity
is released (resp. absorbed) at interface during solidification (resp.
melting). Here for our convention $c_{s}^{co}>c_{l}^{co}$, the quantity
$\Delta c^{co}$ is released at interface during dissolution and absorbed
during precipitation. Finally, the quantity $\chi=1$ plays the role
of specific heat.

\begin{figure*}[t]
\begin{centering}
\subfloat[\label{fig:Interface-position_Ds=00003D0}Interface position $x_{i}(t)$
decreasing with time because of dissolution process: the interface
moves towards negative values of $x$.]{\begin{centering}
\includegraphics[angle=-90,scale=0.33]{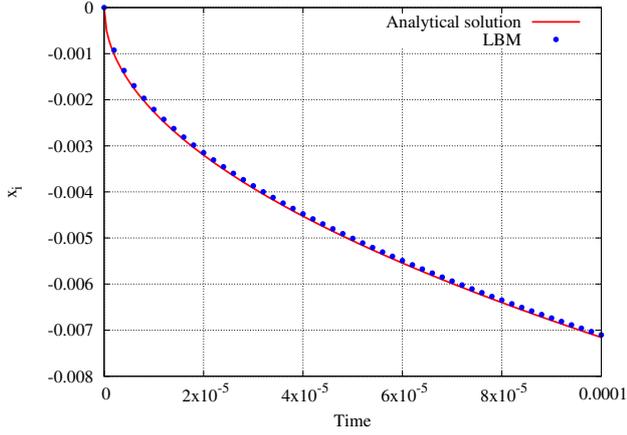}
\par\end{centering}
}~~~~\subfloat[\label{fig:PotChim_Ds=00003D0}Chemical potential $\overline{\mu}_{l}$
w.r.t. $x$ for three times $t_{1}=5\times10^{-5}$ (red), $t_{2}=2.5\times10^{-4}$
(blue) and $t_{3}=5\times10^{-4}$ (cyan). The $L^{2}$-errors decrease
with time: 1.6$\times10^{-3}$ for $t_{1}$, 7.2$\times10^{-4}$ for
$t_{2}$ and 5.4$\times10^{-4}$ for $t_{3}$. The two vertical lines
(black) indicate the interface positions at $x_{i}(t_{1})$ and $x_{i}(t_{3})$. ]{\begin{centering}
\includegraphics[angle=-90,scale=0.33]{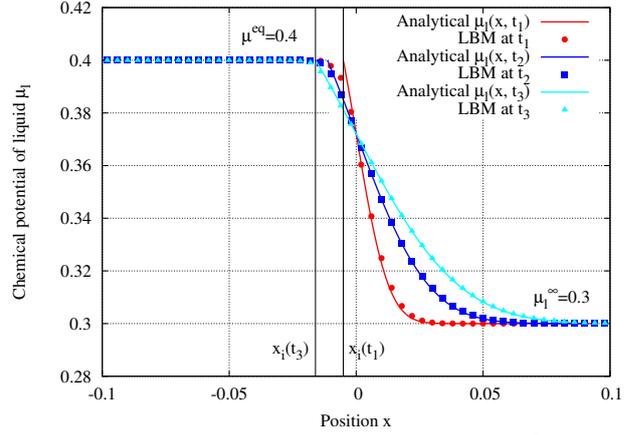}
\par\end{centering}
}
\par\end{centering}
\caption{\label{fig:Validation_Ds0}Analytical solutions Eqs. (\ref{eq:x_i_Ds0}),
(\ref{eq:Analytical_muL_Ds0}), (\ref{eq:Analytical_Cl_Ds0}) (lines)
compared with LBM (symbols) for a case of dissolution with $D_{s}=0$
and $D_{l}=1$. The anti-trapping current $\boldsymbol{j}_{at}$ is
considered in the simulation. The system is initialized with an interface
located at $x=0$. The solid and liquid are respectively on left-
and right-side.}
\end{figure*}

In Eq. (\ref{eq:Cartesian_Transcendental}), the dissolution or precipitation
processes can occur depending on the sign of second term. We keep
$c_{s}^{co}=0.6$ and $c_{l}^{co}=0.5$ (i.e. $\Delta c^{co}>0$),
and we set $\overline{\mu}_{l}^{\infty}=0.3$ meaning that $\overline{\mu}^{eq}-\overline{\mu}_{l}^{\infty}>0$.
The root of this equation is equal to $\alpha=-0.357835$. The interface
position $x_{i}(t)$ is a function of $\alpha$, $t$ and $D_{l}$
which writes:

\begin{subequations}

\begin{equation}
x_{i}(t)=2\alpha\sqrt{D_{l}t}\label{eq:x_i_Ds0}
\end{equation}
Since $\alpha<0$, the interface position moves from $x_{i}(0)=0$
towards negative values of $x$, meaning that a dissolution process
occurs. The chemical potential of liquid is

\begin{equation}
\overline{\mu}_{l}^{as}(x,\,t)=\overline{\mu}_{l}^{\infty}+(\overline{\mu}^{eq}-\overline{\mu}_{l}^{\infty})\frac{\text{erfc}[x/2\sqrt{D_{l}t}]}{\text{erfc}(\alpha)}\label{eq:Analytical_muL_Ds0}
\end{equation}
for $x\in]x_{i}(t),\,L_{x}]$. In the liquid, $\overline{\mu}_{l}^{as}(x,\,t)$
diffuses from the equilibrium value $\overline{\mu}^{eq}$ at the
interface until $\overline{\mu}_{l}^{\infty}$ when $x\rightarrow L_{x}$.
In the solid, the chemical potential $\overline{\mu}_{s}^{as}(x,\,t)$
is constant of value $\overline{\mu}^{eq}$ for $x\in[-L_{x},\,x_{i}(t)[$.

\end{subequations}

For LBM simulations, the parameters of $\phi$-equation are $W=5\times10^{-3}$,
$\lambda^{\star}=230$ and $M_{\phi}=1.2$. In $c$-equation, the
diffusion is interpolated by $\mathcal{D}(\phi)=\phi D_{l}$ and the
anti-trapping current is considered. The initial condition of composition
is imposed by Eq. (\ref{eq:C_init}) with $c_{l}^{\infty}=\overline{\mu}_{l}^{\infty}+m_{l}=0.4$
and $c_{s}^{\infty}=\overline{\mu}_{s}^{\infty}+m_{s}=0.6$ with $m_{s}=0.2$
and $m_{l}=0.1$. The comparisons between the analytical solutions
and the LBM simulation are presented in Fig. \ref{fig:Validation_Ds0}.
Compared to the previous section, now the curve of the interface position
decreases with time (Fig. \ref{fig:Interface-position_Ds=00003D0})
because the dissolution process occurs. The results of LBM are in
good agreement with the analytical solutions for three times $t_{1}=5\times10^{-5}$,
$t_{2}=2.5\times10^{-4}$ and $t_{3}=5\times10^{-4}$ (Fig. \ref{fig:PotChim_Ds=00003D0}).

\begin{figure*}[t]
\begin{centering}
\subfloat[\label{fig:Comparison-C_Without-With_Jat}Profiles of composition
compared between LBM (symbols), analytical (dashed line) and semi-analytical
solutions (solid lines). The $L^{2}$-errors are $3.6\times10^{-3}$
with anti-trapping and $4.8\times10^{-3}$ without.]{\begin{centering}
\includegraphics[angle=-90,scale=0.33]{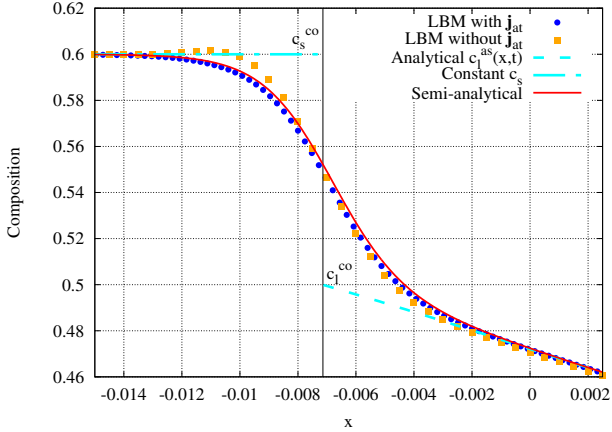}
\par\end{centering}
}~~~~\subfloat[\label{fig:Comparison-mu_Without-With_Jat}Profiles of chemical potential
compared between LBM (symbols) and analytical solution (line).
The $L^{2}$-errors are $3.5\times10^{-3}$ with anti-trapping and
$6.8\times10^{-3}$ without.]{\begin{centering}
\includegraphics[angle=-90,scale=0.33]{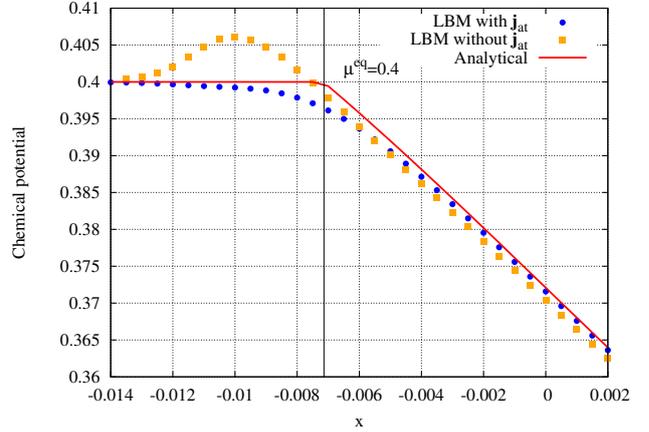}
\par\end{centering}
}
\par\end{centering}
\caption{\label{fig:Effect_Jat}LB simulations with (dots) and without (squares)
$\boldsymbol{j}_{at}$ in $c$-equation. Comparison on profiles of
composition (Fig. \ref{fig:Comparison-C_Without-With_Jat}) and chemical
potential (Fig. \ref{fig:Comparison-mu_Without-With_Jat}). Zoom between
$[-0.015,\,0.002]$ at $t=10^{-4}$.}
\end{figure*}

The anti-trapping effect is compared on the profiles of composition
and chemical potential (Fig. \ref{fig:Effect_Jat}). For composition,
the analytical solution can be derived from Eq. (\ref{eq:Analytical_muL_Ds0})
by adding $m_{l}$ on both sides and by adding and subtracting $m_{l}$
inside $(\overline{\mu}^{eq}-\overline{\mu}_{l}^{\infty})$. We obtain:

\begin{subequations}

\begin{equation}
c_{l}^{as}(x,\,t)=c_{l}^{\infty}+(c_{l}^{co}-c_{l}^{\infty})\frac{\text{erfc}[x/2\sqrt{D_{l}t}]}{\text{erfc}(\alpha)}\label{eq:Analytical_Cl_Ds0}
\end{equation}
where $c_{l}^{co}=\overline{\mu}^{eq}+m_{l}=0.5$. In the solid phase,
the composition is a constant of value $c_{s}(x,\,t)=\overline{\mu}^{eq}+m_{s}=0.6$
corresponding to its value of coexistence $c_{s}^{co}$. The compositions
$c_{l}^{as}(x,\,t)$ and $c_{s}(x,\,t)$ are plotted with dashed lines
on Fig. \ref{fig:Comparison-C_Without-With_Jat}.

The LBM simulations are carried out successively with and without
anti-trapping current. The profiles of composition are reported on
Fig. \ref{fig:Comparison-C_Without-With_Jat} at $t=10^{-4}$ (symbols).
Without anti-trapping, the theory cannot provide a value of $\lambda^{\star}$
because the phase-field model is not strictly equivalent to the sharp
interface one (see Section \ref{subsec:Discussion_Matched-Asymptotics}).
Hence, the value of $\lambda^{\star}=500$ is chosen such as the displacement
of the interface is close to the analytical solution. The simulation
corresponds to the best fit that is possible to obtain when $D_{s}=0$
and $\boldsymbol{j}_{at}=\boldsymbol{0}$ in $c$-equation (squares
on Fig. \ref{fig:Comparison-C_Without-With_Jat}). On that figure,
the semi-analytical solution ($sas$) is plotted for comparison:

\begin{equation}
c^{sas}(\phi)=c_{l}^{as}(x,\,t)\phi(x,\,t)+c_{s}\left[1-\phi(x,\,t)\right]\label{eq:Semi-AnalyticalSol}
\end{equation}
The $sas$-solution corresponds to an interpolation of $c_{l}^{as}(x,\,t)$
and $c_{s}$ with $\phi$. When $\boldsymbol{j}_{at}$ is not considered
in $c$-equation, the compositions fit well far from the interface.
However, inside the interface region, the compositions are over-estimated
on the solid side whereas they are under-estimated on the liquid side.
On the interval $[-L_{x},\,x_{i}(t)[$ (solid), the profile slightly
oscillates above the composition of coexistence. That oscillation
is more visible when we plot the chemical potential (Fig. \ref{fig:Comparison-mu_Without-With_Jat}).
That lack of accuracy slows down the displacement of interface compared
to the analytical solution Eq. (\ref{eq:x_i_Ds0}).

\end{subequations}

\section{\label{sec:Simulations}Dissolution of porous medium: counter term
effect}

In Section \ref{sec:Validations}, the initial conditions of $\phi$
and $c$ are defined by two hyperbolic tangent functions. Here, the
phase-field is initialized with an input datafile which comes from
the characterization of a 3D porous sample with X-ray tomography.
The datafile contains $256\times256\times236$ rows with three indices
of position ($x$, $y$ and $z$) and one additional index describing
the solid (value\textsf{ 0}) or the pore (value \textsf{255}). For
simulating the dissolution, we assume that the poral volume is filled
with a solute of smaller composition than the coexistence composition
of liquid. A two-dimensional slice of size $256\times256$ has been
extracted from the datafile and rescaled to $1024\times1024$ nodes
covering a square of size $[0,\,1]^{2}$ ($\delta x\simeq9.76\times10^{-4}$).
The time-step of discretization is $\delta t=5\times10^{-7}$. The
type of all boundary conditions is zero flux.

For the parameters of $\phi$-equation, the diffusivity is $M_{\phi}=1.2$
and the interface width is set equal to $W=0.02$ (i.e. $\sim20\delta x$).
The value of coupling coefficient $\lambda^{\star}=230$ (corresponding
to $\beta=0$) is computed by using Eq. (\ref{eq:Lambda-Star}) with
values of $\mathscr{K}$, $\mathscr{F}$ and $\mathscr{I}=2/3$ defined
in Tab. \ref{tab:integrals}. For $c$-equation, the coexistence compositions
of solid and liquid are respectively equal to $c_{s}^{co}=0.6$ and
$c_{l}^{co}=0.5$ ($\Delta c^{co}=0.1$) and the chemical potential
of equilibrium is $\overline{\mu}^{eq}=0.4$. The diffusion coefficients
are zero in the solid ($D_{s}=0$) and one in the liquid ($D_{l}=1$).
The anti-trapping current $\boldsymbol{j}_{at}$ is used in the simulations.

\begin{figure*}[t]
\begin{centering}
\subfloat[\label{fig:InitCond_phi}Initialization of phase-field: for solid
$\phi(\boldsymbol{x}_{s},\,t_{0})=\phi_{s}=0$ (black) and for liquid
$\phi(\boldsymbol{x}_{l},\,t_{0})=\phi_{l}=1$ (blue).]{\begin{centering}
\includegraphics[scale=0.31]{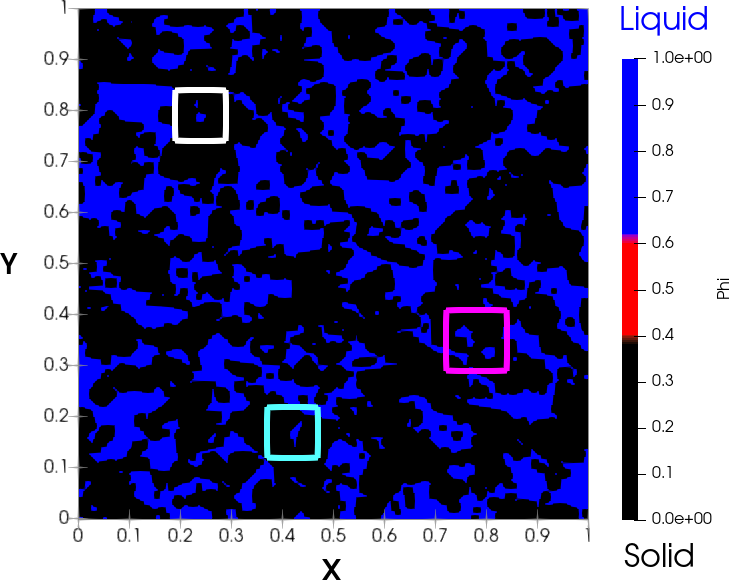}
\par\end{centering}
}~~~~~~\subfloat[\label{fig:InitCond_c}Initialization of composition: for solid, the
composition is set equal to the solid coexistence i.e. $c(\boldsymbol{x}_{s},\,t_{0})=c_{s}^{co}=0.6$
(yellow). For liquid $c(\boldsymbol{x}_{l},\,t_{0})=0.4$ (dark blue)
i.e. below the composition of coexistence $c_{l}^{co}=0.5$.]{\begin{centering}
\includegraphics[scale=0.31]{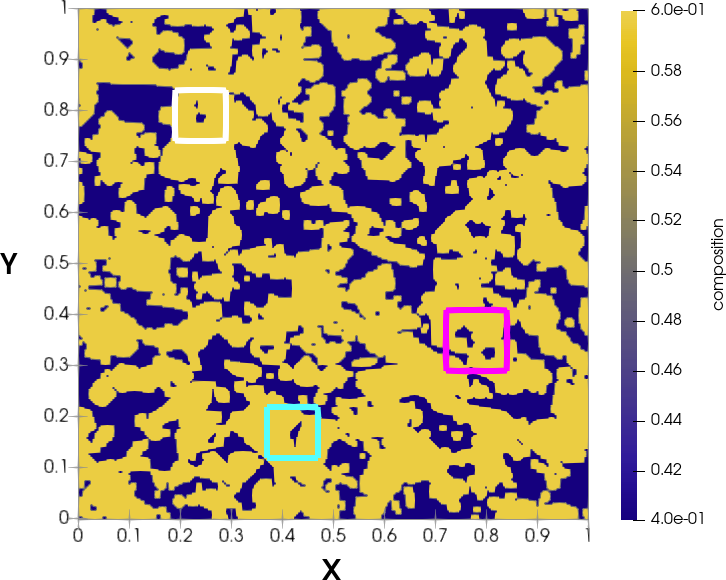}
\par\end{centering}
}
\par\end{centering}
\caption{\label{fig:CondInit}Positions $\boldsymbol{x}_{s}$ and $\boldsymbol{x}_{l}$
of datafile used to define the initial conditions for $\phi$ (Fig.
\ref{fig:InitCond_phi}) and $c$ (Fig. \ref{fig:InitCond_c}). Three
squares are sketched for comparing the evolution of small pores enclosed
inside the solid.}
\end{figure*}

The phase-field is simply initialized at $t_{0}=0$ with two discontinuous
values: for solid $\phi(\boldsymbol{x}_{s},\,t_{0})=\phi_{s}=0$ and
for liquid $\phi(\boldsymbol{x}_{l},\,t_{0})=\phi_{l}=1$. The composition
of the solid phase $c(\boldsymbol{x}_{s},\,t_{0})$ is set equal to
the coexistence composition of solid $c_{s}^{co}$. For liquid, the
initial condition is below its coexistence composition: $c(\boldsymbol{x}_{l},\,t_{0})=0.4<0.5$.
Those initializations are presented on Fig. \ref{fig:InitCond_phi}
for $\phi$ and Fig. \ref{fig:InitCond_c} for $c$. On both figures,
three squares are sketched for comparing the evolution of small pores
which are enclosed inside the solid.

With those initial conditions, the dissolution process occurs until
the composition of the liquid phase is equal to $c_{l}^{co}$. Two
simulations are compared. In the first one, the $\phi$-equation is
Eq. (\ref{eq:ModelA_EqPhi_CounterTerm}) which accounts for the counter
term $-M_{\phi}\kappa\bigl|\boldsymbol{\nabla}\phi\bigr|$. In the
second one, the curvature-driven motion is possible because the $\phi$-equation
is Eq. (\ref{eq:ModelA_EqPhi}). For both simulations, a diffuse interface
replaces at first time-steps the initial discontinuity between the
solid and liquid phases. The code ran 70 seconds on a single GPU (Volta
100) until the steady state is reached after $10^{4}$ time steps.

\begin{figure*}[p]
\begin{centering}
\subfloat[\label{fig:Phase-field_with_CT}Evolution of the phase-field when
the counter term is considered in $\phi$-equation.]{\begin{centering}
\begin{tabular}{ccc}
{\small{}$t_{1}=2\times10^{2}\delta t$} & {\small{}$t_{2}=10^{3}$$\delta t$} & {\small{}$t_{f}=10^{4}\delta t$}\tabularnewline
\includegraphics[scale=0.24]{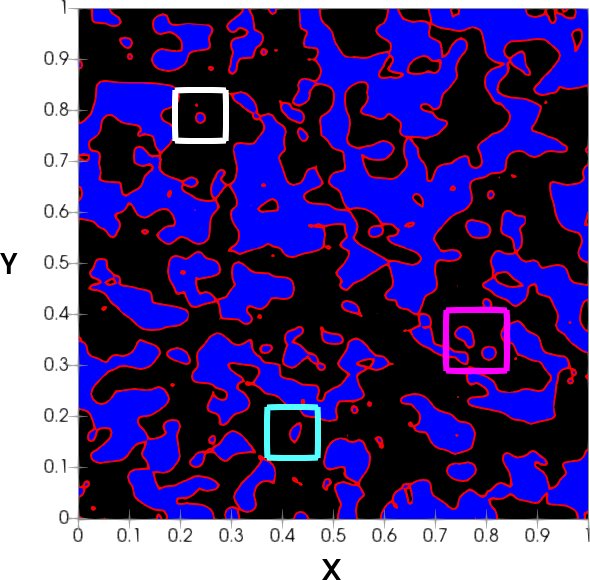} & \includegraphics[scale=0.24]{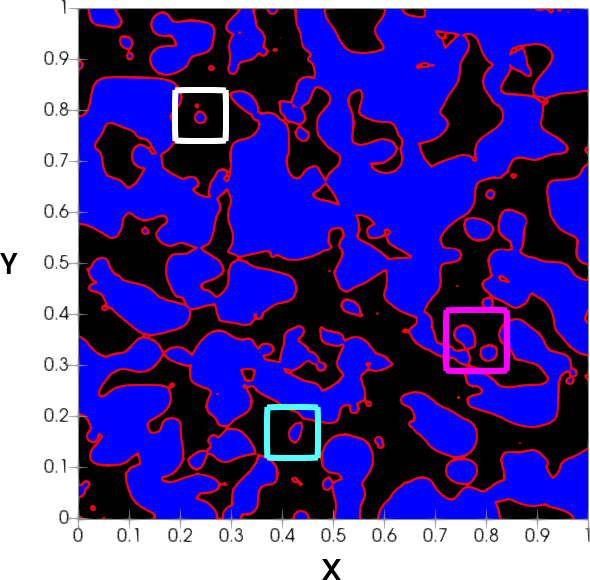} & \includegraphics[scale=0.24]{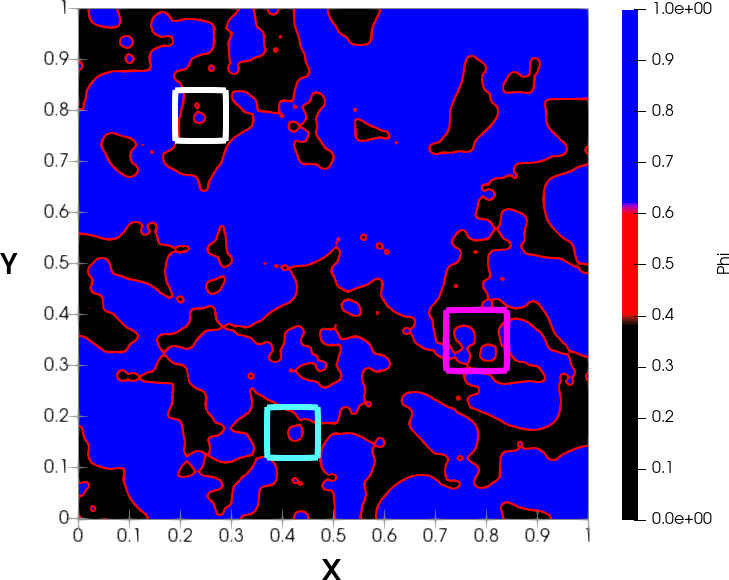}\tabularnewline
\end{tabular}
\par\end{centering}
}
\par\end{centering}
\begin{centering}
\subfloat[\label{fig:Phase-field_withoutCT}Phase-field at same time steps without
counter term. Because of the curvature-driven motion the final shape
of solid is smoother and the small pores inside the squares have disappeared.]{\begin{centering}
\begin{tabular}{ccc}
\includegraphics[scale=0.24]{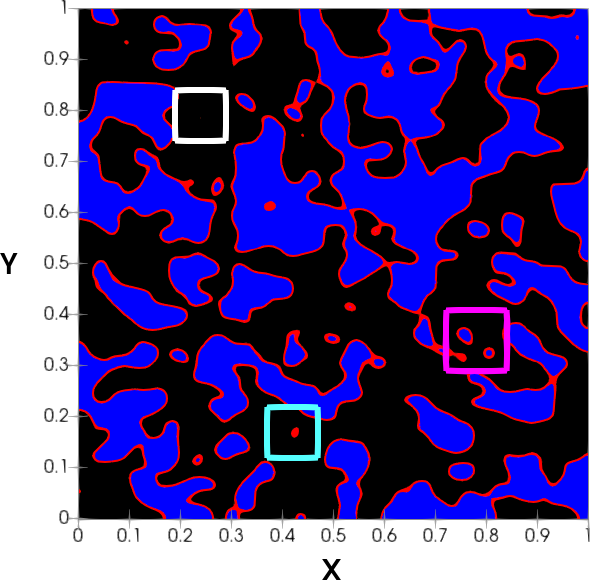} & \includegraphics[scale=0.24]{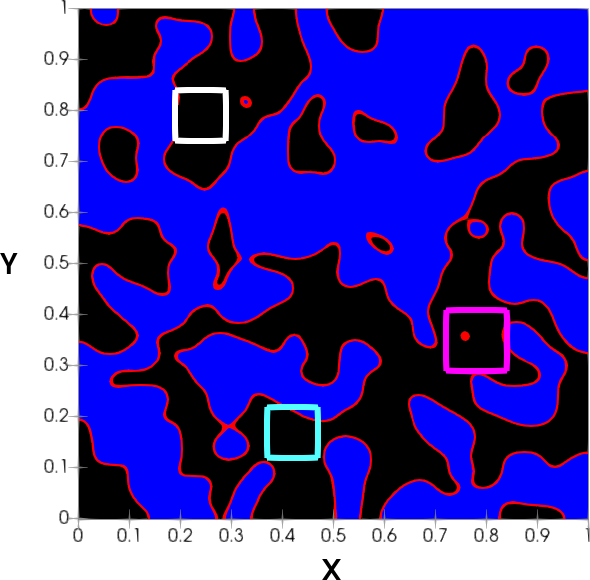} & \includegraphics[scale=0.24]{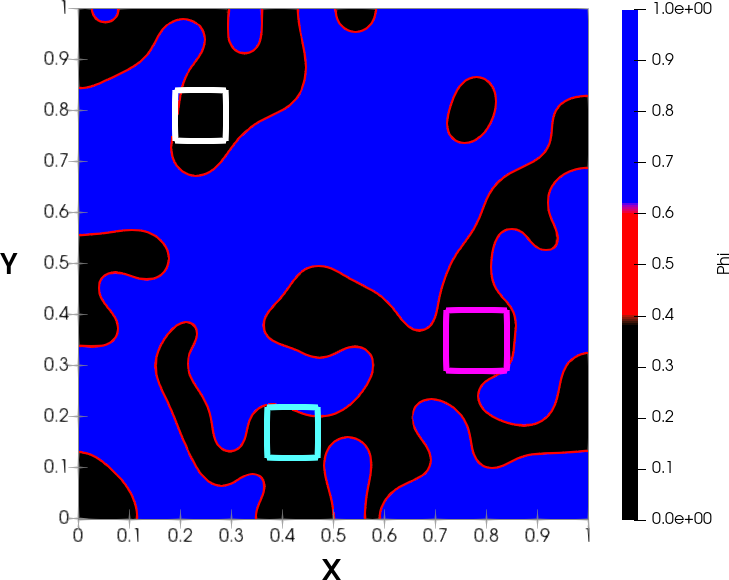}\tabularnewline
\end{tabular}
\par\end{centering}
}
\par\end{centering}
\begin{centering}
\subfloat[\label{fig:Compos_withCT}Fields of composition at same time steps
with counter term in $\phi$-equation. At final time the composition
is homogeneous in the solid phase of value $c_{s}^{co}$.]{\begin{centering}
\begin{tabular}{ccc}
\includegraphics[scale=0.24]{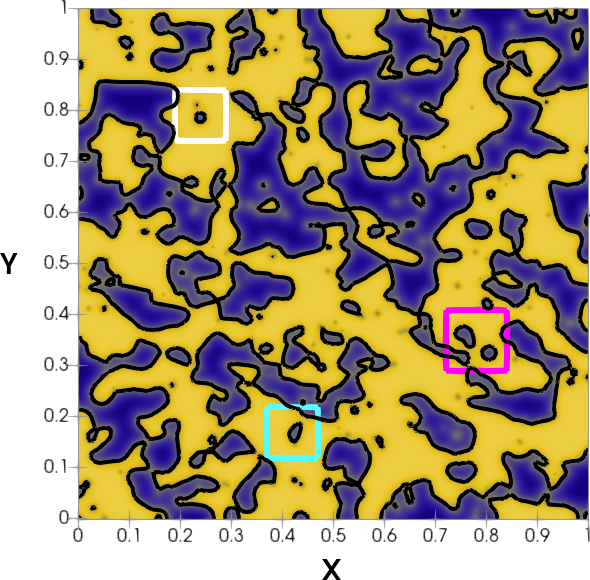} & \includegraphics[scale=0.24]{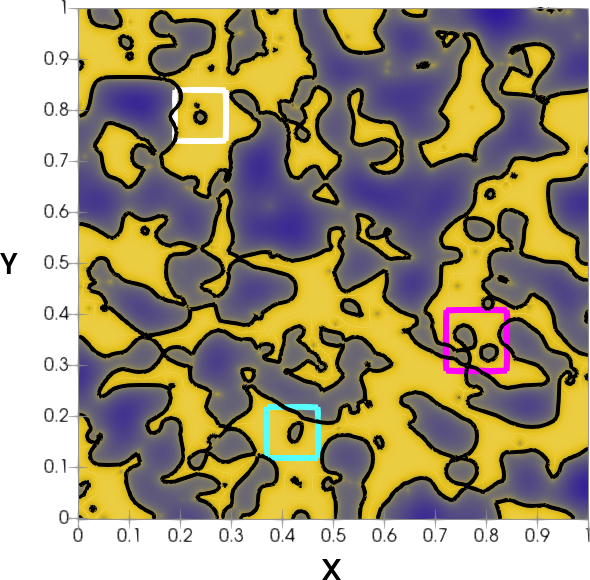} & \includegraphics[scale=0.24]{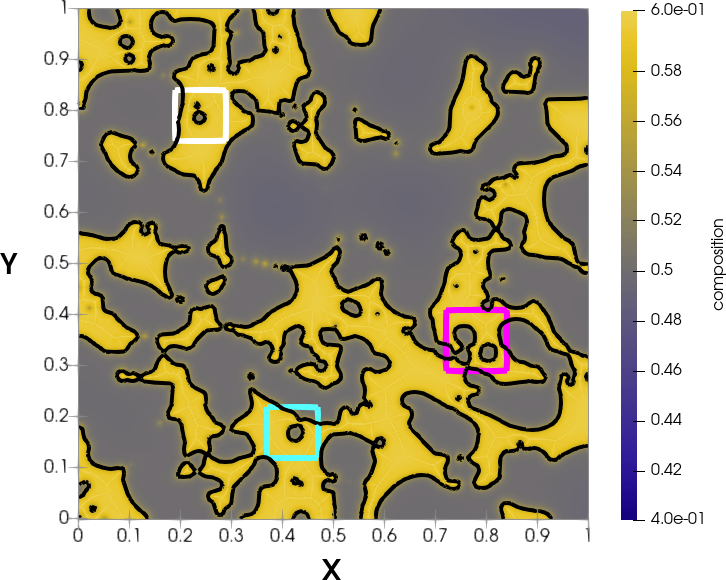}\tabularnewline
\end{tabular}
\par\end{centering}
}
\par\end{centering}
\begin{centering}
\subfloat[\label{fig:Compos_withoutCT}Fields of composition without counter
term at same time steps. The composition of solid is heterogeneous
because the small pores have disappeared and the diffusion is zero.
Several areas of smaller composition than $c_{s}^{co}$ are trapped
in the solid phase (e.g. inside the squares).]{\begin{centering}
\begin{tabular}{ccc}
\includegraphics[scale=0.24]{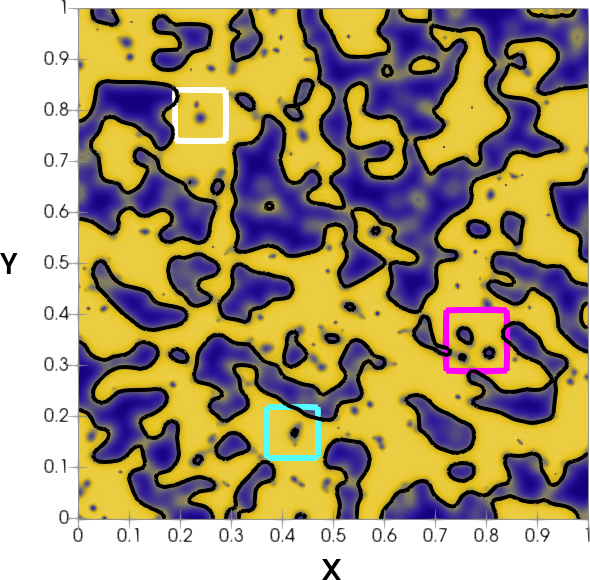} & \includegraphics[scale=0.24]{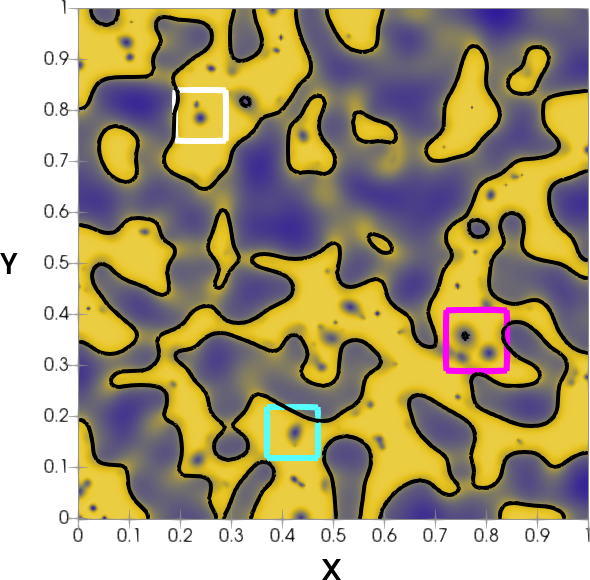} & \includegraphics[scale=0.24]{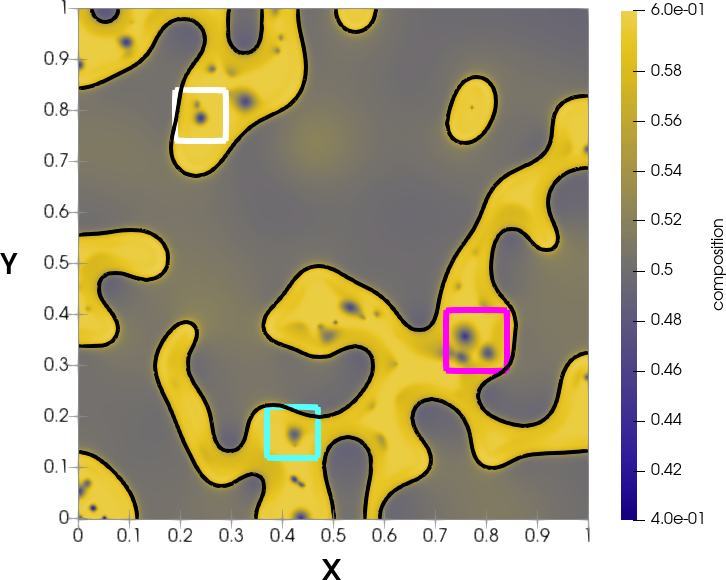}\tabularnewline
\end{tabular}
\par\end{centering}
}
\par\end{centering}
\caption{\label{fig:Counter-term-effect}Dissolution of porous medium simulated
by a phase-field model based on the grand-potential. Snapshots of
phase-field and composition at $t_{1}=2\times10^{2}\delta t$ (left),
$t_{2}=10^{3}\delta t$ (middle) and final time of simulation $t_{f}=10^{4}\delta t$
(right). Two simulations are carried out: in Figs. \ref{fig:Phase-field_with_CT}
and \ref{fig:Compos_withCT} the $\phi$-equation is Eq. (\ref{eq:ModelA_EqPhi_CounterTerm})
whereas in Figs. \ref{fig:Phase-field_withoutCT} and \ref{fig:Compos_withoutCT}
the $\phi$-equation is Eq. (\ref{eq:ModelA_EqPhi}). The iso-contour
$\phi=0.5$ (black line) is super-imposed on Figs. \ref{fig:Compos_withCT}
and \ref{fig:Compos_withoutCT}.}
\end{figure*}

The results are presented in Fig. \ref{fig:Counter-term-effect} for
three times: $t_{1}=10^{2}\delta t$ (left), $t_{2}=10^{3}\delta t$
(middle) and $t_{f}=10^{4}\delta t$ (right). At first sight, the
difference concerns the shapes of the solid phase at the end of simulations.
When the counter term is considered, the interface is much more irregular
(Fig. \ref{fig:Phase-field_with_CT}-right) than that obtained without
counter term (Fig. \ref{fig:Phase-field_withoutCT}-right). The reason
is that, with counter term, the interface motion is only caused by
differences of composition in liquid and solid. The dissolution occurs
in isotropic way until the equilibrium is reached. Without counter
term, the irregularities of solid disappear because of the curvature-driven
motion. Finally, the shape of the solid phase is much smoother.

For both simulations, when the steady state is reached, the composition
of liquid phase is equal to the coexistence composition of liquid
$c(\boldsymbol{x}_{l},\,t_{f})=c_{l}^{co}$ (gray areas in the right
figures of \ref{fig:Compos_withCT} and \ref{fig:Compos_withoutCT}).
However, the composition inside the solid phase is different. When
the curvature-driven motion is canceled, the composition $c(\boldsymbol{x}_{s},\,t_{f})$
is homogeneous of value $c_{s}^{co}$ (see Fig. \ref{fig:Compos_withCT}-right).
When that motion is taken into account, the solid composition $c(\boldsymbol{x}_{s},\,t_{f})$
is heterogeneous as revealed by the presence of areas of composition
lower than $c_{s}^{co}$ (gray areas inside squares in Fig. \ref{fig:Compos_withoutCT}-right).
Those areas correspond to solid phases as confirmed by Fig. \ref{fig:Phase-field_withoutCT}-right.

That heterogeneity of composition is explained by the curvature-driven
motion occurring when the counter term is not considered in $\phi$-equation.
That interface motion makes disappear the small pores embedded in
the solid phase. For instance at $t_{1}$, the small one inside the
white square has disappeared (Fig. \ref{fig:Phase-field_withoutCT}-left)
and the pore inside the cyan square has almost disappeared (red dot).
That same pore has fully disappeared at $t_{2}$ (Fig. \ref{fig:Phase-field_withoutCT}-middle)
and one of the two pores inside the magenta square has also disappeared.
At last both of them have disappeared at $t_{f}$ (Fig. \ref{fig:Phase-field_withoutCT}-right).
With counter term, all those pores still exist at the end of simulation
(Fig. \ref{fig:Phase-field_with_CT}-right).

With the curvature-driven motion a special area which is initially
liquid ($\phi=1$) may become solid ($\phi=0$) even though the local
composition $c(\boldsymbol{x}_{l},\,t)$ is not greater than $c_{l}^{co}$.
That curvature motion acts like a precipitation process. For those
areas, the diffusion coefficient changes from $D_{l}$ to $D_{s}=0$
meaning that the diffusion process does not occur anymore. The value
of composition is ``frozen'' explaining why small islands of lower
composition are embedded in the solid phase.

An in-depth physical analysis is based on the Gibbs-Thomson condition
Eq. (\ref{eq:Gibbs-Thomson_PF}) i.e. $\overline{\mu}=\overline{\mu}^{eq}-d_{0}\kappa$
(with $\beta=0$). When two phases coexist, the interface will move
towards the position where the chemical potential $\overline{\mu}$
is closer to $\overline{\mu}^{eq}$. In the first case, the counter
term cancels the motion $d_{0}\kappa$ whereas in the second case
that motion exists. In our simulations $\overline{\mu}(\boldsymbol{x}_{s},\,t)=\overline{\mu}^{eq}=0.4$
in the solid and $\overline{\mu}(\boldsymbol{x}_{l},\,t)=0.3$ in
the liquid. For small pores trapped in the solid, the interface will
move towards the liquid phase and the physical process acts like precipitation.
The interface disappears because it is the unique way to reach the
equilibrium value $\overline{\mu}^{eq}$. On the contrary, for outgrowths,
the curvature is opposite and the interface will move towards the
solid phase, dissolution occurs.

\section{\label{sec:Conclusion}Conclusion}

In this work we have presented a phase-field model of dissolution
and precipitation. Its main feature lies in its derivation which is
based on the functional of grand-potential $\Omega[\phi,\,\mu]$.
In that theoretical framework, the phase-field $\phi$ and the chemical
potential $\mu$ are the two main dynamical variables. In models based
on free energy, $\phi$ and the composition $c$ are the two main
variables. The benefits of using the grand-potential are twofold.
First, for models based on free-energy, two additional conditions
must be solved inside the diffuse zone in order to ensure the equality
of chemical potential at interface. In grand-potential theory, it
is not necessary because the model includes that assumption in its
formulation. Second, the chemical potential is an intensive thermodynamic
quantity like temperature so that many analogies can be done with
solidification problems. Hence, the analytical solutions of the Stefan
problem can be used for validation by comparing directly the temperature
and the chemical potential. Besides, the matched asymptotic expansions
can be directly inspired from those already performed for solidification
problems. The phase-field model is composed of two PDEs. The first
equation computes the evolution of the interface position $\phi$.
The second one is a mixed formulation using the composition and chemical
potential. Although that equation requires a closure relationship
between $c$ and $\mu$, that formulation improves the mass conservation.

In many simulations of dissolution or precipitation, two main hypotheses
are often considered. First, the interface is assumed to move because
of chemical reactions, which is often considered by a kinetics of
first-order. That assumption means that the curvature-driven motion
is neglected in the Gibbs-Thomson condition. In the phase-field theory,
that motion is always contained in the $\phi$-equation. In order
to cancel it, a counter term $-M_{\phi}\kappa\bigl|\boldsymbol{\nabla}\phi\bigr|$
must be added in $\phi$-equation. The second hypothesis is the diffusion
that is neglected in the solid phase. In that case, the anti-trapping
current $\boldsymbol{j}_{at}$ must be considered in $c$-equation.
Those two terms are not contained in the functional of grand-potential,
they are added for phenomenological reasons. Without them, the phase-field
model is not equivalent to the sharp interface model because several
spurious terms arise from the matched asymptotic expansions.

The model has been implemented with lattice Boltzmann schemes in the
\texttt{LBM\_saclay} code. The one-dimensional validations have been
carried out with two analytical solutions of the Stefan problem. The
test cases present one process of precipitation with $D_{s}\simeq D_{l}$
and another one of dissolution for $D_{s}=0$. The first one is performed
without anti-trapping and compares the profiles of composition. The
jump of composition is well-reproduced by the model at interface.
The diffusive behavior is also perfectly fitted for each phase. The
second test emphasizes the analogy with problems of solidification
(or melting) where the equilibrium chemical potential $\mu^{eq}$
plays the role of melting temperature and $\Delta c^{co}$ is compared
to the latent heat. For that test, the use of anti-trapping current
avoids the oscillations of algorithm and improves the accuracy of
composition profiles.

Finally, the numerical model has been applied for simulating the dissolution
process of a porous medium. The rock sample has been characterized
by X-ray microtomography. The datafile has been used for defining
the initial conditions for $\phi$ and $c$. Two simulations have
compared the impact of counter term on the shape of solid. When the
counter term is not considered in $\phi$-equation, the curvature-driven
motion makes disappear small areas of liquid trapped inside the solid
phase. The main consequence of that effect, acting like precipitation,
is the heterogeneity of composition inside the solid phase. When the
counter term is taken into account, the solid/liquid interface is
much more irregular and the composition is homogeneous inside the
solid phase.

The grand-potential densities $\omega_{\Phi}$ of each phase are defined
by the Legendre transform of free energy densities $f_{\Phi}$. In
this work, the main assumption is that $f_{s}$ and $f_{l}$ are defined
by two parabolas with identical curvature $\epsilon_{s}=\epsilon_{l}$.
That hypothesis simplifies the link between the thermodynamic parameters
$m_{\Phi}$, $\epsilon_{\Phi}$ and $\overline{f}_{\Phi}^{min}$and
the properties of equilibrium i.e. the coexistence compositions $c_{s}^{co}$,
$c_{l}^{co}$, and the equilibrium chemical potential $\mu^{eq}$.
Nevertheless, for real materials the thermodynamics does not fulfill
necessarily that condition. An in-depth study with $\epsilon_{s}\neq\epsilon_{l}$
is planned for future work for binary and ternary mixtures.

\appendix

\section*{Acknowledgments}

The authors wish to thank \noun{Mathis Plapp} for the insightful discussions
on 1) the theoretical framework of grand-potential and 2) the asymptotic
expansions of phase-field model. They also acknowledge the Genden
project (number R0091010339) for computational resources of supercomputer
Jean-Zay (IDRIS, France).

\section{\label{sec:Matched-asymptotic-expansions}Matched asymptotic expansions}

The starting point of this Appendix is the phase-field model composed
of Eqs. (\ref{eq:ModelA_EqPhi})--(\ref{eq:SourceTerm_Coexistence}).
The analysis is performed into two main stages. First, the whole computational
domain is divided into two regions. The first region is located far
from the interface (bulk phases) and called the ``outer domain''.
The second region is the diffuse interface and called the ``inner
domain''. In \ref{subsec:Dimensionless-variables_and_matching},
we define the dimensionless variables and the matching conditions
between both regions. Next, we present the analysis of the outer domain
in \ref{subsec:OuterDom_expansions} and the inner domain in \ref{subsec:InnerDom_expansions}.
Finally a discussion is carried out in \ref{subsec:Discussion-on-error}
to remove the error terms.

\subsection{\label{subsec:Dimensionless-variables_and_matching}Main definitions}

\subsubsection{Outer domain and inner domain}

The unknown are $\phi^{out}$, $\mu^{out}$, $c^{out}$ for the outer
domain, and $\phi^{in}$, $\mu^{in}$, $c^{in}$ for the inner domain.
They are expanded as power of a small parameter ${\displaystyle \varepsilon\ll1}$:

\begin{align*}
\phi^{out} & \simeq\phi_{0}^{out}+\varepsilon\phi_{1}^{out}+\varepsilon^{2}\phi_{2}^{out}, & \phi^{in} & \simeq\phi_{0}^{in}+\varepsilon\phi_{1}^{in}+\varepsilon^{2}\phi_{2}^{in}\\
c^{out} & \simeq c_{0}^{out}+\varepsilon c_{1}^{out}+\varepsilon^{2}c_{2}^{out}, & c^{in} & \simeq c_{0}^{in}+\varepsilon c_{1}^{in}+\varepsilon^{2}c_{2}^{in}\\
\mu^{out} & \simeq\mu_{0}^{out}+\varepsilon\mu_{1}^{out}+\varepsilon^{2}\mu_{2}^{out}, & \mu^{in} & \simeq\mu_{0}^{in}+\varepsilon\mu_{1}^{in}+\varepsilon^{2}\mu_{2}^{in}
\end{align*}

The small parameter of expansions is defined by $\varepsilon=W/d_{0}$,
where $d_{0}=W/(\alpha\lambda)$ is the capillary length, and $\alpha$
is a parameter to be determined. The two coefficients $D_{l}$ and
$d_{0}$ define a characteristic speed $v_{c}=D_{l}/d_{0}=\varepsilon D_{l}/W$,
and a characteristic time $t_{c}=d_{0}^{2}/D_{l}=d_{0}/v_{c}$. For
the expansions, we also assume $1/\kappa\sim d_{0}$ i.e. ${\displaystyle W\kappa\sim\varepsilon}$.
In the phase-field equation, the coefficient $\lambda$ is replaced
by $\varepsilon/\alpha$. Finally, we define the following dimensionless
quantities: the interface speed $\bar{v}_{n}=v_{n}/v_{c}$, the interface
curvature $\bar{\kappa}=d_{0}\kappa$, the time $\bar{t}=t/t_{c}$,
the spatial coordinate $\bar{\boldsymbol{x}}=\boldsymbol{x}/d_{0}$,
and the dimensionless diffusivity $\bar{D}=D_{l}/M_{\phi}$. Finally
$q_{s}=D_{s}/D_{l}$ is the ratio of diffusion coefficients.

For each domain, the model is re-written with the curvilinear coordinates
$r$ and $s$, where $r$ is the signed distance to the level line
$\phi=0.5$, and $s$ the arc length along the interface. The dimensionless
coordinates are also defined by: $\eta=r/d_{0}$ and $\overline{s}=s/d_{0}$.
The spatial operators (divergence, gradient, laplacian) and the time
derivatives are also expressed in this new system of coordinates,
and next expanded in power of $\varepsilon$. Finally, the terms of
same order are gathered and the solutions of all orders can be calculated
with appropriate boundary conditions: the ``matching conditions''.

\subsubsection{\label{subsec:matching_conditions}Matching conditions}

The matching conditions are established by comparing the limits of
inner variables far from the interface with the limits of outer variables
near the interface. For that purpose we define $\xi=r/W$, a ``stretched''
normal coordinate in the inner region. Let us notice that the curvilinear
coordinates $r$ and $s$ have been made dimensionless by introducing
$\eta=r/d_{0}$ and $\overline{s}=s/d_{0}$ in the outer domain, we
observe that $\eta/\xi=\varepsilon\ll1$. This motivates us to compare
the limits of inner variables when $\xi\rightarrow\pm\infty$ with
the limits of outer variables when $\eta\rightarrow0^{\pm}$. For
the phase-fields $\phi_{0}^{in}$ and $\phi_{0}^{out}$, we define
the matching conditions:

\[
\underset{\xi\rightarrow+\infty}{\lim}\phi_{0}^{in}=\underset{\eta\rightarrow0^{+}}{\lim}\phi_{0}^{out}\quad\text{and}\quad\underset{\xi\rightarrow-\infty}{\lim}\phi_{0}^{in}=\underset{\eta\rightarrow0^{-}}{\lim}\phi_{0}^{out}
\]
For the chemical potentials $\mu_{j}^{in}$ and $\mu_{j}^{out}$ ($0\leq j\leq2$),
the matching conditions are:

\begin{align*}
\underset{\xi\rightarrow\pm\infty}{\lim}\mu_{0}^{in} & =\mu_{0}^{out}(0^{\pm})\\
\underset{\xi\rightarrow\pm\infty}{\lim}\mu_{1}^{in} & =\mu_{1}^{out}(0^{\pm})+\partial_{\eta}\mu_{0}^{out}(0^{\pm})\xi\\
\underset{\xi\rightarrow\pm\infty}{\lim}\partial_{\xi}\mu_{2}^{in} & =\partial_{\eta}\mu_{1}^{out}(0^{\pm})+\partial_{\eta\eta}^{2}\mu_{0}^{out}(0^{\pm})\xi
\end{align*}
where $\mu_{j}^{out}(0^{\pm})=\lim_{\eta\rightarrow0^{\pm}}\mu_{j}^{out}$
for $j=0,\,1$ and $\partial_{\eta}\mu_{0}^{out}(0^{\pm})=\lim_{\eta\rightarrow0^{\pm}}\partial_{\eta}\mu_{0}^{out}$.

\subsection{\label{subsec:OuterDom_expansions}Analysis of outer domain}

In the outer domain, the use of curvilinear coordinates is not necessary
because the region is far from the interface. Thus, the model is written
with the dimensionless cartesian coordinates $\overline{\boldsymbol{x}}$,
the dimensionless time $\overline{t}$ and the parameter of expansion
$\varepsilon$. The model writes:

\begin{align}
\bar{D}\varepsilon^{2}\partial_{\bar{t}}\phi^{out} & =\varepsilon^{2}\boldsymbol{\bar{\nabla}}^{2}\phi^{out}-\omega_{dw}^{\prime}(\phi^{out})-\frac{\varepsilon}{\alpha}\partial_{\phi}\omega^{p}(\phi^{out},\mu^{out})\nonumber \\
\partial_{\bar{t}}c & =\boldsymbol{\bar{\nabla}}\cdot\left[q(\phi^{out})\boldsymbol{\bar{\nabla}}\mu^{out}-\boldsymbol{j}_{at}(\phi^{out})\right]\label{eq:Outer_model_diff}
\end{align}
where the notation $\omega^{p}=p(\phi)\omega_{l}(\mu)+[1-p(\phi)]\omega_{s}(\mu)$
and $\boldsymbol{j}_{at}(\phi^{out})=-a(\phi^{out})\Delta c^{co}\partial_{\bar{t}}\phi^{out}\boldsymbol{n}$
with the normal vector $\boldsymbol{n}=\boldsymbol{\bar{\nabla}}\phi^{out}/\left|\boldsymbol{\bar{\nabla}}\phi^{out}\right|$.
The expansions of $\phi$-equation write for each order:

\begin{subequations}

\begin{align}
\mathcal{O}(1):\quad & 0=\omega_{dw}^{\prime}(\phi_{0}^{out})\label{eq:Outer_Oeps0_pf}\\
\mathcal{O}(\varepsilon):\quad & 0=\omega_{dw}^{\prime\prime}(\phi_{0}^{out})\phi_{1}^{out}+\alpha^{-1}\partial_{\phi}\omega^{p}(\phi_{0}^{out},\mu_{0}^{out})\label{eq:Outer_Oeps1_pf}\\
\mathcal{O}(\varepsilon^{2}):\quad & 0=\bar{D}\varepsilon^{2}\partial_{\bar{t}}\phi_{0}^{out}-\boldsymbol{\bar{\nabla}}^{2}\phi_{0}^{out}+\omega_{dw}^{\prime\prime\prime}(\phi_{0}^{out})\phi_{2}^{out}\nonumber \\
 & +\frac{1}{2}\omega_{dw}^{\prime\prime}(\phi_{0}^{out})(\phi_{1}^{out})^{2}+\frac{1}{\alpha}\partial_{\phi\phi}^{2}\omega^{p}(\phi_{0}^{out},\mu_{0}^{out})\phi_{1}^{out}\nonumber \\
 & \qquad\qquad+\alpha^{-1}\partial_{\phi\mu}^{2}\omega^{p}(\phi_{0}^{out},\mu_{0}^{out})\mu_{1}^{out}\label{eq:Outer_Oeps2_pf}
\end{align}

\end{subequations}

According to Eq. (\ref{eq:Outer_Oeps0_pf}), $\omega_{dw}(\phi_{0}^{out})$
takes a minimal value in the outer domain, i.e. $\phi_{0}^{out}$
is either equal to $\phi_{s}=0$ or $\phi_{l}=1$. In addition, $p(\phi)$
must be chosen such that its derivatives vanish in the bulk phases.
Then Eq. (\ref{eq:Outer_Oeps1_pf}) becomes simply $\omega_{dw}^{\prime\prime}(\phi_{0}^{out})\phi_{1}^{out}=0$,
implying $\phi_{1}^{out}=0$. Similarly, Eq. (\ref{eq:Outer_Oeps2_pf})
simplifies to $\omega_{dw}^{\prime\prime}(\phi_{0}^{out})\phi_{2}^{out}=0$
involving $\phi_{2}^{out}=0$. Finally, the analysis of $\phi$-equation
in the outer domain yields
\begin{equation}
\phi_{0}^{out}=\phi_{\Phi}\quad\text{and}\quad\phi_{1}^{out}=\phi_{2}^{out}=0\quad\text{for}\,\,\,\Phi=s,\,l\label{eq:Result_Phi_out}
\end{equation}
We can complete the matching conditions with the additional relation
$\lim_{\xi\rightarrow\pm\infty}\phi_{j}^{in}=0$ for $j\geq1$. Because
of Eq. (\ref{eq:Result_Phi_out}), the term $\partial_{\overline{t}}\phi^{out}$
is null in Eq. (\ref{eq:Outer_model_diff}), and the $c$-equation
simplifies to:

\begin{equation}
\partial_{\bar{t}}c=\boldsymbol{\bar{\nabla}}\cdot\left[q(\phi_{\Phi})\boldsymbol{\bar{\nabla}}\mu^{out}\right]\quad\text{for}\,\,\,\Phi=s,\,l\label{eq:DiffusionEq}
\end{equation}
The analysis of the outer domain recovers the standard diffusion equation
Eq. (\ref{eq:DiffusionEq}) with a constant phase-field Eq. (\ref{eq:Result_Phi_out})
for each bulk phase.

\subsection{\label{subsec:InnerDom_expansions}Analysis of inner domain}

Now we focus on the analysis of the inner domain. After expansions
of each PDE (\ref{subsec:Expanded-equations_Inner}), the solutions
are calculated order-by-order and used to derive the interface conditions
(\ref{subsec:Inner_O(1)}--\ref{subsec:Inner(eps2)}).

\subsubsection{\label{subsec:Expanded-equations_Inner}Expanded equations}

The expansions of $\phi$-equation and $c$-equation write:

\begin{strip}

\begin{subequations}

\begin{eqnarray}
\mathcal{O}(1):\quad & \partial_{\xi\xi}^{2}\phi_{0}^{in}-\omega_{dw}^{\prime}(\phi_{0}^{in}) & =0\label{subeq:Oeps0_pf}\\
\mathcal{O}(\varepsilon):\quad & \partial_{\xi\xi}^{2}\phi_{1}^{in}-\omega_{dw}^{\prime\prime}(\phi_{0}^{in})\phi_{1}^{in} & =-\left(\bar{D}\bar{v}_{n}+\bar{\kappa}\right)\partial_{\xi}\phi_{0}+\alpha^{-1}\partial_{\phi}\omega^{p}\left(\phi_{0}^{in},\mu_{0}^{in}\right)\label{subeq:Oeps1_pf}\\
\mathcal{O}(\varepsilon^{2}):\quad & \partial_{\xi\xi}^{2}\phi_{2}^{in}-\omega_{dw}^{\prime\prime}(\phi_{0}^{in})\phi_{2}^{in} & =(1/2)(\phi_{1}^{in})^{2}\omega_{dw}^{\prime\prime\prime}(\phi_{0}^{in})+\bar{D}\partial_{\overline{t}}\phi_{0}^{in}-\partial_{\overline{s}\overline{s}}^{2}\phi_{0}^{in}-(\bar{D}\bar{v}_{n}+\bar{\kappa})\partial_{\xi}\phi_{1}^{in}\nonumber \\
 &  & \qquad+\xi\bar{\kappa}^{2}\partial_{\xi}\phi_{0}^{in}+\alpha^{-1}\partial_{\phi\phi}^{2}\omega^{p}(\phi_{0}^{in},\mu_{0}^{in})\phi_{1}^{in}+\alpha^{-1}\partial_{\phi\mu}^{2}\omega^{p}(\phi_{0}^{in},\mu_{0}^{in})\mu_{1}^{in}\label{subeq:Oeps2_pf}
\end{eqnarray}

\end{subequations}

\begin{subequations}

\begin{eqnarray}
\mathcal{O}(1):\quad & \partial_{\xi}\left[q(\phi_{0}^{in})\partial_{\xi}\mu_{0}^{in}\right] & =0\label{subeq:Oeps0_diff}\\
\mathcal{O}(\varepsilon):\quad & \partial_{\xi}\left[q(\phi_{0}^{in})\partial_{\xi}\mu_{1}^{in}\right] & =-\partial_{\xi}\left[\phi_{1}^{in}\partial_{\phi}q(\phi_{0}^{in})\partial_{\xi}\mu_{0}^{in}\right]-\bar{v}_{n}\partial_{\xi}c_{0}^{in}-\bar{\kappa}q(\phi_{0}^{in})\partial_{\xi}\mu_{0}^{in}+\partial_{\xi}\left[a(\phi_{0}^{in})\Delta c^{co}\bar{v}_{n}\partial_{\xi}\phi_{0}^{in}\right]\label{subeq:Oeps1_diff}\\
\mathcal{O}(\varepsilon^{2}):\quad & \partial_{\xi}\left[q(\phi_{0}^{in})\partial_{\xi}\mu_{2}^{in}\right] & =\partial_{\overline{t}}c_{0}^{in}+\xi\bar{\kappa}^{2}q(\phi_{0}^{in})\partial_{\xi}\mu_{0}^{in}-\bar{v}_{n}\partial_{\xi}c_{1}^{in}-\bar{\kappa}q(\phi_{0}^{in})\partial_{\xi}\mu_{1}^{in}\nonumber \\
 &  & \qquad-\partial_{\overline{s}}\left[q(\phi_{0}^{in})\partial_{\overline{s}}\mu_{0}^{in}\right]-\bar{\kappa}\phi_{1}^{in}\partial_{\phi}q(\phi_{0}^{in})\partial_{\xi}\mu_{0}^{in}-\partial_{\xi}\left[\phi_{1}^{in}\partial_{\phi}q(\phi_{0}^{in})\partial_{\xi}\mu_{1}^{in}\right]\nonumber \\
 &  & \qquad-\partial_{\xi}\left\{ \left[\phi_{2}^{in}\partial_{\phi}q(\phi_{0}^{in})+(\phi_{1}^{in\,2}/2)\partial_{\phi\phi}^{2}q(\phi_{0}^{in})\right]\partial_{\xi}\mu_{0}^{in}\right\} +\Delta c^{co}\bar{\kappa}a(\phi_{0}^{in})\bar{v}_{n}\partial_{\xi}\phi_{0}^{in}\nonumber \\
 &  & \qquad+\Delta c^{co}\partial_{\xi}\left[a(\phi_{0}^{in})\bar{v}_{n}\partial_{\xi}\phi_{1}^{in}+a^{\prime}(\phi_{0}^{in})\bar{v}_{n}\phi_{1}^{in}\partial_{\xi}\phi_{0}^{in}\right]\label{subeq:Oeps2_diff}
\end{eqnarray}

\end{subequations}

\end{strip}

The closure equation writes (only $\mathcal{O}(1)$ is necessary):

\begin{align}
\mathcal{O}(1): & \qquad c_{0}^{in}=-\partial_{\mu}\omega^{h}(\phi_{0}^{in},\mu_{0}^{in})\label{eq:Oeps0_closure}
\end{align}

In the rest of this section, the superscript ``in'' will be removed
for all inner variables i.e. for $0\leq j\leq2$ we note $\phi_{j}\equiv\phi_{j}^{in}$,
$\mu_{j}\equiv\mu_{j}^{in}$ and $c_{j}\equiv c_{j}^{in}$. The superscript
``out'' is kept for the variables of outer domain.

\subsubsection{\label{subsec:Inner_O(1)}Analysis of terms $\mathcal{O}(1)$}

For $\phi$-equation, Eq. (\ref{subeq:Oeps0_pf}) can be easily solved
by using $\omega_{dw}$. At order zero, we obtain the hyperbolic tangent
profile of the phase-field:
\begin{equation}
\phi_{0}(\xi)=\frac{1+\tanh(2\xi)}{2}\label{eq:Oeps0_phase_field_phi0}
\end{equation}

For $c$-equation, two successive integrations of Eq. (\ref{subeq:Oeps0_diff})
yield $\partial_{\xi}\mu_{0}=A(\overline{s})/q(\phi_{0})$ and $\mu_{0}=B(\overline{s})+A(\overline{s})\int_{0}^{\xi}q^{-1}(\phi_{0})d\xi$.
Both constants of integration $A$ and $B$ can depend on $\overline{s}$.
We can see that $A=0$ is necessary because the integral does not
converge when $\xi\rightarrow+\infty$ (because $q\rightarrow1$ when
$\xi\rightarrow+\infty$). Hence, $\mu_{0}$ depends only on the position
along the interface $\overline{s}$, and the matching conditions yield:

\begin{equation}
\mu_{0}^{out}(0^{+})=\mu_{0}^{out}(0^{-})\label{eq:Continuity_Chempot}
\end{equation}
i.e. the continuity of the chemical potential at order zero.

\subsubsection{\label{subsec:Inner_O(eps)}$\phi$-equation: analysis of terms $\mathcal{O}(\varepsilon)$}

The Gibbs-Thomson condition at order zero arises from the analysis
of terms $\mathcal{O}(\varepsilon)$ of $\phi$-equation. After multiplication
of Eq. (\ref{subeq:Oeps1_pf}) by $\partial_{\xi}\phi_{0}$ and integration
wrt $\xi$ from $-\infty$ to $+\infty$, we obtain:

\begin{align}
\int_{-\infty}^{+\infty}\mathcal{L}(\phi_{1})\partial_{\xi}\phi_{0}d\xi & =-\left(\bar{D}\bar{v}_{n}+\bar{\kappa}\right)\int_{-\infty}^{+\infty}(\partial_{\xi}\phi_{0})^{2}d\xi+\nonumber \\
 & \quad\frac{1}{\alpha}\int_{-\infty}^{+\infty}\partial_{\xi}\phi_{0}\partial_{\phi}\omega^{p}(\phi_{0},\mu_{0})d\xi\label{eq:Oeps1_phi_integrated}
\end{align}
where we set $\mathcal{L}(\phi_{1})=\partial_{\xi\xi}^{2}\phi_{1}-\omega_{dw}^{\prime\prime}(\phi_{0})\phi_{1}$.
The left-hand side (LHS) of Eq. (\ref{eq:Oeps1_phi_integrated}) is
integrated by parts (using $\mathcal{O}(1)$ phase field for the last
equality), we find:
\[
\int_{-\infty}^{+\infty}\partial_{\xi}\phi_{0}\mathcal{L}(\phi_{1})d\xi=-\int_{-\infty}^{+\infty}\partial_{\xi}\phi_{1}\mathcal{L}(\phi_{0})d\xi=0
\]
In the right-hand side (RHS) of Eq. (\ref{eq:Oeps1_phi_integrated}),
the first integral is noted $\mathscr{I}=\int_{-\infty}^{+\infty}(\partial_{\xi}\phi_{0})^{2}d\xi=2/3$.
The second integral can be calculated because $\mu_{0}$ is independent
of $\xi$. Finally Eq. (\ref{eq:Oeps1_phi_integrated}) becomes:
\begin{equation}
0=-(\bar{D}\bar{v}_{n}+\bar{\kappa})\mathscr{I}+\frac{1}{\alpha}\left[\omega_{l}(\mu_{0})-\omega_{s}(\mu_{0})\right]\label{eq:Oeps1_phase_field_adim_gibbs_thomson}
\end{equation}
Using the expression $\omega_{l}(\mu_{0})-\omega_{s}(\mu_{0})=-(c_{l}^{co}-c_{s}^{co})(\mu_{0}-\mu^{eq})$
and the definitions of $\overline{D}$, $\overline{v}_{n}$ and $\overline{\kappa}$,
we obtain:
\begin{equation}
(\mu_{0}-\mu^{eq})\Delta c^{co}=-\frac{\mathscr{I}W}{\lambda}\kappa-\frac{\mathscr{I}W}{M_{\phi}\lambda}v_{n}\label{eq:Oeps1_phase_field_quadratic_gibbs_thomson}
\end{equation}
where $\Delta c^{co}=c_{l}^{co}-c_{s}^{co}$. We identify $d_{0}=\mathscr{I}W/\lambda$,
meaning that $\alpha=1/\mathscr{I}$, and $\beta_{0}=d_{0}/M_{\phi}$.

Another useful result for simplifying the future analyses is the first-order
phase-field $\phi_{1}$. The bilinear form $(\psi,\,\phi)\rightarrow\int\psi\mathcal{L}(\phi)$
is continuous and coercive where $\psi$ and $\phi$ are two functions
vanishing at $+\infty$ and $-\infty$. This means that the operator
$\mathcal{L}$ is invertible in this space. Combined with Eq. (\ref{eq:Oeps1_phase_field_adim_gibbs_thomson}),
this indicates that $\phi_{1}$ is determined by:
\begin{align*}
\phi_{1} & =\frac{\Delta\omega(\mu_{0})}{\alpha}\mathcal{L}^{-1}\left[p^{\prime}(\phi_{0})-\frac{1}{\mathscr{I}}\partial_{\xi}\phi_{0}\right]
\end{align*}
With the derivatives of $p(\phi)$ and $\phi_{0}$ defined in Tab
\ref{tab:Interpolation-functions-used} and $\mathscr{I}=2/3$, the
term inside the brackets vanishes. The unique solution is $\phi_{1}=0$
meaning that all terms depending on $\phi_{1}$ in Eqs. (\ref{subeq:Oeps2_pf}),
(\ref{subeq:Oeps1_diff}) and (\ref{subeq:Oeps2_diff}) can be removed.

\subsubsection{$c$-equation: analysis of terms $\mathcal{O}(\varepsilon)$}

The Stefan condition at order zero arises from the analysis of terms
$\mathcal{O}(\varepsilon)$ of $c$-equation. However, the analysis
also reveals two spurious terms for the chemical potential: the chemical
potentials of first-order $\mu_{1}^{out}(0^{+})$ and $\mu_{1}^{out}(0^{-})$
are not identical on both sides of the interface.

We start with a simplification of $c$-equation Eq. (\ref{subeq:Oeps1_diff})
with the previous result $\partial_{\xi}\mu_{0}=0$:
\[
\partial_{\xi}\left[q(\phi_{0})\partial_{\xi}\mu_{1}\right]=-\bar{v}_{n}\partial_{\xi}c_{0}+\partial_{\xi}\left[\Delta c^{co}a(\phi_{0})\bar{v}_{n}\partial_{\xi}\phi_{0}\right]
\]
After one integration, we obtain:
\begin{equation}
q(\phi_{0})\partial_{\xi}\mu_{1}=-\bar{v}_{n}c_{0}+A+\Delta c^{co}a(\phi_{0})\bar{v}_{n}\partial_{\xi}\phi_{0}\label{eq:Flux_mu1}
\end{equation}
Considering the limit $\xi\rightarrow-\infty$, the integration constant
$A$ is found equal to $A=\Lambda_{s}+\bar{v}_{n}c_{s}(\mu_{0})$
where $\Lambda_{s}=q_{s}\partial_{\eta}\mu_{0}^{out}(0^{-})$. The
closure relation Eq. (\ref{eq:Oeps0_closure}) writes $c_{0}=h(\phi_{0})c_{l}(\mu_{0})+\left[1-h(\phi_{0})\right]c_{s}(\mu_{0})$.
Eq. (\ref{eq:Flux_mu1}) becomes

\begin{equation}
\partial_{\xi}\mu_{1}=\frac{1}{q(\phi_{0})}\left[-\bar{v}_{n}h(\phi_{0})\Delta c_{0}+\Lambda_{s}+\Delta c^{co}a(\phi_{0})\bar{v}_{n}\partial_{\xi}\phi_{0}\right]\label{eq:dmu1indxi}
\end{equation}
where $\Delta c_{0}=c_{l}(\mu_{0})-c_{s}(\mu_{0})$. Because of the
choice of free energies, we also have $\Delta c_{0}=\Delta c^{co}$
(see Eqs. (\ref{eq:Cl_coexistence})-(\ref{eq:Cs_coexistence})).
Integrating once again from $0$ to $\xi$, we obtain
\begin{align}
\mu_{1} & =\Upsilon-\bar{v}_{n}\Delta c_{0}\int_{0}^{\xi}\frac{h(\phi_{0})}{q(\phi_{0})}\nonumber \\
 & \qquad+\Lambda_{s}\int_{0}^{\xi}\frac{1}{q(\phi_{0})}+\bar{v}_{n}\int_{0}^{\xi}\frac{a(\phi_{0})\Delta c^{co}\partial_{\xi}\phi_{0}}{q(\phi_{0})}\label{eq:mu1in_exp}
\end{align}
where $\Upsilon$ is a constant of integration that will be determined
in \ref{subsec:phi-eq_Oeps2}. Now the matching conditions are used
when $\xi\rightarrow\pm\infty$. When $\xi\rightarrow+\infty$, we
ensure the convergence of the first two integrals
by adding $+1-1$ i.e. the first integral considered is $\int_{0}^{\xi}h(\phi_{0})/q(\phi_{0})+1-1$
and the second one is $\Lambda_{s}\int_{0}^{\xi}q^{-1}(\phi_{0})+1-1$.
This yields:

\begin{subequations}
\begin{align}
\mu_{1}^{out}(0^{+})+\partial_{\eta}\mu_{0}^{out}(0^{+})\xi=\Upsilon+\Delta c_{0}\bar{v}_{n}\mathscr{F}_{l}\qquad\quad\nonumber \\
-\bar{v}_{n}\Delta c_{0}\xi+q_{s}\partial_{\eta}\mu_{0}^{out}(0^{-})[\mathscr{G}_{l}+\xi]\label{subeq:Oeps1_diff_mu1in_+inf}\\
\mu_{1}^{out}(0^{-})+\partial_{\eta}\mu_{0}^{out}(0^{-})\xi=\Upsilon+\Delta c_{0}\bar{v}_{n}\mathscr{F}_{s}\qquad\quad\nonumber \\
+q_{s}\partial_{\eta}\mu_{0}^{out}(0^{-})[\mathscr{G}_{s}+(\xi/q_{s})]\label{subeq:Oeps1_diff_mu1in_-inf}
\end{align}
where $\mathscr{F}_{l}$, $\mathscr{F}_{s}$, $\mathscr{G}_{l}$ and
$\mathscr{G}_{s}$ are four integrals which are defined in Tab. \ref{tab:integrals}.
In Eqs. (\ref{subeq:Oeps1_diff_mu1in_+inf})-(\ref{subeq:Oeps1_diff_mu1in_-inf}),
the $\xi$-terms are considered separately and gathered in one additional
equation. The three equations write:

\end{subequations}

\begin{subequations}

\begin{align}
\mu_{1}^{out}(0^{+}) & =\Upsilon+\bar{v}_{n}\Delta c_{0}\mathscr{F}_{l}+q_{s}\partial_{\eta}\mu_{0}^{out}(0^{-})\mathscr{G}_{l}\label{subeq:Oeps1_diff_mu0out+}\\
\mu_{1}^{out}(0^{-}) & =\Upsilon+\bar{v}_{n}\Delta c_{0}\mathscr{F}_{s}+q_{s}\partial_{\eta}\mu_{0}^{out}(0^{-})\mathscr{G}_{s}\label{subeq:Oeps1_diff_mu0out-}\\
\partial_{\eta}\mu_{0}^{out}(0^{+}) & =-\bar{v}_{n}\Delta c_{0}+q_{s}\partial_{\eta}\mu_{0}^{out}(0^{-})\label{subeq:Oeps1_diff_dmu0out+}
\end{align}
The last relationship Eq. (\ref{subeq:Oeps1_diff_dmu0out+}) is the
Stefan condition at order zero:

\end{subequations}

\begin{align}
\partial_{\eta}\mu_{0}^{out}(0^{+})-q_{s}\partial_{\eta}\mu_{0}^{out}(0^{-}) & =-\bar{v}_{n}\Delta c_{0}\label{eq:Oeps1_diff_flux_conservation_zeroth_order}
\end{align}
Eq. (\ref{subeq:Oeps1_diff_mu0out+}) minus Eq. (\ref{subeq:Oeps1_diff_mu0out-})
yields the jump of the first-order chemical potential:

\begin{equation}
\mu_{1}^{out}(0^{+})-\mu_{1}^{out}(0^{-})=\bar{v}_{n}\Delta c_{0}\Delta\mathscr{F}+q_{s}\partial_{\eta}\mu_{0}^{out}(0^{-})\Delta\mathscr{G}\label{eq:Oeps1_diff_chempot_jump_first_order}
\end{equation}
where $\Delta\mathscr{F}=\mathscr{F}_{l}-\mathscr{F}_{s}$ and $\Delta\mathscr{G}=\mathscr{G}_{l}-\mathscr{G}_{s}$.
The first-order chemical potentials are not the same on both side
of the interface. The discontinuity contains two terms proportional
to $\bar{v}_{n}\Delta c_{0}$ and $q_{s}\partial_{\eta}\mu_{0}^{out}(0^{-})$.

\subsubsection{\label{subsec:phi-eq_Oeps2}$\phi$-equation: analysis of terms $\mathcal{O}(\varepsilon^{2})$}

The analysis of terms $O(\varepsilon^{2})$ of $\phi$-equation reveals
that the discontinuity of the first-order chemical potential adds
an error term in the Gibbs-Thomson condition. Since $\phi_{1}=0$
and $\phi_{0}$ does not depend on $\bar{t}$ and $\overline{s}$,
Eq. (\ref{subeq:Oeps2_pf}) simplifies to:
\begin{align*}
\partial_{\xi\xi}^{2}\phi_{2}-\omega_{dw}^{\prime\prime}(\phi_{0})\phi_{2} & =\alpha^{-1}\partial_{\phi\mu}^{2}\omega^{p}(\phi_{0},\mu_{0})\mu_{1}+\xi\bar{\kappa}^{2}\partial_{\xi}\phi_{0}
\end{align*}
We use the same method applied to $\phi$-eq of order $\mathcal{O}(\varepsilon)$:
the LHS is noted $\mathcal{L}(\phi_{2})$ and we multiply the two
sides by $\partial_{\xi}\phi_{0}$ before integrating over $\xi$
varying from $-\infty$ to $+\infty$:

\begin{align}
\int_{-\infty}^{\infty}\partial_{\xi}\phi_{0}\mathcal{L}(\phi_{2}) & d\xi=\alpha^{-1}\int_{-\infty}^{+\infty}\partial_{\xi}\phi_{0}\partial_{\phi\mu}^{2}\omega^{p}(\phi_{0},\mu_{0})\mu_{1}d\xi\nonumber \\
 & \qquad+\int_{-\infty}^{\infty}\bar{\kappa}^{2}\xi(\partial_{\xi}\phi_{0})^{2}d\xi\label{eq:Oeps2_phase_field_full_integral}
\end{align}
With the same arguments as the $\mathcal{O}(\varepsilon)$-equation,
the LHS of Eq. (\ref{eq:Oeps2_phase_field_full_integral}) is null.
The last integral on the RHS vanishes because the integrand is odd.
Eq. (\ref{eq:Oeps2_phase_field_full_integral}) is reduced to:
\begin{equation}
\int_{-\infty}^{+\infty}\left[\partial_{\xi}\phi_{0}\partial_{\phi\mu}^{2}\omega^{p}(\phi_{0},\mu_{0})\mu_{1}\right]d\xi=0\label{eq:Oeps2_phase_field_nonzero_terms}
\end{equation}
From this point, the analysis is straightforward. Comparatively, during
the analysis of a KKS-type formulation of the model (like in \citep[Appendix A]{Provatas-Elder_Book_2010}),
a similar integral would have been obtained, but with the interpolation
of grand potentials $\omega^{p}$ which is replaced by the free energy
of an interpolation of compositions. A lengthy analysis involving
the closure equation \ref{eq:Oeps0_closure} expanded at first-order
would have been required to get to the same point. Here, the use of
the grand-potential formulation significantly simplifies the analysis.
It can also be noted that such a simplification has already been used
for the $\mathcal{O}(\varepsilon)$-terms of $\phi$-equation in Sec.
\ref{subsec:Inner_O(eps)} because the integration of $\partial_{\xi}\phi_{0}\partial_{\phi}\omega^{p}(\phi_{0},\,\mu_{0})$
is much easier than $\partial_{\xi}\phi_{0}\partial_{\phi}f(\phi_{0},\,c_{0})$
(since $\mu_{0}$ is constant and not $c_{0}$). In Eq. (\ref{eq:Oeps2_phase_field_nonzero_terms}),
we replace $\mu_{1}$ by its expression Eq. (\ref{eq:mu1in_exp}):

\begin{align*}
\int_{-\infty}^{+\infty}\partial_{\xi}\phi_{0}\partial_{\phi\mu}^{2}\omega^{p}(\phi_{0},\mu_{0})\mu_{1}=-\int_{-\infty}^{+\infty}\left[\partial_{\xi}\phi_{0}p^{\prime}(\phi_{0})\Delta c_{0}\right]\times\\
\left\{ \Upsilon-\bar{v}_{n}\Delta c_{0}\int_{0}^{\xi}\left[\frac{h(\phi_{0})}{q(\phi_{0})}-\frac{a(\phi_{0})\partial_{\xi}\phi_{0}}{q(\phi_{0})}\right]+\Lambda_{s}\int_{0}^{\xi}\frac{1}{q(\phi_{0})}\right\} \\
=\bar{v}_{n}\left(\Delta c_{0}\right)^{2}\mathscr{K}-\Upsilon\Delta c_{0}\qquad\,\,\quad\\
\qquad-\Lambda_{s}(\mathscr{G}_{l}+\tilde{\mathscr{F}}_{l}-\tilde{\mathscr{F}}_{s})\Delta c_{0}
\end{align*}
where the three integrals $\mathscr{K}$, $\tilde{\mathscr{F}}_{l}$
and $\tilde{\mathscr{F}}_{s}$ are defined in Tab. \ref{tab:integrals}.
For that result we used the two relations $\int_{-\infty}^{+\infty}\partial_{\xi}\phi_{0}p^{\prime}(\phi_{0})=p(1)-p(0)=1$
and $\int_{-\infty}^{+\infty}\partial_{\xi}\phi_{0}p^{\prime}(\phi_{0})\int_{0}^{\xi}q^{-1}(\phi_{0})=\mathscr{G}_{l}+\tilde{\mathscr{F}}_{l}-\tilde{\mathscr{F}}_{s}$.
From Eq. (\ref{eq:Oeps2_phase_field_nonzero_terms}) the integral
is zero, so $\Upsilon$ is given by:
\begin{equation}
\Upsilon=\bar{v}_{n}\Delta c_{0}\mathscr{K}-\Lambda_{s}(\mathscr{G}_{l}+\tilde{\mathscr{F}}_{l}-\tilde{\mathscr{F}}_{s})\label{eq:Constant_of_Integration}
\end{equation}
That relation is used to replace $\Upsilon$ in Eqs. (\ref{subeq:Oeps1_diff_mu0out+})-(\ref{subeq:Oeps1_diff_mu0out-}):

\begin{subequations}

\textbf{
\begin{align}
\mu_{1}^{out}(0^{+}) & =\bar{v}_{n}\Delta c_{0}[\mathscr{K}+\mathscr{F}_{l}]-\Lambda_{s}\Delta\tilde{\mathscr{F}}\label{subeq:Oeps2_phase_field_first_order_gibbs_thomson+}\\
\mu_{1}^{out}(0^{-}) & =\bar{v}_{n}\Delta c_{0}[\mathscr{K}+\mathscr{F}_{s}]-\Lambda_{s}[\Delta\mathscr{G}+\Delta\tilde{\mathscr{F}}]\label{subeq:Oeps2_phase_field_first_order_gibbs_thomson-}
\end{align}
}where we have set $\Delta\tilde{\mathscr{F}}=\tilde{\mathscr{F}}_{l}-\tilde{\mathscr{F}}_{s}$
and $\Delta\mathscr{G}=\mathscr{G}_{l}-\mathscr{G}_{s}$.

\end{subequations}

By summing Eq. (\ref{eq:Oeps1_phase_field_quadratic_gibbs_thomson})
with $\varepsilon\times$Eqs. (\ref{subeq:Oeps2_phase_field_first_order_gibbs_thomson+})-(\ref{subeq:Oeps2_phase_field_first_order_gibbs_thomson-})
such as $\mu^{out}(0^{\pm})=\mu_{0}(0^{\pm})+\varepsilon\mu_{1}^{out}(0^{\pm})$,
we get the discontinuity of chemical potential at interface:
\begin{equation}
\mu_{l}-\mu_{s}=\varepsilon v_{n}\frac{\mathscr{I}W}{D_{l}\lambda}\Delta\mathscr{F}+q_{s}W\partial_{r}\mu_{0}^{out}(0^{-})\Delta\mathscr{G}\label{eq:Jump_ChemPot_Eps1}
\end{equation}
where $\mu_{l}\equiv\mu^{out}(0^{+})$ and $\mu_{s}\equiv\mu^{out}(0^{-})$.
That jump involves two Gibbs-Thomson conditions, one for each side
of the interface:

\begin{subequations}

\begin{align}
\mu_{l}-\mu^{eq} & =-\frac{d_{0}}{\Delta c^{co}}\kappa-\frac{\beta_{l}}{\Delta c^{co}}v_{n}+\mathbb{E}_{1}\Delta\tilde{\mathscr{F}}\label{eq:Gibbs-Thomson_Liquid}\\
\mu_{s}-\mu^{eq} & =-\frac{d_{0}}{\Delta c^{co}}\kappa-\frac{\beta_{s}}{\Delta c^{co}}v_{n}+\mathbb{E}_{1}\left[\Delta\tilde{\mathscr{F}}-\Delta\mathscr{G}\right]\label{eq:Gibbs-Thomson_Solid}
\end{align}
where the first error term is noted $\mathbb{E}_{1}=q_{s}W\partial_{r}\mu_{0}^{out}(0^{-})$
which cancels if $q_{s}=0$ i.e. $D_{s}=0$. Those two relations are
simply written for $\Phi=s,\,l$:

\end{subequations}

\begin{subequations}

\begin{equation}
\mu_{\Phi}-\mu^{eq}=-\frac{d_{0}}{\Delta c^{co}}\kappa-\frac{\beta_{\Phi}}{\Delta c^{co}}v_{n}+\mathbb{E}_{1}\left[\Delta\tilde{\mathscr{F}}-\Delta\mathscr{G}_{\Phi}\right]\label{eq:Gibbs-Thomson_Final}
\end{equation}
where $\Delta\mathscr{G}_{\Phi}=\mathscr{G}_{l}-\mathscr{G}_{\Phi}$.
The capillary length $d_{0}$ and the kinetic coefficient $\beta_{\Phi}$
are defined by:
\begin{align}
d_{0} & =\frac{\mathscr{I}W}{\lambda}\label{eq:CapillaryLength_Appendix}\\
\beta_{\Phi} & =\frac{\mathscr{I}W}{\lambda M_{\phi}}\left[1-\lambda\frac{M_{\phi}}{D_{l}}\frac{\mathscr{K}+\mathscr{F}_{\Phi}}{\mathscr{I}}(\Delta c^{co})^{2}\right]\label{eq:KineticCoeff_Appendix}
\end{align}

\end{subequations}

\subsubsection{\label{subsec:Inner(eps2)}$c$-equation: analysis of terms $\mathcal{O}(\varepsilon^{2})$ }

For $c$-equation, two spurious terms in the Stefan condition arise
from the analysis of terms $O(\varepsilon^{2})$. After considering
$\phi_{1}=0$, $\partial_{\xi}\mu_{0}=0$, and $\partial_{\overline{t}}c_{0}=0$
(because $c_{0}$ is constant), Eq. (\ref{subeq:Oeps2_diff}) becomes:

\begin{align*}
\partial_{\xi}\left[q(\phi_{0})\partial_{\xi}\mu_{2}\right] & =-\bar{v}_{n}\partial_{\xi}c_{1}-\bar{\kappa}q(\phi_{0})\partial_{\xi}\mu_{1}-q(\phi_{0})\partial_{\overline{s}\overline{s}}^{2}\mu_{0}\\
 & \qquad+\bar{\kappa}a(\phi_{0})\Delta c^{co}\bar{v}_{n}\partial_{\xi}\phi_{0}
\end{align*}
In the RHS, the second term $q(\phi_{0})\partial_{\xi}\mu_{1}$ is
replaced by its expression Eq. (\ref{eq:dmu1indxi}). After integration
w.r.t. $\xi$, we obtain:

\begin{align*}
q(\phi_{0})\partial_{\xi}\mu_{2} & =-\bar{v}_{n}c_{1}+\Delta c_{0}\bar{\kappa}\bar{v}_{n}\int_{0}^{\xi}h(\phi_{0})-\bar{\kappa}\Lambda_{s}\xi\\
 & \qquad-\partial_{\overline{s}\overline{s}}^{2}\mu_{0}\int_{0}^{\xi}q(\phi_{0})+B(\overline{s})
\end{align*}
where $B(\overline{s})$ is a constant of integration.
Then we use the matching conditions and look at the limits when $\xi$
tends to $+\infty$ and $-\infty$. For $\xi\rightarrow+\infty$,
we obtain: 

\begin{align*}
\partial_{\eta}\mu_{1}^{out}(0^{+})+\partial_{\eta\eta}^{2}\mu_{1}^{out}(0^{+})\xi=-\bar{v}_{n}c_{1}(0^{+})\\
+\Delta c_{0}\bar{\kappa}\bar{v}_{n}\int_{0}^{\infty}[h(\phi_{0})-1]+\Delta c_{0}\bar{\kappa}\bar{v}_{n}\xi+B(\overline{s})\\
-\bar{\kappa}\Lambda_{s}\xi-\partial_{\overline{s}\overline{s}}^{2}\mu_{0}\int_{0}^{\infty}\{[q(\phi_{0})-1]+\xi\}
\end{align*}
and for $\xi\rightarrow-\infty$:

\begin{align*}
q_{s}\partial_{\eta}\mu_{1}^{out}(0^{-})+q_{s}\partial_{\eta\eta}^{2}\mu_{1}^{out}(0^{-})\xi=-\bar{v}_{n}c_{1}(0^{-})\\
+\Delta c_{0}\bar{\kappa}\bar{v}_{n}\int_{0}^{-\infty}h(\phi_{0})-\bar{\kappa}\Lambda_{s}\xi+B(\overline{s})\\
-\partial_{\overline{s}\overline{s}}^{2}\mu_{0}\int_{0}^{-\infty}\{[q(\phi_{0})-q_{s}]+q_{s}\xi\}
\end{align*}
We focus on the terms independent of $\xi$:
\begin{align*}
\partial_{\eta}\mu_{1}^{out}(0^{+}) & =-\bar{v}_{n}c_{1}(0^{+})-\Delta c_{0}\bar{\kappa}\bar{v}_{0}\mathscr{H}_{l}+B(\overline{s})-\mathscr{J}_{l}\partial_{\overline{s}\overline{s}}^{2}\mu_{0}\\
q_{s}\partial_{\eta}\mu_{1}^{out}(0^{-}) & =-\bar{v}_{n}c_{1}(0^{-})-\Delta c_{0}\bar{\kappa}\bar{v}_{0}\mathscr{H}_{s}+B(\overline{s})-\mathscr{J}_{s}\partial_{\overline{s}\overline{s}}^{2}\mu_{0}
\end{align*}
where the integrals $\mathscr{H}_{l}$, $\mathscr{H}_{s}$, $\mathscr{J}_{l}$
and $\mathscr{J}_{s}$ are defined in Tab. \ref{tab:integrals}. The
first equation minus the second one yields:
\begin{align}
\partial_{\eta}\mu_{1}^{out}(0^{+})-q_{s}\partial_{\eta}\mu_{1}^{out}(0^{-}) & =-\bar{v}_{n}\Delta c_{1}-\Delta c_{0}\bar{\kappa}\bar{v}_{n}\Delta\mathscr{H}\nonumber \\
 & \qquad-\partial_{\overline{s}\overline{s}}^{2}\mu_{0}\Delta\mathscr{J}\label{eq:Oeps2_diff_stefan_condition_order1}
\end{align}
where $\Delta\mathscr{H}=\mathscr{H}_{l}-\mathscr{H}_{s}$ and $\Delta\mathscr{J}=\mathscr{J}_{l}-\mathscr{J}_{s}$.
To obtain the Stefan condition, we recombine Eq. (\ref{eq:Oeps1_diff_flux_conservation_zeroth_order})
with $\varepsilon\times$Eq. (\ref{eq:Oeps2_diff_stefan_condition_order1}):

\begin{align}
D_{l}\partial_{r}\mu|_{l}-D_{s}\partial_{r}\mu|_{s} & =-v_{n}\Delta c^{co}-\mathbb{E}_{2}\Delta\mathscr{H}-\mathbb{E}_{3}\Delta\mathscr{J}\label{eq:Stefan_Final}
\end{align}
where $\mathbb{E}_{2}=W\kappa v_{0}\Delta c_{0}$ and $\mathbb{E}_{3}=WD_{l}\partial_{ss}^{2}\mu_{0}$.

\subsection{\label{subsec:Discussion-on-error}Removal of error terms $\mathbb{E}_{1},\,\mathbb{E}_{2},\,\mathbb{E}_{3}$}

Three error terms appear in the sharp interface model which is recovered
by the analysis: $\mathbb{E}_{1}$ in the Gibbs-Thomson condition
Eq. (\ref{eq:Gibbs-Thomson_Final}) and $\mathbb{E}_{2}$, $\mathbb{E}_{3}$
in the Stefan condition Eq. (\ref{eq:Stefan_Final}). The errors $\mathbb{E}_{2}$,
$\mathbb{E}_{3}$ disappear provided that $\Delta\mathscr{H}=0$ and
$\Delta\mathscr{J}=0$. The condition $\Delta\mathscr{H}=0$ is fulfilled
for $h(\phi)$ defined in Tab. \ref{tab:Interpolation-functions-used}.
As a matter of fact, that stays true as long as $h(\phi)$ is an odd
function of $\phi$. For $\Delta\mathscr{J}$, the same property of
the interpolation function $q(\phi)$: $\Delta\mathscr{J}=0$ as long
as an odd function is used to interpolate the diffusivities.

The condition $\Delta\mathscr{F}=0$ cannot be ensured in the model
without antitrapping current (i.e. $a(\phi)=0$). However, the condition
can be respected by choosing appropriately the function $a(\phi)$
of $\boldsymbol{j}_{at}$. We already know that the integrand $h(\phi_{0})$
verifies $\Delta\mathscr{H}=0$. To fulfill the condition $\Delta\mathscr{F}=0$,
the minimal requirement is to equalize the integrand of $\Delta\mathscr{F}$
with $h(\phi_{0})$:
\[
\frac{h(\phi_{0})-a(\phi_{0})\partial_{\xi}\phi_{0}}{q(\phi_{0})}=h(\phi_{0})
\]
Knowing $\partial_{\xi}\phi_{0}=4\phi_{0}(1-\phi_{0})$, we have to
define the function $a(\phi_{0})$ as:
\[
a(\phi_{0})=\frac{\left[1-q(\phi_{0})\right]}{4\phi_{0}(1-\phi_{0})}h(\phi_{0})=\frac{1-q_{s}}{4}
\]

As already mentioned, the error term $\mathbb{E}_{1}$ is proportional
to $q_{s}$ and vanishes if $q_{s}=0$ (i.e. $D_{s}=0).$ If $q_{s}\neq0$
then the error term disappears if $\Delta\tilde{\mathscr{F}}=0$ and
$\Delta\mathscr{G}=0$. Cancelling the term $\Delta\mathscr{G}$ is
possible by using one harmonic interpolation for the diffusion coefficients
instead of the linear interpolation $q(\phi)$ of Tab. \ref{tab:Interpolation-functions-used}.
However, in that case, $\Delta\mathscr{J}$ is no longer zero. To
cancel both, one solution is to use a tensorial diffusivity so as
to set the interpolation as linear in the direction tangent to the
interface and harmonic in the normal direction (see \citep{Nicoli-Plapp-Henry_PRE2011,Ohno_etal_PRE2016}).
In our model, $\mathscr{G}_{l}=\mathscr{G}_{s}=0$ if $q_{s}=1$ (see
Tab. \ref{tab:integrals}) i.e. for identical diffusivities of bulk
phases. Finally the error term $\mathbb{E}_{1}$ disappears for two
cases: $D_{s}=0$ and $D_{s}=D_{l}$.

\bibliographystyle{elsarticle-num}
\bibliography{PF_Dissolution_Accept2fev2022}

\begin{thebibliography}{10}
\expandafter\ifx\csname url\endcsname\relax
  \def\url#1{\texttt{#1}}\fi
\expandafter\ifx\csname urlprefix\endcsname\relax\def\urlprefix{URL }\fi
\expandafter\ifx\csname href\endcsname\relax
  \def\href#1#2{#2} \def\path#1{#1}\fi

\bibitem{Book_LBM2017}
T.~Kr\"uger, H.~Kusumaatmaja, A.~Kuzmin, O.~Shardt, G.~Silva, E.~Viggen, The
  {L}attice {B}oltzmann {M}ethod. Principles and Practice, Springer, 2017.
\newblock \href {https://doi.org/10.1007/978-3-319-44649-3}
  {\path{doi:10.1007/978-3-319-44649-3}}.

\bibitem{Pan-Luo-Miller_CF2006}
C.~Pan, L.-S. Luo, C.~T. Miller, {An evaluation of lattice Boltzmann schemes
  for porous medium flow simulation}, Computers \& Fluids 35~(8) (2006)
  898--909, proceedings of the First International Conference for Mesoscopic
  Methods in Engineering and Science.
\newblock \href {https://doi.org/10.1016/j.compfluid.2005.03.008}
  {\path{doi:10.1016/j.compfluid.2005.03.008}}.

\bibitem{Genty-Pot_TRT_TiPM2013}
A.~Genty, V.~Pot, {Numerical Simulation of 3D Liquid-Gas Distribution in Porous
  Media by a Two-Phase TRT Lattice Boltzmann Method}, Transport in Porous Media
  96 (2013) pp. 271--294.
\newblock \href {https://doi.org/10.1007/s11242-012-0087-9}
  {\path{doi:10.1007/s11242-012-0087-9}}.

\bibitem{Pot_etal_AdWR2015}
V.~Pot, S.~Peth, O.~Monga, L.~Vogel, A.~Genty, P.~Garnier, L.~Vieubl\'e-Gonod,
  M.~Ogurreck, F.~Beckmann, P.~Baveye, {Three-dimensional distribution of water
  and air in soil pores: Comparison of two-phase two-relaxation-times
  lattice-Boltzmann and morphological model outputs with synchrotron X-ray
  computed tomography data}, Advances in Water Resources 84 (2015) 87 -- 102.
\newblock \href {https://doi.org/10.1016/j.advwatres.2015.08.006}
  {\path{doi:10.1016/j.advwatres.2015.08.006}}.

\bibitem{Review_LBM_PorousMed_IJHMT2019}
Y.-L. He, Q.~Liu, Q.~Li, W.-Q. Tao, {Lattice Boltzmann methods for single-phase
  and solid-liquid phase-change heat transfer in porous media: A review},
  International Journal of Heat and Mass Transfer 129 (2019) 160--197.
\newblock \href {https://doi.org/10.1016/j.ijheatmasstransfer.2018.08.135}
  {\path{doi:10.1016/j.ijheatmasstransfer.2018.08.135}}.

\bibitem{Kang_etal_WRR2007}
Q.~Kang, P.~C. Lichtner, D.~Zhang, {An improved lattice Boltzmann model for
  multicomponent reactive transport in porous media at the pore scale}, Water
  Resources Research 43~(12) (2007) W12S14 1--12.
\newblock \href {https://doi.org/10.1029/2006WR005551}
  {\path{doi:10.1029/2006WR005551}}.

\bibitem{KANG20141049}
Q.~Kang, L.~Chen, A.~J. Valocchi, H.~S. Viswanathan, Pore-scale study of
  dissolution-induced changes in permeability and porosity of porous media,
  Journal of Hydrology 517 (2014) 1049--1055.
\newblock \href {https://doi.org/10.1016/j.jhydrol.2014.06.045}
  {\path{doi:10.1016/j.jhydrol.2014.06.045}}.

\bibitem{Zhang_etal_ChemEngSci2022}
Y.~Zhang, F.~Jiang, T.~Tsuji, {Influence of pore space heterogeneity on mineral
  dissolution and permeability evolution investigated using lattice Boltzmann
  method}, Chemical Engineering Science 247 (2022) 117048.
\newblock \href {https://doi.org/10.1016/j.ces.2021.117048}
  {\path{doi:10.1016/j.ces.2021.117048}}.

\bibitem{Xu-Meakin_PhaseField-Preci_JChemPhys2008}
Z.~Xu, P.~Meakin, Phase-field modeling of solute precipitation and dissolution,
  The Journal of Chemical Physics 129~(1) (2008) 014705.
\newblock \href {https://doi.org/10.1063/1.2948949}
  {\path{doi:10.1063/1.2948949}}.

\bibitem{Mai_etal_CorroSci2016}
W.~Mai, S.~Soghrati, R.~G. Buchheit, A phase field model for simulating the
  pitting corrosion, Corrosion Science 110 (2016) 157 -- 166.
\newblock \href {https://doi.org/10.1016/j.corsci.2016.04.001}
  {\path{doi:10.1016/j.corsci.2016.04.001}}.

\bibitem{Bringedal_etal_MMS2020}
C.~Bringedal, L.~von Wolff, I.~S. Pop, {Phase Field Modeling of Precipitation
  and Dissolution Processes in Porous Media: Upscaling and Numerical
  Experiments}, Multiscale Modeling \& Simulation 18~(2) (2020) 1076--1112.
\newblock \href {https://doi.org/10.1137/19M1239003}
  {\path{doi:10.1137/19M1239003}}.

\bibitem{Gao_etal_JCAM2020}
H.~Gao, L.~Ju, R.~Duddu, H.~Li, An efficient second-order linear scheme for the
  phase field model of corrosive dissolution, Journal of Computational and
  Applied Mathematics 367 (2020) 112472.
\newblock \href {https://doi.org/10.1016/j.cam.2019.112472}
  {\path{doi:10.1016/j.cam.2019.112472}}.

\bibitem{Karma-Rappel_PRE1998}
A.~Karma, W.-J. Rappel, Quantitative phase-field modeling of dendritic growth
  in two and three dimensions, Physical Review E 57~(4) (1998) pp. 4323--4349.
\newblock \href {https://doi.org/10.1103/PhysRevE.57.4323}
  {\path{doi:10.1103/PhysRevE.57.4323}}.

\bibitem{Echebarria_etal_PhysRevE.70.061604}
B.~Echebarria, R.~Folch, A.~Karma, M.~Plapp, Quantitative phase-field model of
  alloy solidification, Phys. Rev. E 70 (2004) 061604.
\newblock \href {https://doi.org/10.1103/PhysRevE.70.061604}
  {\path{doi:10.1103/PhysRevE.70.061604}}.

\bibitem{Ramirez_etal_BinaryAlloy_PRE2004}
J.~C. Ramirez, C.~Beckermann, A.~Karma, H.-J. Diepers, Phase-field modeling of
  binary alloy solidification with coupled heat and solute diffusion, Physical
  Review E 69~(051607) (2004) 1--16.
\newblock \href {https://doi.org/10.1103/PhysRevE.69.051607}
  {\path{doi:10.1103/PhysRevE.69.051607}}.

\bibitem{KKS_model_PhysRevE.60.7186}
S.~G. Kim, W.~T. Kim, T.~Suzuki, Phase-field model for binary alloys, Phys.
  Rev. E 60 (1999) 7186--7197.
\newblock \href {https://doi.org/10.1103/PhysRevE.60.7186}
  {\path{doi:10.1103/PhysRevE.60.7186}}.

\bibitem{Provatas-Elder_Book_2010}
N.~Provatas, K.~Elder, {Phase-Field Methods in Materials Science and
  Engineering}, Wiley-VCH, 2010.

\bibitem{Plapp_PhysRevE.84.031601}
M.~Plapp, Unified derivation of phase-field models for alloy solidification
  from a grand-potential functional, Phys. Rev. E 84 (2011) 031601.
\newblock \href {https://doi.org/10.1103/PhysRevE.84.031601}
  {\path{doi:10.1103/PhysRevE.84.031601}}.

\bibitem{Choudhury-Nestler_PRE2012}
A.~Choudhury, B.~Nestler, Grand-potential formulation for multicomponent phase
  transformations combined with thin-interface asymptotics of the
  double-obstacle potential, Phys. Rev. E 85 (2012) 021602.
\newblock \href {https://doi.org/10.1103/PhysRevE.85.021602}
  {\path{doi:10.1103/PhysRevE.85.021602}}.

\bibitem{Cogswell_GrdPot_PRE2015}
D.~A. Cogswell, Quantitative phase-field modeling of dendritic
  electrodeposition, Phys. Rev. E 92 (2015) 011301.
\newblock \href {https://doi.org/10.1103/PhysRevE.92.011301}
  {\path{doi:10.1103/PhysRevE.92.011301}}.

\bibitem{Aagesen_etal_GrdPot_PRE2018}
L.~K. Aagesen, Y.~Gao, D.~Schwen, K.~Ahmed, Grand-potential-based phase-field
  model for multiple phases, grains, and chemical components, Phys. Rev. E 98
  (2018) 023309.
\newblock \href {https://doi.org/10.1103/PhysRevE.98.023309}
  {\path{doi:10.1103/PhysRevE.98.023309}}.

\bibitem{Simon_etal_GrdPot_CompMatSci2020}
P.-C.~A. Simon, L.~K. Aagesen, A.~T. Motta, M.~R. Tonks, The effects of
  introducing elasticity using different interpolation schemes to the grand
  potential phase field model, Computational Materials Science 183 (2020)
  109790.
\newblock \href {https://doi.org/10.1016/j.commatsci.2020.109790}
  {\path{doi:10.1016/j.commatsci.2020.109790}}.

\bibitem{Book_Ratke-Voorhees_2002}
{L. Ratke and P. W. Voorhees}, {Growth and Coarsening}, Springer Berlin
  Heidelberg, 2002.
\newblock \href {https://doi.org/10.1007/978-3-662-04884-9}
  {\path{doi:10.1007/978-3-662-04884-9}}.

\bibitem{Folch_etal_PRE1999}
R.~Folch, J.~Casademunt, A.~Hern\'andez-Machado, L.~Ram\'{\i}rez-Piscina,
  {Phase-field model for Hele-Shaw flows with arbitrary viscosity contrast. I.
  Theoretical approach}, Phys. Rev. E 60 (1999) 1724--1733.
\newblock \href {https://doi.org/10.1103/PhysRevE.60.1724}
  {\path{doi:10.1103/PhysRevE.60.1724}}.

\bibitem{Sun-Beckermann_JCP2007}
Y.~Sun, C.~Beckermann, Sharp interface tracking using the phase-field equation,
  Journal of Computational Physics 220~(2) (2007) 626 -- 653.
\newblock \href {https://doi.org/10.1016/j.jcp.2006.05.025}
  {\path{doi:10.1016/j.jcp.2006.05.025}}.

\bibitem{Chiu-Lin_JCP2011}
P.-H. Chiu, Y.-T. Lin, A conservative phase field method for solving
  incompressible two-phase flows, Journal of Computational Physics 230~(1)
  (2011) 185 -- 204.
\newblock \href {https://doi.org/10.1016/j.jcp.2010.09.021}
  {\path{doi:10.1016/j.jcp.2010.09.021}}.

\bibitem{Verdier_etal_CMAME2020}
W.~Verdier, P.~Kestener, A.~Cartalade, {Performance portability of lattice
  Boltzmann methods for two-phase flows with phase change}, Computer Methods in
  Applied Mechanics and Engineering 370 (2020) 113266.
\newblock \href {https://doi.org/10.1016/j.cma.2020.113266}
  {\path{doi:10.1016/j.cma.2020.113266}}.

\bibitem{Choudhury_etal_COSSMS2015}
A.~Choudhury, M.~Kellner, B.~Nestler, A method for coupling the phase-field
  model based on a grand-potential formalism to thermodynamic databases,
  Current Opinion in Solid State and Materials Science 19~(5) (2015) 287--300.
\newblock \href {https://doi.org/10.1016/j.cossms.2015.03.003}
  {\path{doi:10.1016/j.cossms.2015.03.003}}.

\bibitem{Introini_etal_JNM2021}
C.~Intro{\"i}ni, J.~Sercombe, I.~Rami\`ere, R.~{L}e {T}ellier, Phase-field
  modeling with the {TAF-ID} of incipient melting and oxygen transport in
  nuclear fuel during power transients, Journal of Nuclear Materials 556 (2021)
  153173.
\newblock \href {https://doi.org/10.1016/j.jnucmat.2021.153173}
  {\path{doi:10.1016/j.jnucmat.2021.153173}}.

\bibitem{Sundman_etal_IMMI2015}
B.~Sundman, U.~R. Kattner, M.~Palumbo, S.~G. Fries, {OpenCalphad} - a free
  thermodynamic software, Integrating Materials and Manufacturing Innovation
  4~(1) (2015) 1--15.
\newblock \href {https://doi.org/10.1186/s40192-014-0029-1}
  {\path{doi:10.1186/s40192-014-0029-1}}.

\bibitem{Sundman_etal_CompMatSci2015}
B.~Sundman, X.-G. Lu, H.~Ohtani, The implementation of an algorithm to
  calculate thermodynamic equilibria for multi-component systems with non-ideal
  phases in a free software, Computational Materials Science 101 (2015)
  127--137.
\newblock \href {https://doi.org/10.1016/j.commatsci.2015.01.029}
  {\path{doi:10.1016/j.commatsci.2015.01.029}}.

\bibitem{Karma_AntiTrapping_PRL2001}
A.~Karma, {Phase-Field Formulation for Quantitative Modeling of Alloy
  Solidification}, Phys. Rev. Lett. 87 (2001) 115701.
\newblock \href {https://doi.org/10.1103/PhysRevLett.87.115701}
  {\path{doi:10.1103/PhysRevLett.87.115701}}.

\bibitem{Lee-Lin_JCP2005}
T.~Lee, C.-L. Lin, A stable discretization of the lattice {B}oltzmann equation
  for simulation of incompressible two-phase flows at high density ratio,
  Journal of Computational Physics 206~(1) (2005) 16--47.
\newblock \href {https://doi.org/10.1016/j.jcp.2004.12.001}
  {\path{doi:10.1016/j.jcp.2004.12.001}}.

\bibitem{Zheng_etal_LargeDensityRatio_JCP2006}
H.~Zheng, C.~Shu, Y.~Chew, {A lattice Boltzmann model for multiphase flows with
  large density ratio}, Journal of Computational Physics 218 (2006) pp.
  353--371.
\newblock \href {https://doi.org/10.1016/j.jcp.2006.02.015}
  {\path{doi:10.1016/j.jcp.2006.02.015}}.

\bibitem{Fakhari_etal_JCP2017}
A.~Fakhari, D.~Bolster, L.-S. Luo, {A weighted multiple-relaxation-time lattice
  Boltzmann method for multiphase flows and its application to partial
  coalescence cascades}, Journal of Computational Physics 341 (2017) 22 -- 43.
\newblock \href {https://doi.org/10.1016/j.jcp.2017.03.062}
  {\path{doi:10.1016/j.jcp.2017.03.062}}.

\bibitem{Bayle_PhD2020}
R.~Bayle, {Simulation des m{\'e}canismes de changement de phase dans des
  m{\'e}moires PCM avec la m{\'e}thode multi-champ de phase}, Phd thesis,
  {Institut Polytechnique de Paris},
  \url{https://tel.archives-ouvertes.fr/tel-03043958} (Jul. 2020).

\bibitem{Brenner-Boussinot_PRE2012}
E.~A. Brener, G.~Boussinot, Kinetic cross coupling between nonconserved and
  conserved fields in phase field models, Phys. Rev. E 86 (2012) 060601.
\newblock \href {https://doi.org/10.1103/PhysRevE.86.060601}
  {\path{doi:10.1103/PhysRevE.86.060601}}.

\bibitem{Fang-Mi_PRE2013}
A.~Fang, Y.~Mi, {Recovering thermodynamic consistency of the antitrapping
  model: A variational phase-field formulation for alloy solidification}, Phys.
  Rev. E 87 (2013) 012402.
\newblock \href {https://doi.org/10.1103/PhysRevE.87.012402}
  {\path{doi:10.1103/PhysRevE.87.012402}}.

\bibitem{Fife_1988}
P.~C. Fife, {Dynamics of Internal Layers and Diffusive Interfaces}, Society for
  Industrial and Applied Mathematics, 1988.
\newblock \href {https://doi.org/10.1137/1.9781611970180}
  {\path{doi:10.1137/1.9781611970180}}.

\bibitem{Caginalp_PRA1989}
G.~Caginalp, {Stefan and Hele-Shaw type models as asymptotic limits of the
  phase-field equations}, Phys. Rev. A 39 (1989) 5887--5896.
\newblock \href {https://doi.org/10.1103/PhysRevA.39.5887}
  {\path{doi:10.1103/PhysRevA.39.5887}}.

\bibitem{Almgren_SIAM1999}
R.~F. Almgren, {Second-Order Phase Field Asymptotics for Unequal
  Conductivities}, SIAM Journal on Applied Mathematics 59~(6) (1999)
  2086--2107.
\newblock \href {https://doi.org/10.1137/S0036139997330027}
  {\path{doi:10.1137/S0036139997330027}}.

\bibitem{McFadden-Wheeler-Anderson_PhysD2000}
G.~McFadden, A.~Wheeler, D.~Anderson, Thin interface asymptotics for an
  energy/entropy approach to phase-field models with unequal conductivities,
  Physica D: Nonlinear Phenomena 144~(1) (2000) 154--168.
\newblock \href {https://doi.org/10.1016/S0167-2789(00)00064-6}
  {\path{doi:10.1016/S0167-2789(00)00064-6}}.

\bibitem{Ohno-Matsuura_PRE2009}
M.~Ohno, K.~Matsuura, Quantitative phase-field modeling for dilute alloy
  solidification involving diffusion in the solid, Phys. Rev. E 79 (2009)
  031603.
\newblock \href {https://doi.org/10.1103/PhysRevE.79.031603}
  {\path{doi:10.1103/PhysRevE.79.031603}}.

\bibitem{Ohno_etal_PRE2016}
M.~Ohno, T.~Takaki, Y.~Shibuta, Variational formulation and numerical accuracy
  of a quantitative phase-field model for binary alloy solidification with
  two-sided diffusion, Phys. Rev. E 93 (2016) 012802.
\newblock \href {https://doi.org/10.1103/PhysRevE.93.012802}
  {\path{doi:10.1103/PhysRevE.93.012802}}.

\bibitem{Hester_etal_ProcRSA2020}
E.~W. Hester, L.-A. Couston, B.~Favier, K.~J. Burns, G.~M. Vasil, Improved
  phase-field models of melting and dissolution in multi-component flows,
  Proceedings of the Royal Society A: Mathematical, Physical and Engineering
  Sciences 476~(2242) (2020) 20200508.
\newblock \href {https://doi.org/10.1098/rspa.2020.0508}
  {\path{doi:10.1098/rspa.2020.0508}}.

\bibitem{Jamet-Misbah_PRE2008}
D.~Jamet, C.~Misbah, {Thermodynamically consistent picture of the phase-field
  model of vesicles: Elimination of the surface tension}, Phys. Rev. E 78
  (2008) 041903.
\newblock \href {https://doi.org/10.1103/PhysRevE.78.041903}
  {\path{doi:10.1103/PhysRevE.78.041903}}.

\bibitem{Cartalade_etal_CAMWA2016}
A.~Cartalade, A.~Younsi, M.~Plapp, {Lattice Boltzmann simulations of 3D crystal
  growth: Numerical schemes for a phase-field model with anti-trapping
  current}, Computers \& Mathematics with Applications 71~(9) (2016)
  1784--1798.
\newblock \href {https://doi.org/10.1016/j.camwa.2016.02.029}
  {\path{doi:10.1016/j.camwa.2016.02.029}}.

\bibitem{Lee_Parasitic_CAMWA2009}
T.~Lee, {Effects of incompressibility on the elimination of parasitic currents
  in the lattice Boltzmann equation method for binary fluids}, Computers and
  Mathematics with Applications 58 (2009) pp. 987--994.
\newblock \href {https://doi.org/10.1016/j.camwa.2009.02.017}
  {\path{doi:10.1016/j.camwa.2009.02.017}}.

\bibitem{Lee-Liu_DropImpact_LBM_JCP2010}
T.~Lee, L.~Liu, {Lattice Boltzmann simulations of micron-scale drop impact on
  dry surfaces}, Journal of Computational Physics 229 (2010) 8045--8063.
\newblock \href {https://doi.org/10.1016/j.jcp.2010.07.007}
  {\path{doi:10.1016/j.jcp.2010.07.007}}.

\bibitem{Fakhari-Rahimian_PRE2010}
A.~Fakhari, M.~Rahimian, {Phase-field modeling by the method of lattice
  Boltzmann equations}, Physical Review E 81 (2010) 036707.

\bibitem{Hahn-Ozisik_2012}
D.~W. Hahn, M.~N. {\"O}zi{\c{s}}ik, {Heat Conduction}, John Wiley {\&} Sons,
  Inc., 2012.
\newblock \href {https://doi.org/10.1002/9781118411285}
  {\path{doi:10.1002/9781118411285}}.

\bibitem{Maugis_etal_ActaMater1997}
P.~Maugis, W.~Hopfe, J.~Morral, J.~Kirkaldy, Multiple interface velocity
  solutions for ternary biphase infinite diffusion couples, Acta Materialia
  45~(5) (1997) 1941--1954.
\newblock \href {https://doi.org/10.1016/S1359-6454(96)00321-7}
  {\path{doi:10.1016/S1359-6454(96)00321-7}}.

\bibitem{Nicoli-Plapp-Henry_PRE2011}
M.~Nicoli, M.~Plapp, H.~Henry, Tensorial mobilities for accurate solution of
  transport problems in models with diffuse interfaces, Phys. Rev. E 84 (2011)
  046707.
\newblock \href {https://doi.org/10.1103/PhysRevE.84.046707}
  {\path{doi:10.1103/PhysRevE.84.046707}}.

\end{thebibliography}

\end{document}